\documentclass[final,12p,times,twocolumn]{elsarticle}
\usepackage{lineno,hyperref}
\usepackage{graphicx}
\usepackage[linesnumbered,ruled,lined]{algorithm2e}
\usepackage{float}
\usepackage{tikz}
\usepackage{epstopdf}
\usepackage{amssymb}
\usepackage{amsmath}
\usepackage{caption}
\usepackage{subcaption}

\usetikzlibrary{arrows.meta}
\journal{Journal of Hydrodynamics}
\newcommand{\tr}{\mathop{\mathrm{tr}}}

\newcommand{\br}{\mathbf{r}}

\makeatletter
\renewcommand\paragraph{\@startsection{paragraph}{4}{\z@}%
	{-2.5ex\@plus -1ex \@minus -.25ex}%
	{1.25ex \@plus .25ex}%
	{\normalfont\normalsize\bfseries}}
\makeatother
\pdfoutput=1
\begin{document}
	\begin{frontmatter}
		\title{Review on Smoothed Particle Hydrodynamics: Methodology development and recent achievement}
		\author[myfirstaddress]{Chi Zhang }
		\ead{c.zhang@tum.de}
		\author[myfirstaddress,mysecondaryaddress]{Yujie Zhu}
		\ead{yujie.zhu@tum.de}
		\author[myfirstaddress]{Dong Wu}
		\ead{dong.wu@tum.de}
		\author[myfirstaddress]{Xiangyu Hu \corref{mycorrespondingauthor}}
		\ead{xiangyu.hu@tum.de}
		\address[myfirstaddress]{TUM School of Engineering and Design, Technical University of Munich, 85748 Garching, Germany}
		\address[mysecondaryaddress]{Xi'an research institute of Hi-Tech, 
			70025 Xi'an, China}
		\cortext[mycorrespondingauthor]{Corresponding author.}
		\begin{abstract}
			Since its inception, the full Lagrangian meshless
			smoothed particle hydrodynamics (SPH) method 
			has experienced a tremendous enhancement in methodology and 
			impacted a range of multi-physics applications in science and engineering. 
			The paper presents a concise review on latest developments and achievements of the SPH method, 
			including (1) brief review of theory and fundamental with kernel corrections, 
			(2) the Riemann-based SPH method with dissipation limiting and high-order data reconstruction by using MUSCL, WENO and MOOD schemes, 
			(3) particle neighbor searching with particle sorting and efficient dual-criteria time stepping schemes, 
			(4) total Lagrangian formulation with stablized, dynamics relaxation and hourglass control schemes, 
			(5) fluid-structure interaction scheme with interface treatments and multi-resolution discretizations, 
			(6) novel applications of particle relaxation for mesh and particle generations. 
			Last but not least, benchmark tests for validating computational accuracy, convergence, 
			robustness and efficiency are also supplied accordingly. 
		\end{abstract}
	\end{frontmatter}
%
%
\section{Introduction}\label{sec:introduction}
As a fully Lagrangian meshless method, 
whereby a set of particles are introduced to discretize the continuum media 
and their interactions determined by a Gaussian-like kernel function to approximate the mechanics, 
the smooth particle hydrodynamics (SPH) \cite{lucy1977numerical, gingold1977smoothed} 
has been demonstrated to be a compromising alternative of mesh-based methods and received significant interest in the past decades \cite{monaghan1992smoothed, randles1996smoothed, liu2003smoothed, liu2010smoothed}.
Thanks to its Lagrangian feature, 
the SPH method has shown peculiar advantages in handling free-surface flows \cite{monaghan1994simulating, ferrari2009new} 
involving violent impact and breaking events \cite{luo2021particle}, 
structure analysis with crack propagation and large deformation \cite{randles1996smoothed, bonet1998simple, monaghan2000sph} 
and multi-physics problems \cite{zhang2021sphinxsys} 
including fluid-structure interactions (FSI) \cite{antoci2007numerical, liu2019smoothed, zhang2017smoothed}, 
multi-phase flows \cite{colagrossi2003numerical, wang2016overview, rezavand2020weakly}, 
additive manufacturing \cite{russell2018numerical, meier2021novel} and cardiac modeling \cite{lluch2019breaking, zhang2020integrative, zhang2021multi}, 
and comprehensive reviews can be found in recent 
Refs. \cite{luo2021particle,liu2019smoothed, violeau2016smoothed, shadloo2016smoothed, ye2019smoothed, zhang2019smoothed, gotoh2021entirely}. 
From the methodological point of view, 
tremendous efforts have been devoted to address the improvement of 
convergence, consistency and stability, 
the treatment of boundary conditions, 
the adaptive discretization and extension to multi-physics applications,  
as highlighted by Refs. \cite{zhang2021sphinxsys, lind2020review, vacondio2021grand}

This paper aims at providing a concise description on the state-of-the-art 
methodology development and achievement for the SPH method.
In particular, 
attempts are devoted to address the perspectives of the Riemann-based SPH method with dissipation limiting and high-order data reconstruction,
efficient particle-interaction configuration updating, 
stablized scheme, dynamics relaxation and hourglass control for total Lagrangian formulation, 
novel applications for mesh and particle generations, 
and FSI schemes. 
This paper is organized as follows.
In Section \ref{sec:sph},
we will briefly summarize the theory and fundamental of the SPH method. 
In Section \ref{sec:fluid}, 
the traditional and Riemann-based SPH discretizations of the fluid dynamics is presented 
with special attention devoted to the Riemann solver with dissipation limiting, 
high-order reconstruction and efficient update of particle-interaction configuration. 
In Section \ref{sec:solid}, 
we will present the total Lagrangian SPH formulation with stablized scheme, 
dynamics relaxation and hourglass control.  
Section \ref{sec:fsi} reports the interface treatments and multi-resolution discretization 
for FSI. 
Section \ref{sec:generator} focuses on the novel applications of 
the particle relaxation for high-quality particle and unstructure mesh generations.
Finally, 
concluding remarks are given in Section \ref{sec:conclusion}. 
%
%
\section{SPH methodology}\label{sec:sph}
\subsection{Theory and fundamental of SPH}\label{sec:basicsph}
In SPH method, 
a set of Lagrangian particles whose interactions determined by a Gaussian-like kernel function
is introduced to discretize the continuum media.
Then, 
the particle-average based discretization of a variable field $f(\mathbf{r})$ 
can be defined as   
\begin{equation}\label{eq:kernel-approximation}
	f_i  = \int f(\br) W(\br_i - \br, h)d \br. 
\end{equation}
Here, $i$ is the particle index, $f_i$ the discretized particle-average variable and
$\br_{i}$ the particle position.
For the compact-support kernel function $W(\br_{i} - \br, h)$, 
$h$ is the smoothing length determining a radially symmetric support domain with respect to $\br_{i}$. 
As the mass of each particle $m_i$ is known and invariant (indicating mass conservation),
one has the particle volume $V_i = m_i/\rho_i$ with $\rho_i$ denoting the particle-average density.

By introducing particle summation, 
Eq. (\ref{eq:kernel-approximation}) can be approximated as
\begin{equation} 	\label{eq:particle-approximation}
	f(\br) \approx \sum_ j  V_j f_j W(\br - \br_j, h) = \sum_j  \frac{m_j}{\rho_j} f_j W(\br - \br_{j}, h) .
\end{equation}
Here, 
the summation is over all the neighboring particles $j$ located in the support domain of the particle of particle $i$.
Substituting the variable $f(\br)$ with density, 
one gets the approximation of the particle-average density 
\begin{equation}
	\rho_i \approx \sum_{j}  \frac{m_j}{\rho_j} W(\mathbf r_i - \mathbf r_j, h) \rho_j = \sum_j  m_j W_{ij}, 
	\label{eq:particle-density-reconstuction}
\end{equation}
which is an alternative way to write the continuity equation for updating the density.

Similarly, 
the approximation of the spatial derivative of $f(\br)$ can be derived as
\begin{equation}
\footnotesize
	\label{eq:gradsph}
	\begin{split}
		\nabla f(\br) & \approx \int_{\Omega} \nabla f (\br) W(\br_j - \br, h) dV  \\
		& =  - \int_{\Omega} f (\br) \nabla W(\br_j - \br, h) dV \approx  - \sum_j  V_j  f_j \nabla_j W(\br_j - \br, h) . 
	\end{split}
\end{equation}
Furthermore, 
Eq. (\ref{eq:gradsph}) can be rewritten into a strong form
\begin{equation}
	\label{eq:gradsph-strong}
	\nabla f_i = f_i \nabla 1 + \nabla f_i \approx   \sum_j V_j f_{ij} \nabla_i W_{ij}, 
\end{equation}
where the inter-particle difference value $f_{ij} = f_{i} - f_{j}$, 
and $\nabla_i W_{ij} = \mathbf{e}_{ij} \frac{\partial W_{ij}}{\partial r_{ij}}$ with $r_{ij}$ and $\mathbf{e}_{ij}$ 
are the distance and unit vector of the particle pair $(i,j)$, respectively.
The strong-form approximation of the derivative is used to determine the local structure of a field.
On the other hand, 
with a slight different modification, Eq. (\ref{eq:gradsph}) can be rewritten into a weak form as
\begin{equation}
	\label{eq:gradsph-weak}
	\nabla f_i = \nabla f_i - f_{i}\nabla 1 \approx   -2\sum_{j}  V_j \widetilde{f}_{ij} \nabla_i W_{ij}, 
\end{equation}
where the inter-particle average value $\widetilde{f}_{ij} = \left(f_{i} + f_{j}\right)/2$. 
The weak-form approximation of derivative is used to compute 
the surface integration with respect to a variable for solving its conservation law.
Thanks to its anti-symmetric property, 
i.e., 
$\nabla_i W_{ij} = - \nabla _j W_{ji}$, 
the momentum conservation of the particle system is implied.
\subsection{Kernel correction}\label{sec:highordersph}
When particle is close to boundary or particle distribution is irregular, 
the 0-order and 1st-order consistency of particle approximation of Eqs. \eqref{eq:particle-approximation} and \eqref{eq:gradsph}
are not satisfied, respectively.
To remedy this issue, 
a number of correction techniques has been proposed in the literature 
\cite{randles1996smoothed, liu2010smoothed, takeda1994numerical, johnson1996normalized, liu1995reproducing}. 
Here, 
we briefly review the two mostly applied techniques, 
i.e., 
kernel correction and kernel gradient correction which are also known as 
Shepard filter and renormalization formulation \cite{randles1996smoothed, vila2005sph}, 
respectively.
\paragraph{Kernel correction}
Following Refs. \cite{randles1996smoothed, takeda1994numerical, liu1995reproducing, bonet1999variational, zhu2021consistency}, 
the improved partition of unity, 0-order consistency, 
can be achieved by interpolating with a correction kernel $\widetilde W$ as 
\begin{equation}
	f(\br) = \sum_j V_j f_j \widetilde W(\br - \br_j) = \sum_j V_j f_j \alpha(\br) W_j(\br), 
\end{equation}
where the parameter $\alpha_i$ is evaluated by enforcing that any constant distribution is exactly interpolated, 
that is
\begin{equation}
C_0 = \sum_j V_j C_0 \widetilde W(\br - \br_j). 
\end{equation}
Therefore, 
the following condition must be satisfied with the corrected kernel
\begin{equation}
\sum_j V_j f_j \widetilde W(\br - \br_j) = 1, 
\end{equation}
and this gives 
\begin{equation} \label{eq:sph-shepard}
 \alpha(\br) = \frac{1}{\sum_j V_j W_j(\br)}.
\end{equation}
The scalar parameter of $\alpha(\br)$ is also known as Shepard filter \cite{randles1996smoothed, takeda1994numerical} 
and provides a much improved partition of unity, 
in particular for particles near the domain boundary \cite{bonet1999variational, zhu2021consistency}. 
\paragraph{Kernel gradient correction}
To assess the consistency order of the the particle approximation of Eq. \eqref{eq:gradsph}, 
we can Taylor-expand $f_j$ around $\mathbf r$
\begin{equation}
f_j \approx f(\mathbf r) + (\mathbf r_j - \mathbf r) \cdot \nabla f(\mathbf r) + O(\mathbf r^2) ,
\end{equation}
and substitute it to Eq. \eqref{eq:gradsph}
\begin{equation}
\footnotesize
	\nabla f(\br) \approx - \sum_j V_j \left\lbrace f(\mathbf r) + (\mathbf r_j - \mathbf r) \cdot \nabla f(\mathbf r) \right\rbrace \otimes \nabla_j W(\br_i - \br, h) .
\end{equation}
Accurate approximation of $\nabla f(\br)$ requires that 
\begin{equation} \label{eq:grad-unit}
\sum_j V_j  \nabla_j W(\br_i - \br, h) \approx 0, 
\end{equation}
and 
\begin{equation} \label{eq:grad-matrix}
\sum_j V_j \left( \mathbf r_j - \mathbf r\right)  \nabla_j W(\br_i - \br, h) \approx \mathbb I, 
\end{equation}
where $\mathbb I$ is the unit matrix. 
Eq. \eqref{eq:grad-unit} is an gradient expression of the partition of unity requirement. 
To satisfy Eq. \eqref{eq:grad-matrix}, 
a correction matrix $\mathbb B$ is introduced to modify the kernel gradient as 
\begin{equation}
	\widetilde \nabla_i W_{ij} = \mathbb B_i \nabla_i W_{ij},
\end{equation}
and inserted to Eq. \eqref{eq:grad-matrix} 
\begin{equation} \label{eq:grad-matrix-correction}
\sum_j V_j \left( \mathbf r_j - \mathbf r\right)  \mathbb B_i \nabla_i W_{ij} = \mathbb I. 
\end{equation}
This gives
\begin{equation} \label{eq:kernel-matrix-correction}
\mathbb B_i = \left( \sum_j V_j \left( \mathbf r_j - \mathbf r_i \right)  \nabla_i W_{ij} \right) ^{-1} .
\end{equation}
The use of the kernel gradient correction can improve the gradient approximation, 
in particular for irregular distributed particles \cite{randles1996smoothed, vila2005sph, vila1999particle}. 
However, 
extra computational efforts are induced due to the interpolation and matrix inverse 
for each particle at every time step. 
On the other hand, 
this kernel gradient correction is widely applied in the total Lagrangian formulation 
as it only to be calculated once at the initial reference configuration \cite{bonet2000correction, vignjevic2006sph, zhang2021integrative} 
which will be presented in details in Section \ref{sec:solid}.
%
%
\section{Fluid dynamics}\label{sec:fluid}
\subsection{Governing equations} \label{sec:fluidgoverning}
For inviscid flow, 
the conservation of mass and momentum in the Lagrangian frame can be written as
\begin{equation} 
	\begin{cases}\label{eq:fluid-governing}
		\frac{\text d \rho}{\text d t}  =  - \rho \nabla \cdot \mathbf v \\
		\frac{\text d \mathbf v}{\text d t}  =   - \frac{1}{\rho}\nabla p  + \mathbf g + \mathbf f^s
	\end{cases},
\end{equation}
where $\mathbf{v}$ is the velocity, $\rho$ the density, 
$p$ the pressure, $\mathbf g$ the gravity, 
$\frac{\text d}{\text d t}=\frac{\partial}{\partial t} + \mathbf v \cdot \nabla$ stands for the material derivative 
and $\mathbf f^s$ presents the force exerting on the fluid due to the existence of 
solid, 
which can be the solid wall or flexible structure. 

To close the system of Eq. \eqref{eq:fluid-governing}, 
the pressure can be calculated with an artificial isothermal equation of state (EoS) in the form 
\begin{equation} \label{eq:fluid-eos}
	p = \frac{\rho^0 c^2}{\gamma} \left[\left(\frac{\rho}{\rho^0} \right)^\gamma -1 \right],
\end{equation}
where $\gamma = 1~\mathrm{or}~7$ is an empirically determined constant, 
$\rho^0$ the reference density and $c$ the speed of sound \cite{monaghan1994simulating, macdonald1966some}. 
Following the weakly-compressible assumption \cite{morris1997modeling}, 
an artificial sound speed of $c = 10 U_{max}$ , 
where $U_{max}$ denoting the maximum anticipated flow speed, 
is employed for density fluctuation to be approximately $1\%$, 
implying the Mach number $M \approx 0.1$. 
\subsection{Traditional SPH method} \label{sec:fluid-sph}
Having Eqs. \eqref{eq:gradsph-strong} and \eqref{eq:gradsph-weak}, 
the discretization of Eq. \eqref{eq:fluid-governing} in the traditional SPH formulation reads 
\cite{liu2010smoothed, vila1999particle, monaghan2012smoothed, hu2006multi} 
\begin{equation}\label{eq:fluid-sph}
\begin{cases}
	\frac{\mathrm d \rho_i}{\mathrm d t} = \rho_i \sum_j\frac{m_j}{\rho_j} \mathbf v_{ij}\cdot \nabla_i W_{ij} \\
	\frac{\mathrm d \mathbf v_i}{\mathrm d t}  = -\sum_j m_j \left( \frac{p_i + p_j}{\rho_i \rho_j}\right)  \nabla_i W_{ij}  + \mathbf g + \mathbf f^s
\end{cases}, 
\end{equation}
where $ \mathbf v_{ij} = \mathbf v_i -  \mathbf v_j$ is the relative velocity.
To dampen the pressure oscillation and prevent instability in the particle motion, 
where single particle moves in a rather chaotic way, 
Monaghan and Gingold \cite{monaghan1983shock} introduced a Neumann–Richtmeyer type artificial viscosity term 
\begin{equation}\label{eq:artificialvosicosity}
\begin{split}
\Pi_i =  -\sum_j m_j \alpha \frac{h \overline c}{\overline\rho} \frac{\mathbf v_{ij} \cdot \mathbf r_{ij} }{r_{ij}^2}\nabla_{i} W_{ij}, 
\end{split} 
\end{equation}
where $\overline c \left( c_i + c_j\right) /2$, 
$\overline \rho = \left( \rho_i + \rho_j\right) /2$ 
and $\alpha \leq 1.0$ is a tunable parameter. 
While a moderate artificial viscosity is able to stabilize the computation, it may lead to
excessive dissipation which affects the physical flow characteristics \cite{vila1999particle, ferrari2008new}. 
Another weakness is that the tunable parameter 
$\alpha$ requires careful numerical calibrations 
and its values usually are case dependent \cite{ferrari2009new}.
\subsubsection{Density reinitialization} \label{sec:fluid-sph-rhoinit}
Implementing density reinitialization, 
i.e., the density is integrated by the continuity equation and periodically reinitialized by applying proper formulation, 
is an efficient way to address the high frequency density oscillations. 
The straight forward formulation reads \cite{randles1996smoothed, colagrossi2003numerical, ren2015nonlinear}
\begin{equation} \label{eq:rho-reinit-shepard}
\rho_i = \frac{\sum_j m_j W_{ij}}{\sum_j V_j W_{ij}} ,
\end{equation}
by using the Shepard filter \cite{randles1996smoothed} of Eq. \eqref{eq:sph-shepard}, 
resulting first-order accuracy. 
Colagrossi and Landrini \cite{colagrossi2003numerical} suggested to consider a mean-least-squares (MLS) kernel interpolation
\begin{equation} \label{eq:rho-reinit-mls}
\rho_i = \sum_j m_j W_j^{MLS}\left( \mathbf r_j\right) ,
\end{equation}
where $W_j^{MLS}$ is the MLS kernel \cite{colagrossi2003numerical} given by 
\begin{equation} \label{eq:msl_kernel}
	\begin{cases}
	W^{MLS}\left(\mathbf r_j \right) = \mathbf M_i^{-1}\mathbf e_1 \cdot \mathbf b_{ij} W\left(\mathbf r_j \right)  \\
	\mathbf b^T_{ij} = \left[ 1, (x_j - x_i), (y_j - y_i), (z_j - z_i)\right]; \mathbf e_1^T = [1,0,0] \\
	\mathbf M_j = \sum_j \mathbf b_{ij} \otimes \mathbf b_{ij} W(\mathbf r_j) dV_j \\
	\end{cases} .
\end{equation}
This formulation achieves second-order accuracy and shows good results while is computationally rather expensive \cite{antuono2012numerical}. 

More recently, 
Zhang et al. \cite{zhang2020dual} and 
Rezavand et al. \cite{rezavand2021generalised} 
proposed new density reinitialization formulations read
\begin{equation} \label{eq:rho-reinit-freesurface}
\rho_i = \rho^0 \frac{ \sum W_{ij} }{\sum W^0_{ij} } +  \max(0, ~\rho^* - \rho^0 \frac{ \sum W_{ij} }{\sum W^0_{ij} }) \frac{\rho^0}{\rho^*},
\end{equation}
and 
\begin{equation} \label{eq:rho-reinit-internal} 
\rho_i =  \rho^0 \frac{ \sum W_{ij}}{\sum W^0_{ij} } , 
\end{equation}
for free-surface and internal flow, respectively.
Here, 
$\rho^*$ denotes the density before reinitialization and superscript $0$ represents the initial reference value. 
Note that the density reinitialization is applied every $n$-time steps, 
for example $n = 20 $ in Refs.\cite{colagrossi2003numerical, ren2015nonlinear} 
and Zhang et al. \cite{zhang2020dual} apply it every advection time step 
which consists of several acoustic time steps for particle relaxation.  
\subsubsection{Diffusive term in the continuity equation} \label{sec:fluid-sph-rho-diffusion} 
Another approach to reduce the density oscillation is to introduce 
a diffusive term into the continuity equation of Eq. \ref{eq:fluid-sph}, 
resulting smooth pressure field and stable time integration. 
Inspired by the Riemann-based SPH method \cite{vila1999particle, ben2000convergence}, 
Ferrari et al. \cite{ferrari2009new} modified the original SPH discretization of the continuity equation as
\begin{equation}\label{eq:fluid-sph-diffusivedensity}
	\frac{\mathrm d \rho_i}{\mathrm d t} = \rho_i \sum_j\frac{m_j}{\rho_j} \mathbf v_{ij}\cdot \nabla_i W_{ij} + \mathfrak{D}_i ,
\end{equation} 
by introducing a Rusanov diffusive term $\mathfrak{D}_i $ defined as
\begin{equation}\label{eq:fluid-sph-diffusivedensity-farari}
	\mathfrak{D}_i  = - \max\left(c_i, c_j \right)  \sum_{j} \frac{m_j}{\rho_j} \left(\rho_i - \rho_j \right) \frac{\partial W_{ij}}{\partial r_{ij}} .
\end{equation} 
As highlighted by Ref. \cite{ferrari2009new}, 
the Rusanov diffusive term $\mathfrak{D}_i$ is independent of any tunable parameter 
and no artificial viscosity of Eq. \eqref{eq:artificialvosicosity} 
is required in the momentum equation. 
However, 
this scheme is not compatible with the hydrostatic solution, 
exhibiting unphysical free-surface motion and expansion in long term simulations due to the inconsistency 
induced by the singularity of the density approximation at the surface \cite{antuono2012numerical, cercos2016diffusive}.

Molten and Colagrossi \cite{molteni2009simple} pursued a similar idea of introducing diffusive term
but it still suffers the incompatibility with the hydrostatic solution \cite{antuono2010free}. 
To decrease such artifacts, 
Antuono et al. \cite{antuono2012numerical} proposed an improvement and the diffusive term reads
\begin{equation}\label{eq:fluid-sph-diffusivedensity-delta}
\mathfrak{D}_i  = - \delta h c  \sum_{j} \frac{m_j}{\rho_j} \left(2\left( \rho_i - \rho_j \right)\frac{\mathbf r_{ij}}{r_{ij}^2 } + \mathfrak F_{ij} \right)  \cdot \nabla_i W_{ij} ,
\end{equation} 
where $\delta$ is a non-dimensional parameter, 
and $\mathfrak F_{ij} = \widetilde \nabla \rho_i + \widetilde \nabla \rho_j$  
is a renormalized correction term to prevent the singularity at the surface \cite{marrone2011delta}. 
This corrected $\delta$-term is compatible with the hydrostatic solution, 
whereas induces extra computational efforts due to the correction term of $\mathfrak F_{ij}$.
Note that the parameter $\delta$ is not freely tunale and usually set as $0.1$ \cite{marrone2011delta}, 
and a small amount of artificial viscosity defined by Eq. \eqref{eq:artificialvosicosity} 
with parameter of $\alpha = 0.02$ is still applied for numerical stability. 
\subsection{Riemann-based SPH method} \label{sec:fluid-riesph}
As a variant of the SPH method, 
the Riemann-based SPH method solves a one-dimensional Riemann problem along each particle pair to determine its interaction.
Compared with the traditional SPH method presented in Section \ref{sec:fluid-sph}, 
the Riemann-based SPH method introduces implicit numerical dissipation other than use explicit artificial viscosity, 
and achieves the dissipation in a more accurate manner \cite{vila1999particle, monaghan1997sph, moussa2006convergence, rafiee2012comparative}. 
Monaghan \cite{monaghan1997sph} pointed out that the artificial viscosity \cite{monaghan1983shock} 
is analogous to the dissipative terms of the Riemann solver, 
which scales with the wave speed and the velocity jump between interacting particles, 
whereas showed that using the exact, or well approximated, Riemann solution can obtain more accurate results in capturing shock wave. 
The pioneering work of developing Riemann-based SPH method can be tracked back to Vila \cite{vila1999particle}, 
where a Riemann-based ALE-SPH scheme was proposed 
by discretizing the Euler equations in conservative form and
calculating the fluxes between particles with a Riemann solver. 
It is worth noting that the particle in the Riemann-based ALE-SPH scheme represents 
a volume of the considered discretized fluid medium 
and it may move with the fluid velocity (Lagrangian description), 
remain still (Eulerian description) or move in any arbitrary way. 
Instead of following the ALE description, 
Parshikov et al. \cite{parshikov2000improvements} and Parshikov and Stanislav \cite{parshikov2002smoothed} 
proposed a Riemann-based SPH formulation in purely Lagrangian framework by 
using a first-order Riemann solution to describe the contact interaction between particles. 
To improve the accuracy of capturing strong shocks, 
Inutsuka \cite{inutsuka2002reformulation} has reformulated the Riemann-based SPH formulation with second-order Riemann solution.
Then, 
Cha and Whitworth \cite{cha2003implementations} derived different versions of Riemann-based SPH schemes, 
and performed a von Neumann stability analysis and concluded that the Riemann-based SPH is stable for all wavelengths, 
while the traditional SPH is unstable for certain wavelengths. 
Subsequently, 
enormous progress has been made toward accomplishing 
high-order data reconstruction \cite{avesani2014new, zhang2019weakly, wang2021new, avesani2021alternative}, 
dissipation limiting \cite{zhang2017weakly, meng2020multiphase} and 
solid boundary treatment \cite{marongiu2010free, zhang2017weakly}, 
which will be reviewed in the following parts. 

To derive the Riemann-based SPH discretization of Eq. \ref{eq:fluid-governing}, 
we rewrite its traditional SPH discretization as 
\begin{equation}\label{eq:fluid-sph-averagevariables}
\begin{cases}
\frac{\mathrm{d} \rho_i}{\mathrm{d} t}  = 2\rho_i \sum_j\frac{m_j}{\rho_j} \left(\mathbf v_{i} - \overline{\mathbf v}_{ij}\right) \cdot \nabla_{i} W_{ij} \\
\frac{\mathrm{d} \mathbf{v}_i}{\mathrm{d} t} = - 2\sum_j  m_j\left( \frac{\overline p_{ij}}{\rho_i \rho_j}\right)   \nabla_i W_{ij} + \mathbf g + \mathbf f^s 
\end{cases}, 
\end{equation}
by introducing the inter-particle average velocity 
$\overline{\mathbf v}_{ij} =\left( \mathbf v_i +  \mathbf v_j\right) /2$ and pressure
$\overline{p}_{ij} = \left(  p_i + p_j\right) /2$. 
Then, 
the inter-particle average variables are replaced by the Riemann solution, 
i.e., $U^{\ast} $ and $P^{\ast}$, resulting
\begin{equation}\label{eq:fluid-riesph}
\begin{cases}
\frac{\mathrm{d} \rho_i}{\mathrm{d} t}  = 2 \rho_i \sum_j \frac{m_j}{\rho_j} \left(\mathbf v_i - \mathbf v^{\ast}\right)  \cdot \nabla_{i} W_{ij} \\
\frac{\mathrm{d} \mathbf{v}_i}{\mathrm{d} t} =  - 2\sum_j m_j \left(\frac{ P^{\ast}}{\rho_i \rho_j} \right)  \nabla_i W_{ij}
\end{cases}, 
\end{equation}
where $\mathbf v^{\ast} = U^{\ast} \mathbf e_{ij} +  \left( \overline{\mathbf v}_{ij} - \overline U \mathbf e_{ij} \right)$. 
In this case, 
the inter-particle average variables are replaced by solutions of the Riemann problem, 
implying that numerical dissipation, 
i.e., density regularization and numerical viscosity, 
is implicitly present.

\begin{figure}[htb]
	\centering
	\includegraphics[trim = 1cm 2cm 1cm 1.5cm, clip, width=.49\textwidth]{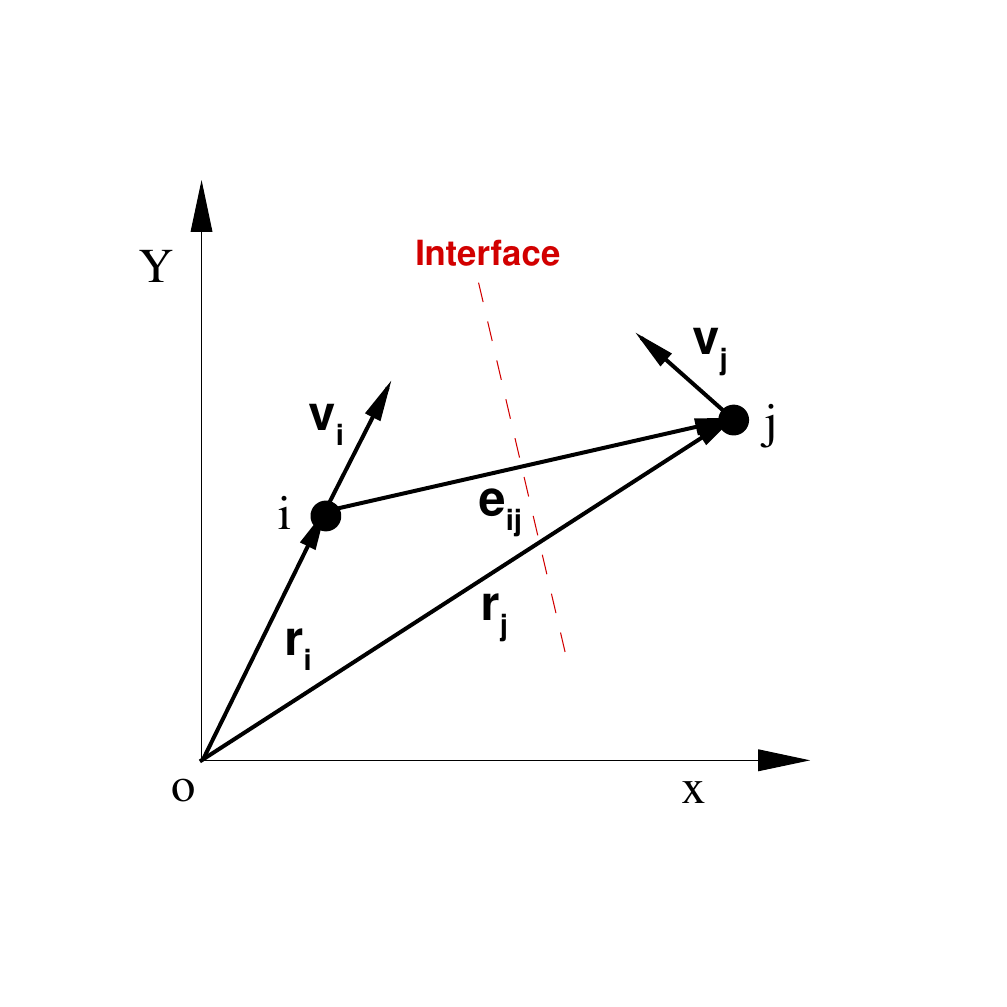}
	\caption{Construction of Riemann problem along the interacting line of particles $i$ and $j$. }
	\label{figs:particle}
\end{figure}

In Riemann-based SPH the solution to the Riemann problem 
is reduced to a one-dimensional problem constructed along the interaction line of particles. 
Then, 
the first step is to construct the $L$ and $R$ states between each pair of interacting particles. 
Following the Godunov-type method, 
which applies piece-wise constant assumption, 
i.e., 
first-order reconstruction, 
the $L$ and $R$ states are defined as
\begin{equation}\label{eqrievector}
\begin{cases}
\left( \rho_L, U_L, P_L\right)  = \left( \rho_i, \mathbf v_i \cdot \mathbf e_{ij}, p_i\right)      \\
\left( \rho_R, U_R, P_R\right)  = \left( \rho_j, \mathbf v_j \cdot \mathbf e_{ij}, p_j\right) 
\end{cases} .
\end{equation}
In this case, 
the initial Riemann left and right states are on particles $i$ and $j$, respectively,
and the discontinuity is at the middle point $\overline{\mathbf r}_{ij} = \frac{1}{2}\left( \mathbf r_i + \mathbf r_j\right)$, 
as shown in Figure \ref{figs:particle}. 
%
\subsubsection{Riemann solver with dissipation limiter}
Since its inception, 
the Riemann-based SPH method has been applied to solve strong shocks problems \cite{puri2014comparison, puri2014approximate, sirotkin2013smoothed}, 
solid mechanics problems \cite{parshikov2000improvements, parshikov2002smoothed, mehra2006high}, 
interface instability \cite{cha2010kelvin, borgani2012hydrodynamic} 
and magnetohydrodynamics (MHD) \cite{iwasaki2011smoothed} problems. 
However,  
it is generally too dissipative to reliably reproduce violent free-surface flows 
involving violent events such as impact and breaking \cite{vila2005sph, roubtsova2006sph, koukouvinis2013improved}.
To cope with the excessive dissipation introduced by directly applying a Riemann solver, 
Zhang et al. \cite{zhang2017weakly} proposed a simple low-dissipation limiter to the classic linearized Riemann solver \cite{toro1991linearized,toro2013riemann}, 
ensuring no or decreased numerical dissipation for expansion or compression waves, respectively. 
This method is compatible with the hydrostatic solution 
and able to resolve violent wave breaking and impact events accurately, 
produces very small damping of mechanical energy, smooth pressure fields and predicts reasonable pressure peaks.
Then, 
this method was extended by 
Rezavand et al. \cite{rezavand2020weakly} for modeling multiphase flow with high density ratio 
and Zhang et al. \cite{rezavand2020weakly} for FSI problems. 
Inspired by Ref. \cite{zhang2017weakly}, 
Meng et al. \cite{meng2020multiphase} proposed a dissipation limiter 
to Roe's approximated Riemann solver \cite{roe1981approximate, rider1994review} to develop 
a multiphase SPH model for complex interface flows. 
Furthermore, 
Meng et al. \cite{meng2020multiphase} derived the equivalent relation between the intrinsic dissipation of the Riemann solver 
and the Reynolds number for accurate modeling of viscous flow. 
\paragraph{Linearized Riemann solver} \label{paragraph:linearizedriesolver}
\begin{figure}[htb]
	\centering
	\includegraphics[trim = 1mm 1cm 1mm 1cm, clip, width=.49\textwidth]{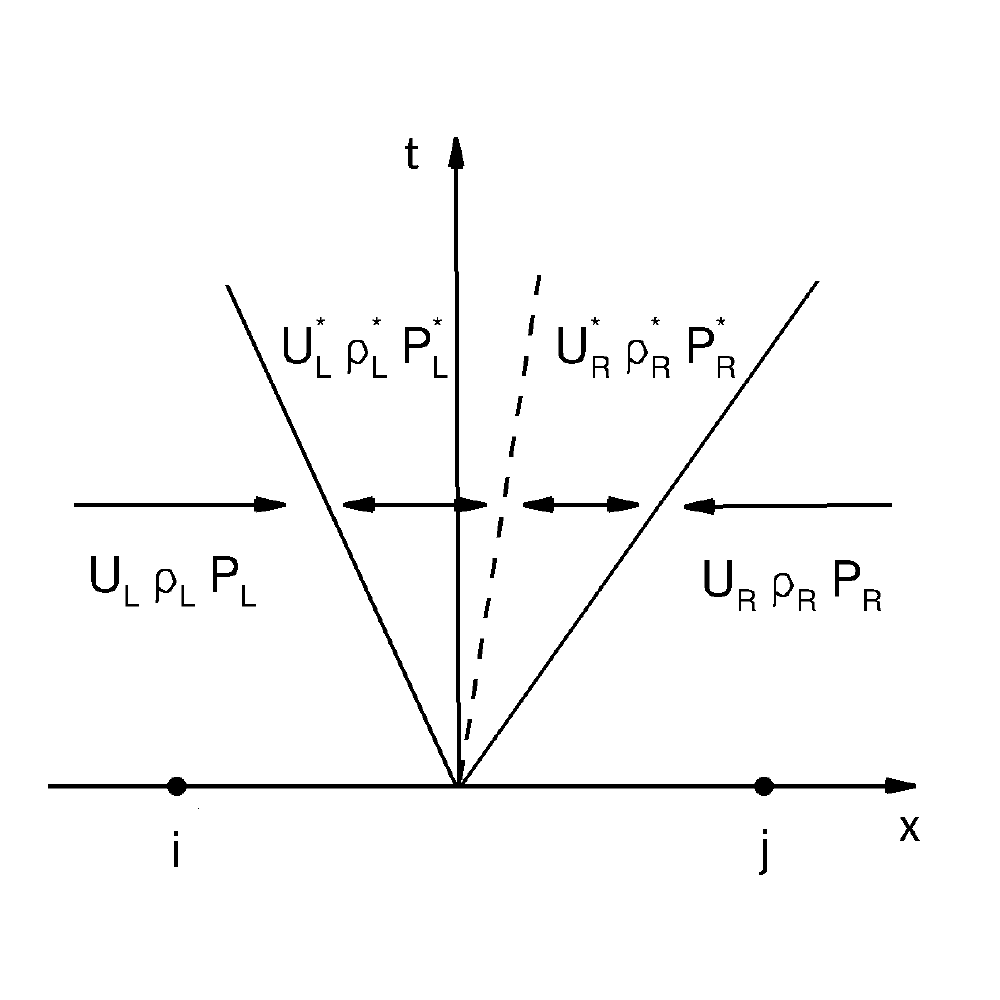}
	\caption{Simplified Riemann fan with two intermediate states.}
	\label{figs:riemann}
\end{figure}
According to Refs. \cite{toro2013riemann, toro2009riemann}, 
the solution of a Riemann problem results in three waves emanating from the discontinuity, 
denoted by $(\rho_L^{\ast}, U_L^{\ast},P_L^{\ast})$ and $(\rho_R^{\ast}, U_R^{\ast},P_R^{\ast})$
as shown in Figure \ref{figs:riemann}. 
Two waves, which can be shock or rarefaction wave,  travel with the smallest or largest wave speed. 
The middle wave is always a contact discontinuity and separates two intermediate states. 
By assuming that the intermediate state satisfies $U_L^{\ast} = U_R^{\ast} =U^{\ast}$ and $P_L^{\ast} = P_R^{\ast} =P^{\ast}$, 
a linearized Riemann solver \cite{toro2009riemann,toro1991linearized} for smooth flows or with only moderately strong shocks can be written as
\begin{equation}\label{eq:linearizedrie}
\begin{cases}
U^\ast = \frac{\rho_L c_L U_L + \rho_R c_R U_R + P_L - P_R}{\rho_L c_L + \rho_R c_R} \\
P^\ast = \frac{\rho_L c_L P_R + \rho_R c_R P_L + \rho_L c_L \rho_R c_R \beta(U_L - U_R)}{\rho_L c_L + \rho_R c_R}
\end{cases},
\end{equation}
where the limiter is defined as
\begin{equation} \label{eq:rie-limiter}
\beta = \min \left(  \eta \max\left(  \frac{U_L - U_R}{c_{LR}}, 0\right) , 1.0 \right), 
\end{equation}
with $c_{LR} = \frac{\rho_L c_L + \rho_R c_R}{\rho_L + \rho_R}$. 
Here, 
the dissipation limiter $\beta$ ensures that there is no dissipation when the fluid is under the action of an expansion wave, 
i.e. $U_L < U_R$,
and that the parameter $\eta$ is used to modulate dissipation when the fluid is under the action of a compression wave, 
i.e. $U_L \geq U_R$. 
In Ref. \cite{zhang2017weakly}, 
constant parameter $\eta = 3$ is suggested according to numerical experiments. 
Meng et al. \cite{meng2020multiphase} presented that the relation between $\eta$ and the parameter $\alpha$ 
in artificial viscosity of Eq.\eqref{eq:artificialvosicosity} is given by 
\begin{equation} \label{eq:limiter-artificialviscosity}
\eta = \frac{\alpha h \left(c_L + c_R \right) }{2\left(U_L - U_R \right) \left|\mathbf r_{ij} \right| }. 
\end{equation}
Having the equivalent relation of the artificial viscosity and the physical kinematics viscosity 
\cite{parshikov2000improvements, adami2012generalized}, 
Meng et al. \cite{meng2020multiphase} derived the relation between the parameter $\eta$ and the Reynolds number as 
\begin{equation} \label{eq:limiter-re}
\eta = \frac{2 \left(d + 2 \right) U_c L_c }{Re\left(U_L - U_R \right) \left|\mathbf r_{ij} \right| }, 
\end{equation}
where $d$ is the dimension, 
$Re$ the Reynolds number, 
$U_c$ and $L_c$ are the characteristic velocity and length, respectively. 

\begin{figure*}[htb]
	\centering
	\includegraphics[width=0.6\textwidth]{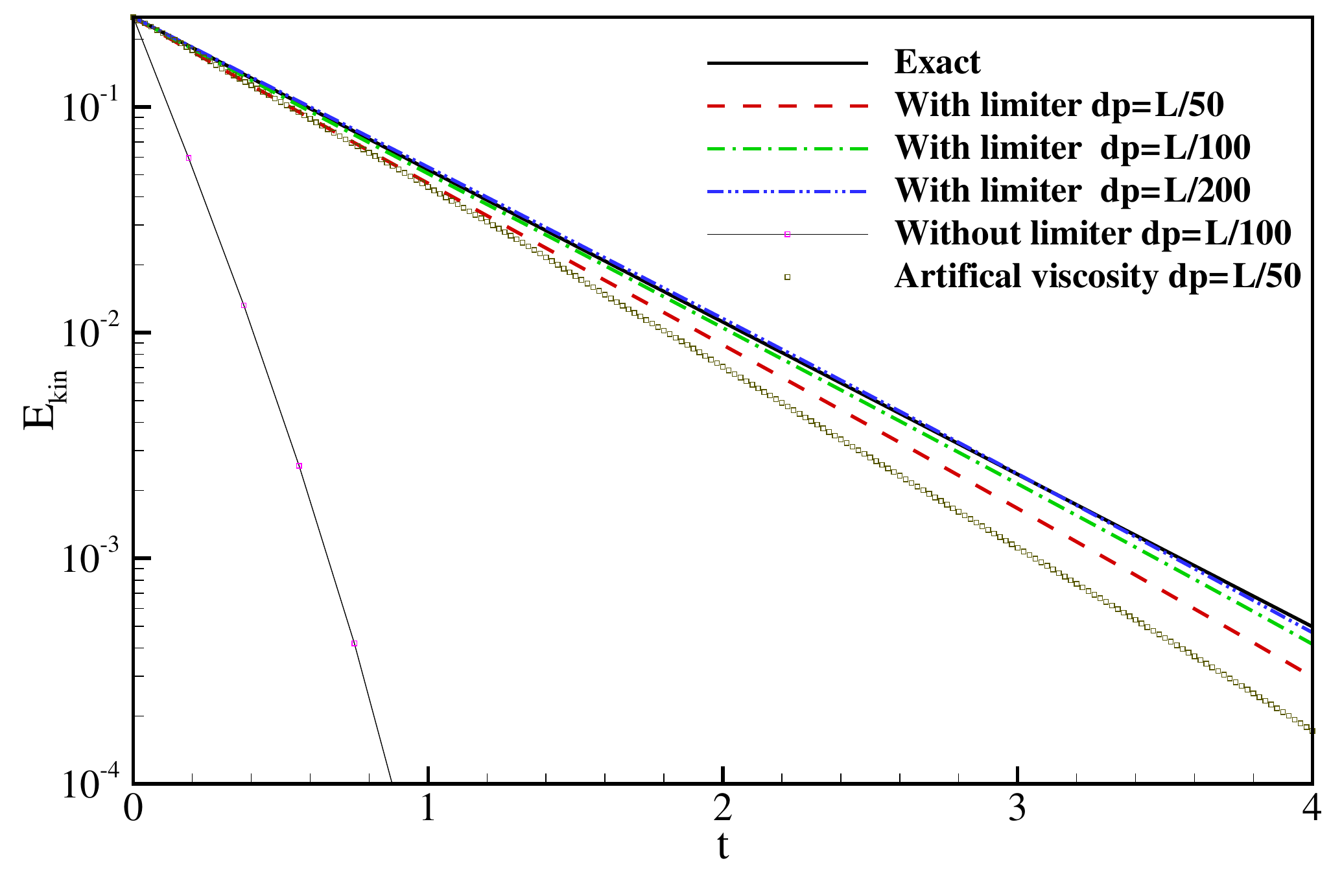}
	\caption{Taylor-Green vortex flow with $Re=100$: 
		Decay of the kinetic energy by using Riemann-based SPH method with the Linearized Riemann solver with and without the 
			dissipation limiter, and the traditional SPH method with artificial viscosity ($\alpha = 0.02$).}
	\label{figs:tgv-kin}
\end{figure*}

\begin{figure*}
	\centering
	\begin{subfigure}[b]{\textwidth}
		\includegraphics[trim = 1mm 1mm 1mm 1mm, clip,width=0.495\textwidth]{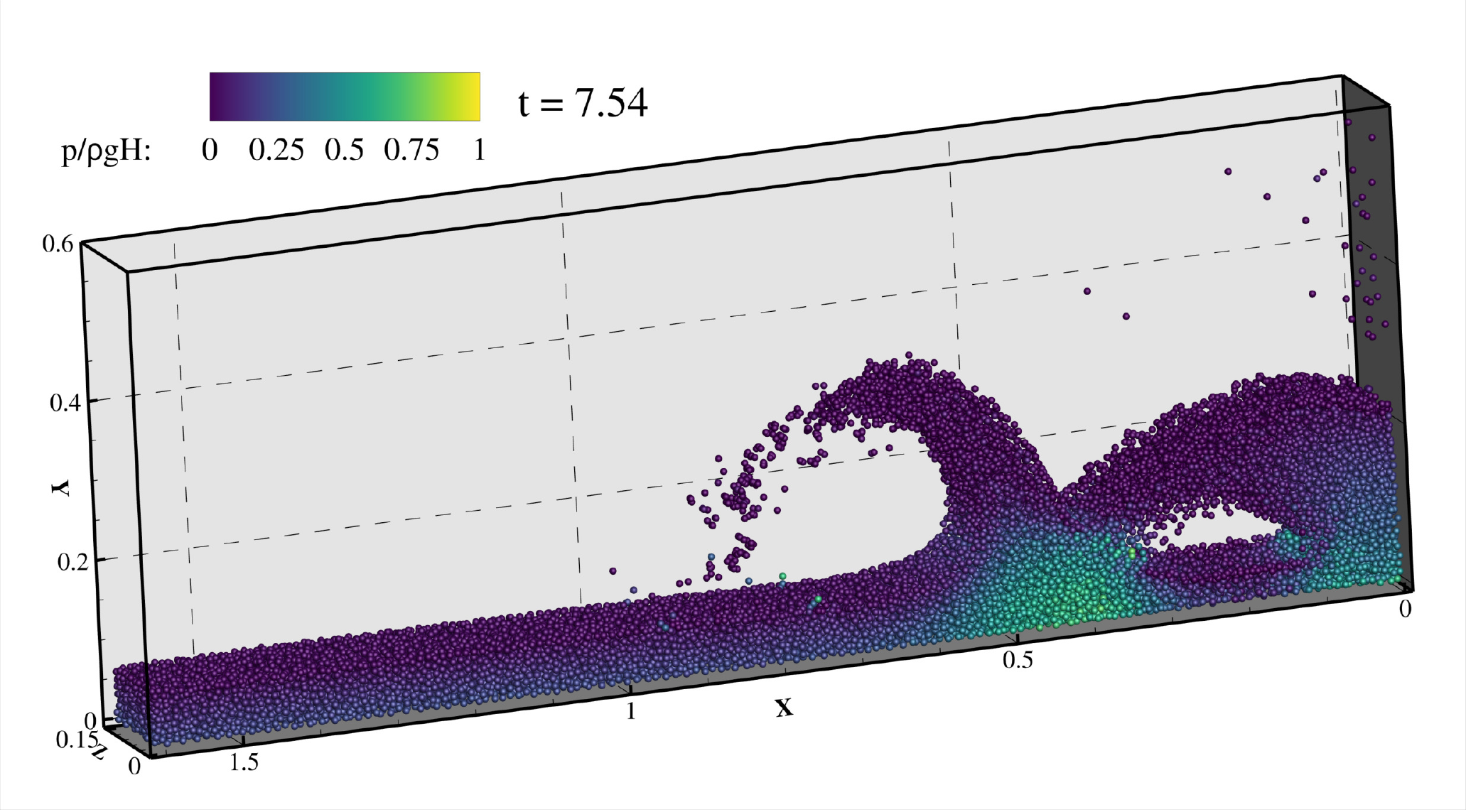}
		\includegraphics[trim = 0mm 0mm 0mm 0mm, clip,width=0.495\textwidth]{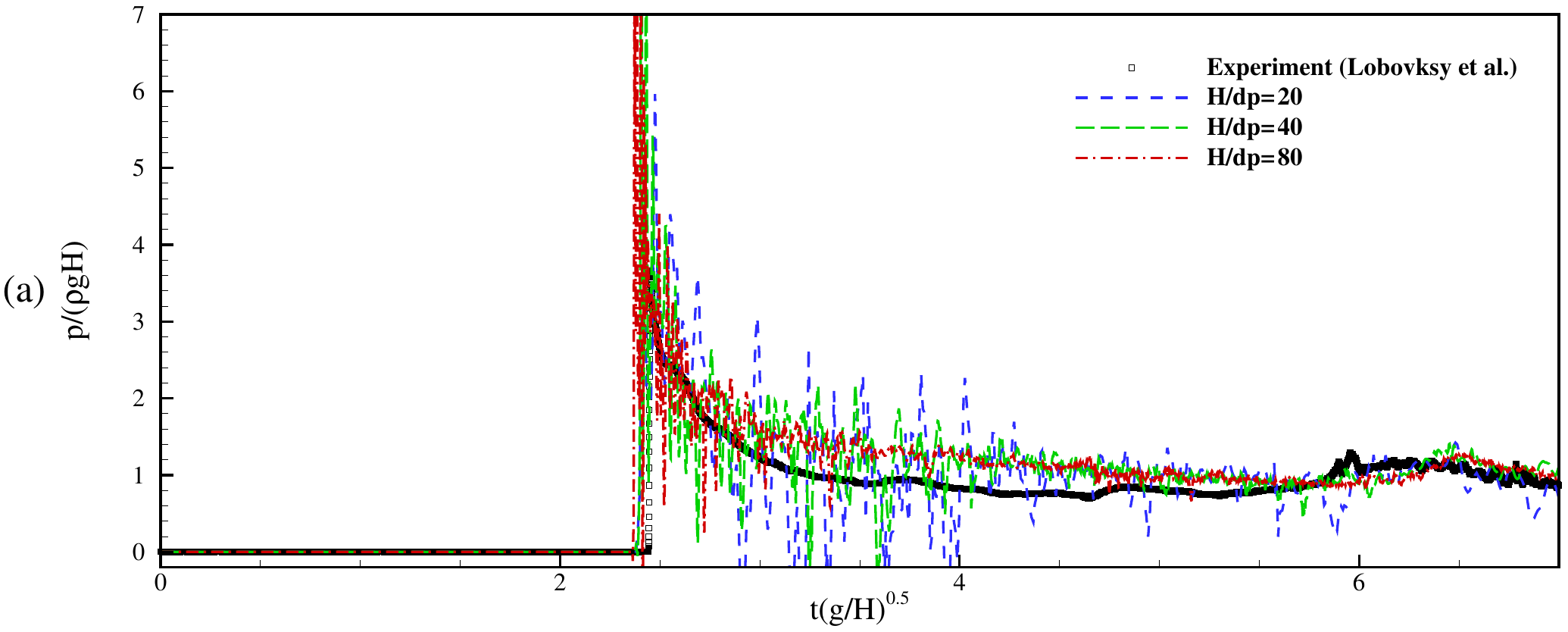}
		\caption{Dam-break flow.}
		\label{figs:sub-dam}
	\end{subfigure}
	\newline
		\begin{subfigure}[b]{\textwidth}
		\centering
		\includegraphics[trim = 0mm 1mm 1mm 0mm, clip,width=0.495\textwidth]{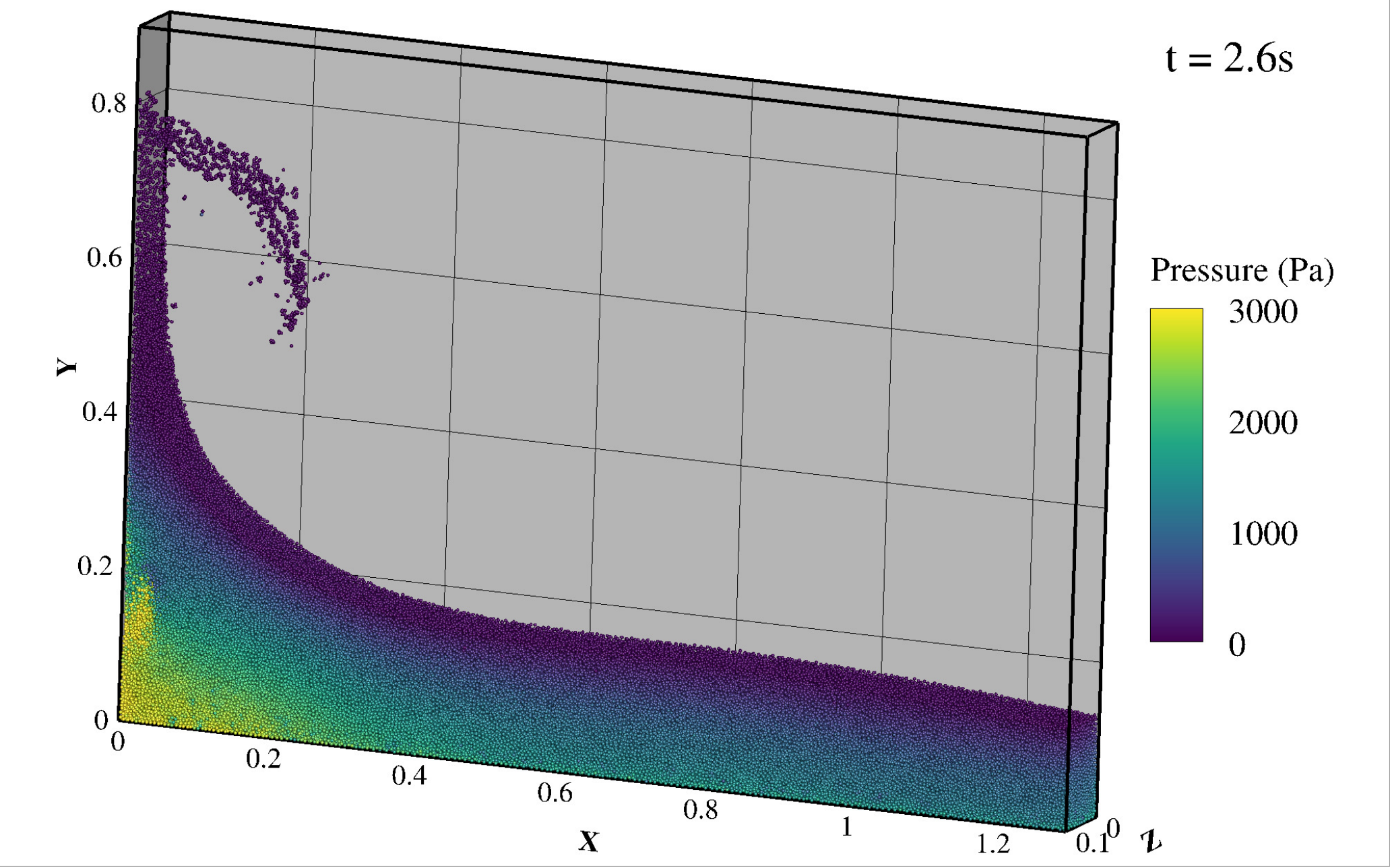}
		\includegraphics[trim = 8mm 0mm 0mm 0mm, clip,width=0.495\textwidth]{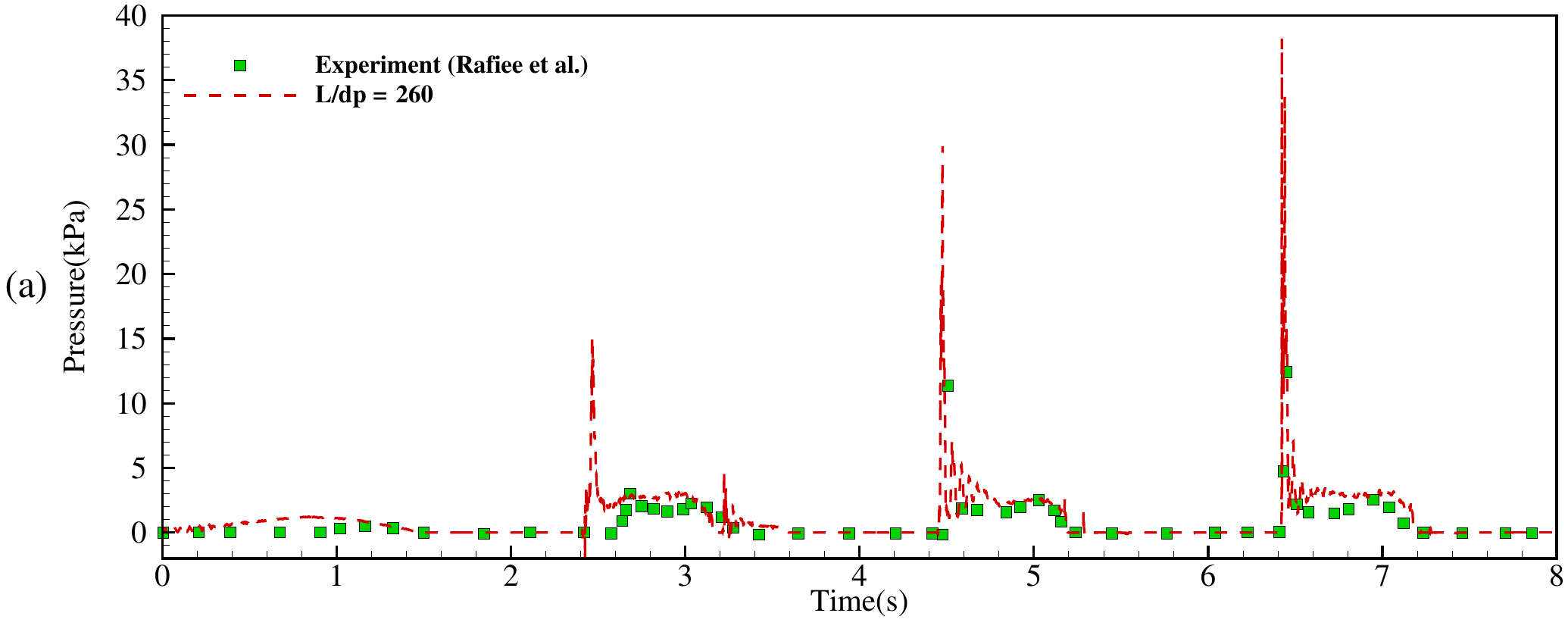}
		\caption{Sloshing flow.}
		\label{figs:sub-sloshing}
	\end{subfigure}
	\caption{Numerical modeling of free-surface flows by using the Riemann-based SPH method 
		with linearized Riemann solver and dissipation limiter:
		Snapshots of particle distribution with pressure contour (left panel) and 
		the time history of the numerical and experimental pressure signals (right panel) 
		for dam-break (a) and sloshing (b) flows.}
	\label{figs:freesurface}
\end{figure*}

Figure \ref{figs:tgv-kin} presents the validation of the linearized Riemann solver with dissipation limiter 
by studying the Taylor-Green vortex flow \cite{zhang2017weakly}. 
Without the dissipation limiter, 
the Riemann-based SPH is too dissipative to predict a reasonable kinetic energy decay. 
While with the limiter, 
low-dissipation feature and better agreement with the analytical solution 
are achieved in comparison with the traditional SPH method with the artificial viscosity ($\alpha = 0.02$) scheme.  
Also, 
the Riemann-based SPH achieves 2nd-order convergence for the total kinetic energy with increasing particle resolution. 
The performance of the Riemann-based SPH with dissipation limiter for modeling 
free-surface flows exhibiting violent events such as impact and breaking 
is addressed in Figure \ref{figs:freesurface} where dam-break and sloshing flows 
are investigated by comparing with experimental data. 
Both tests demonstrate that the Riemann-based SPH can accurately predict the violent motion of free surface and meanwhile 
capture the impacting pressure reasonably. 
%
\paragraph{HLLC Riemann solver}\label{paragraph:hllcriesolver}
The HLLC Riemann solver proposed by Toro \cite{toro2009riemann} has been applied in Riemann-based SPH method 
for capturing strong shock waves \cite{puri2014comparison, puri2014approximate} and energetic flows \cite{rafiee2012comparative}, 
exhibiting excessive numerical dissipation with the  piece-wise constant assumption \cite{rafiee2012comparative}. 
Following Ref. \cite{toro2013riemann}, 
the wave speeds estimate respectively from Left $(L)$ and Right $(R)$ regions, $s_L$ and $s_R$, are
\begin{equation}
s_L = U_L - c_L, \quad s_R = U_R + c_R,
\end{equation}
and then the intermediate wave speed $s^{\ast}$ is then calculated as
\begin{equation}
\footnotesize
s^{\star}=\frac{P_R-P_L+\rho_L U_L\left( s_L-U_L\right)  - \rho_R U_R\left( s_R-U_R\right) }{\rho_L(s_L-U_L)-\rho_R\left( s_R-U_R\right)} .
\end{equation}
Subsequently, 
the intermediate states of pressure can be obtained accordingly by 
\begin{equation}
	P^\ast_K= P_K +\rho_K \left(U_K -s_K \right) \left(U_K-s^\ast\right),\quad  K = R, L. 
\end{equation}
Finally, the HLLC solution to the Riemann problem  is then expressed by
\begin{equation}
	F^{hllc} = 
	\left\{ \begin{array}{ll}
		F_L,  &  0 \le s_L\\
		F^\star_L,  &   s_L\le 0 \le s^{\ast}\\
		F^\star_R, &  s^{\ast}\le 0 \le s_R\\
		F_R,  &   0 \ge s_L
	\end{array}  \quad K = L, R \right. 
\end{equation}
\subsubsection{High-order reconstruction}
Another key feature of the Riemann-based SPH is the possibility of implementing 
high-order data reconstruction \cite{van1977towards, van1979towards, barth1989design, harten1987uniformly, shu1998essentially, pirozzoli2011numerical}, 
which is widely applied in Eulerian Godunov-method \cite{shu1998essentially, pirozzoli2011numerical, shu2020essentially}
to decrease the dissipation and improve the accuracy \cite{vila1999particle, vila2005sph}. 
Vila \cite{vila2005sph} first introduced the MUSCL scheme for second-order data reconstruction into the Riemann-base ALE-SPH method. 
Similar approach was developed by Inutsuka et al. \cite{inutsuka2002reformulation} for reformulating a second-order Riemann-based SPH method.  
Since then, 
more attempts have been aimed at implementing different limiting functions for MUSCL scheme, 
e.g., van Leer limiter \cite{iwasaki2011smoothed, murante2011hydrodynamic}, 
SuperBee limiter \cite{koukouvinis2013improved, rafiee2012comparative, rogers2010simulation}
and Barth-Jespersen-type limiter \cite{hopkins2015new},
to reduce the numerical dissipation and increase spatial order of Riemann-based SPH method. 
More recently, 
increasing attentions are drawn to implement WENO scheme \cite{shu1998essentially, shu2020essentially} 
ant its variants \cite{hu2010adaptive, fu2016family, wang2018wenois}.  
Zhang et al. \cite{zhang2013contact} have considered a fifth-order WENO reconstruction 
for computing one-dimensional problems,
however, its multidimensional extension is not straightforward. 
The first WENO reconstruction for computing multi-dimensional problems is proposed 
by Avesani et al. \cite{avesani2014new},
in which the directionally-biased multi-dimensional candidate stencils 
with high-order Moving-Least-Squares (MLS) reconstructions
are combined with the WENO weighting strategy.
Although this method achieves higher accuracy than those using linear reconstructions, 
it exhibits much lower computational efficiency due to 
a large number of multi-dimensional candidate-stencil evaluations.
Nogueira et al. \cite{nogueira2016high} proposed 
a SPH-MOOD-MLS method 
which uses a MLS-based approximation 
and  a posteriori Multidimensional Optimal Order Detection (MOOD) 
approach for numerical stability. 
This method shows considerable improvement for  
modeling compressible flows with shock and blast waves. 
Different with Ref. \cite{avesani2014new}, 
Zhang et al. \cite{zhang2019weakly} proposed an efficient one-dimensional 4-point stencil incremental stencil WENO (IS-WENO) reconstruction \cite{wang2021new}
along the interaction line of each particle pair, 
where the variable calculation of the missing points is based on SPH derivative approximation as that in MUSCL scheme, 
for the Riemann-based SPH method to increase accuracy by decreasing the numerical dissipation 
other than increasing the formal approximation order. 
This method preserves the capability of producing smooth and accurate pressure fields of the original method 
and now achieves also very small numerical dissipation.
Similar with Ref. \cite{zhang2019weakly}, 
Meng et al. \cite{meng2021targeted} developed 5-point stencil WENO reconstruction \cite{shu1998essentially, shu2020essentially}
along the particle interacting line, while the missing variables are evaluated 
by firstly searching their nearest fluid particles 
and then adopting the first-order Taylor expansion. 
The tests showed that this method is robust and able to accurately capture shockwaves. 
Benefiting from the low-dissipation property, 
it also has a good performance in resolving small-scale structures in flows.
Similar with Refs. \cite{zhang2019weakly, wang2021new}, 
Meng et al. \cite{meng2021targeted} implemented the TENO scheme \cite{fu2016family} 
to capture the shocks and small-scale structures in some compressible flows, 
and obtain superior accuracy in some incompressible vortex flows and free surface flows. 
\paragraph{MUSCL reconstruction} \label{paragraph:muscl} 
The MUSCL scheme was developed by van Leer \cite{van1977towards, van1979towards} 
to replace the piecewise constant approximation of Godunov's scheme 
by reconstructing left and right states with piecewise linar approximations to calculate fluxes in the Eulerian methods \cite{toro2013riemann}. 
In MUSCL scheme, 
the spatial derivatives of field variable are used for data reconstruction, 
while direct usage results unstable scheme due to spurious oscillations near high gradients \cite{toro2013riemann}. 
To remedy the spurious oscillation, 
slope limiter or flux limiter, 
is applied to limit the approximated gradient near shocks or discontinuities. 

Similar as in Eulerian method, 
the reconstructed, limited left and right states are used as input 
to the Riemann solver in the Riemann-based SPH method \cite{rafiee2012comparative, inutsuka2002reformulation}, 
or to obtain the fluxes in the Riemann-based ALE-SPH \cite{vila1999particle, hopkins2015new}. 
With the piecewise linear approximation, 
the left and right states of the Riemann problem are reconstructed from 
\begin{equation}\label{eq:riesph-muscl}
	\begin{cases}
		\Phi_L = \Phi_i + \frac{1}{2} \phi (\lambda_i ) \nabla \Phi _i \cdot \mathbf r_{ji} \\
		\Phi_R = \Phi_j - \frac{1}{2} \phi (\lambda_j ) \nabla \Phi _j \cdot \mathbf r_{ij} 
	\end{cases},
\end{equation}
where $\phi$ is the limiting function and $\lambda$ the ratio of successive gradients defined as 
$\lambda_i = \frac{\Phi_i - \nabla\Phi_i \cdot \mathbf r_{ij}}{\Phi_j - \Phi_i}$ 
and $\lambda_j = \frac{\Phi_j - \Phi_i}{\nabla\Phi_j \cdot \mathbf r_{ji} - \Phi_j}$, respectively. 
Here, the $\nabla\Phi_i$ and $\nabla\Phi_j$ 
are the corresponding gradients calculated from the SPH approximation as
\begin{equation}\label{eq:sph-grad}
\nabla \Phi_i =  \sum_j \frac{m_j}{\rho_j} (\Phi_j - \Phi_i) \nabla_{i} W_{ij}, 
\end{equation}
or with kernel grad correction defined in Eq. \eqref{eq:grad-matrix-correction}. 
Here, 
we briefly summarize several widely used slope limiters. 
In the Minmod limiter \cite{roe1986characteristic}, the limiting function is defined as
\begin{equation}\label{eq:limiter-minmod}
\phi_{mm} \left( \lambda\right)  = \max \left[ 0 , \min \left( 1 , \lambda \right) \right] .
\end{equation}
For the SuperBee \cite{roe1986characteristic} limiter, the limiting function is given by
\begin{equation}\label{eq:limiter-superbee}
\phi_{sb} \left( \lambda\right)= \max \left[ 0, \min \left( 2 \lambda , 1 \right), \min \left( r, 2 \right) \right] .
\end{equation}
As for the van Leer \cite{van1974towards} limiter, the limiting function is defined as
\begin{equation}\label{eq:limiter-vanleer}
\phi_{vl} \left( \lambda\right)= \frac{\lambda + \left| \lambda \right| }{1 +  \left| \lambda \right| } .
\end{equation}
Base on the Barth-Jespersen-type limiter \cite{barth1989design}, 
Hopkins \cite{hopkins2015new} proposed a limiting function defined as
\begin{equation}\label{eq:limiter-hopins}
\phi_{bj} = \min\left[ 1, \beta \min\left(\frac{\Phi^{max}_{ij, ngb} - \Phi_i}{\Phi^{max}_{ij, mid} - \Phi_i}, \frac{\Phi_i - \Phi^{min}_{ij, ngb}}{\Phi_i - \Phi^{min}_{ij, mid} }\right)  \right], 
\end{equation}
where $\beta$ is constant, 
$\Phi^{max}_{ij, ngb}$ and $\Phi^{min}_{ij, ngb}$ are the maximum and minimum values 
of $\Phi_j$ among all neighbor particles $j$ of particle $i$, 
and $\Phi^{max}_{ij, mid}$ and $\Phi^{min}_{ij, mid}$ are the
maximum and minimum values (over all pairs $ij$ of the $j$ neighbours
of $i$) reconstructed on the ‘$i$ side’ of the interface between
particles $i$ and $j$ 
(i.e., $\Phi^{max}_{ij, mid} = \max[ \Phi_i + \nabla \Phi _i \cdot \mathbf r_{ji}]$).
\paragraph{WENO reconstruction} \label{paragraph:weno}
Different with the classical point-wise one-dimensional WENO reconstruction from structured mesh data 
\cite{shu1998essentially, shu2020essentially, jiang1996efficient}, 
the reconstruction from scattered data, 
i.e., cell average data using unstructured mesh or meshless particle data, 
is numerically very critical challenging as it requires solving interpolation problems 
\cite{abgrall1994essentially, friedrich1998weighted}, 
in particular when the reconstruction order is high, or when the scattered data are very unevenly distributed \cite{kaser2005ader, dumbser2007quadrature}.
The mostly common procedure applied in the unstructured mesh method is to construct a set of reconstruction stencils for each element by 
dividing its neighbor elements to different groups \cite{dumbser2007arbitrary}. 
Concerning WENO date reconstruction in the Riemann-based SPH method, 
two types of stencil reconstruction, 
i.e., multi-dimensional \cite{avesani2014new} and one-dimensional \cite{zhang2019weakly} stencils, 
are developed. 
\begin{figure*}[htb!]
	\centering
	\includegraphics[trim = 3cm 2cm 3cm 2cm, clip, width=.9\textwidth]{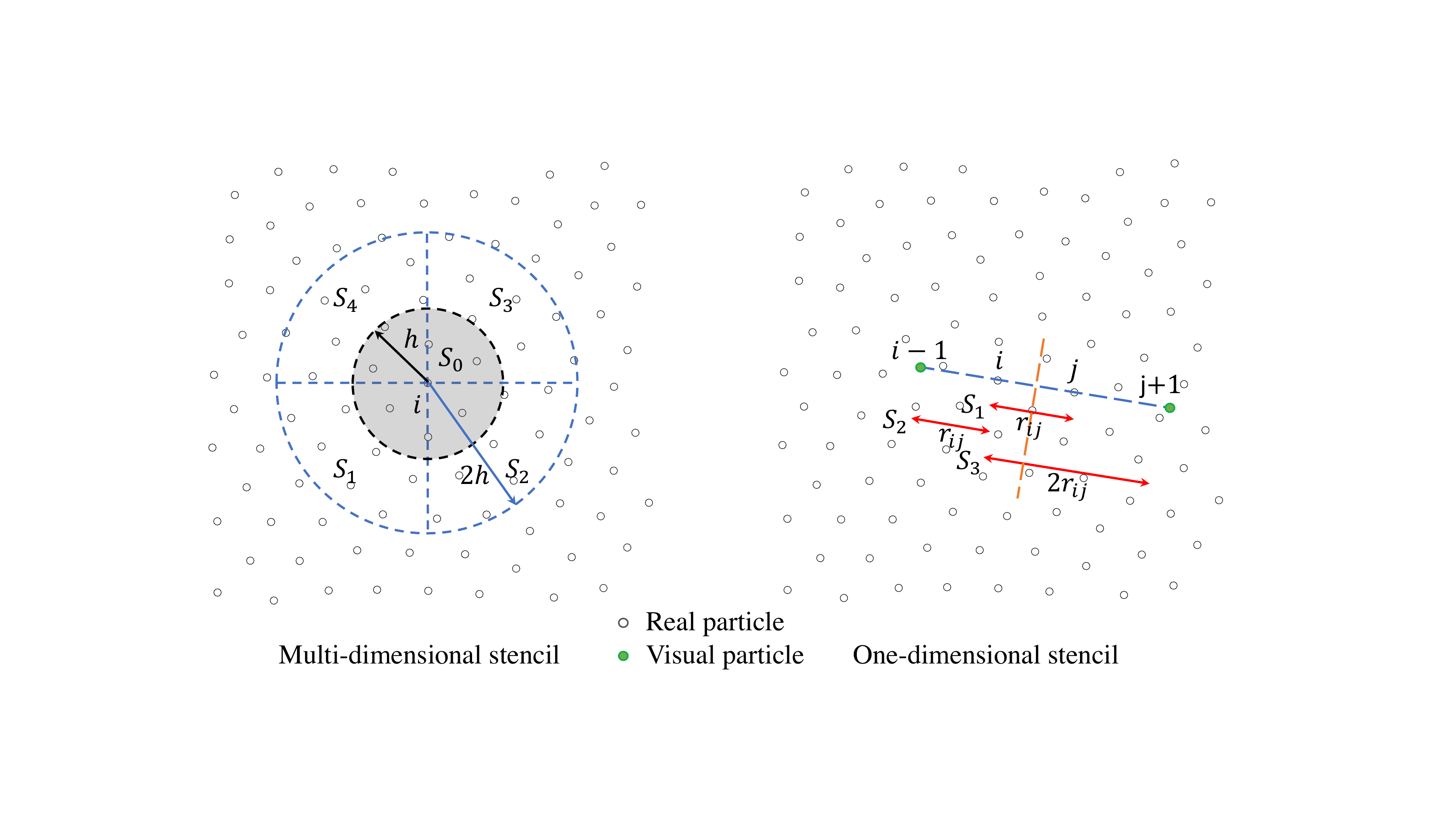}
	\caption{Sketch of multi-phase particles interacting with solid particles along the normal vector 
		through the one-side Riemann problem.}
	\label{figs:stencil}
\end{figure*}

Following Refs. \cite{kaser2005ader, dumbser2007arbitrary}, 
Avesania et al. \cite{avesani2014new} proposed a new class of MLS-ALE-SPH methods 
by first producing each particle a set of high-order MLS reconstructions 
based on multi-dimensional reconstructed stencils and then applying a nonlinear WENO technique
to combine reconstructions with each other. 
For each particle $i$, the reconstructed stencils are defined as
\begin{equation} \label{eq:weno-avesani}
	\begin{cases}
		\mathit S_{i,0} = \bigcup\limits_j \mathcal P_0  \quad |\mathbf r_{ij} |  \leq h \\
		\mathit S_{i,\kappa} =\bigcup\limits_j \mathcal P_\kappa \quad 
			\begin{cases}
				|\mathbf r_{ij} |  \leq 2h \quad \mathrm{and} \quad \kappa \in \left[1, s\right]  \\
				\theta \in\left[ 2\left(\kappa -1 \right)\pi/s , 2\kappa\pi / s\right]
			\end{cases}
	\end{cases},
\end{equation}
where $\mathit S_{i,0}$ is the central stencil containing neighboring particle union $\mathcal P_0$ with distance less than $h$, 
and $\mathit S_{i,\kappa}$ with $\kappa \in \left[1, s\right]$ are the one-sided stencils consisting of neighboring particle union $\mathcal P_\kappa$ 
with distance less than $2h$ and located at specified Circular sector determined by the angle $\theta$ 
formed by the vector $\mathbf r_{ij}$ and the x-axis, 
as shown in Figure \ref{figs:stencil}. 
Note that more one-sided stencils can be constructed by diving the cutoff region of particle $i$ into more Circular sectors. 
Having the definition of the constructed stencils, 
the Moving-Least-Squares interpolation is applied for each particle by assuming the reconstruction polynomials in the form 
\begin{equation}
q_{i,\kappa} \left(\mathbf r \right) = \Phi_i +\sum_{\chi=1}^{N}C_i^\chi B_i^\chi\left(\mathbf r - \mathbf r_i \right), 
\end{equation}
for each stencil defined in Eq. \eqref{eq:weno-avesani}.
Here, 
$N$ is the size of the polynomial basis
(that depends on the polynomial degree and on the space dimension)
$B_i^\chi\left(\mathbf r - \mathbf r_i \right)$ are the associated basis functions, and $C_i^\chi$ are the
(unknown) polynomial coefficients, 
more details are referred to Ref. \cite{avesani2014new}.
With the constructed polynomials, 
the classical WENO scheme can be applied to obtain the final polynomial defined as 
\begin{equation}
q_i \left(\mathbf r \right) = \sum_\kappa w_\kappa  q_{i,\kappa} \left(\mathbf r \right),
\end{equation}
with the normalized nonlinear weights given by 
\begin{equation}
w_\kappa = \frac{\alpha_\kappa}{\sum_\kappa \alpha_\kappa}, \quad \alpha_k = \frac{\alpha_\kappa \lambda_\kappa}{\left(\beta_\kappa + \epsilon \right)^4 }. 
\end{equation}
Here, 
the constant parameter $\epsilon = 10^{-6}$, 
and $\lambda_0 = 1.0$ and $\lambda_\kappa = 10^{-5}$ 
for central and sided stencils, respectively. 
For the calculation of the smoothness indicator, 
Avesania et al. \cite{avesani2014new} proposed the following equation
\begin{equation}
	\beta_\kappa =  \sum_\chi \left( C_i^\chi\right)^2,
\end{equation}

The MLS-ALE-SPH method was further improved by Nogueira et al. \cite{nogueira2016high} using the MOOD paradigm 
to improve the accuracy and the robustness 
and Avesani et al. \cite{avesani2021alternative} adopting ADER approach (Arbitrary Derivative in space and time) 
to guarantee a high order space–time reconstruction.
The MLS-ALE-SPH method and its improvements are able to capture the discontinuities and to maintain accuracy and low numerical dissipation
in smooth regions, 
while they are generally excessive computational expensive due to particle search for all the stencils of each particle 
and the corresponding MLS interpolations. 

To improve the computational efficiency of applying WENO reconstruction, 
Zhang et al. \cite{zhang2019weakly} developed a one-dimensional stencil reconstruction along 
the interacting line of each particle pair as shown in Figure \ref{figs:stencil}. 
They first introduced an IS-WENO reconstruction, 
by which the full 4-point stencil as shown in Figure \ref{figs:stencil} is constructed following the concept of Refs. \cite{fu2016family, wang2018wenois}. 
To construct the 4-point stencil for each interacting particle pair, 
such as particle $i$ and $j$, 
the values at the stencil points are calculated as
\begin{equation}\label{eq:is-weno-stencil}
\begin{cases}
q_{i-1} = \Phi_i - \nabla\Phi_i \cdot \boldsymbol{r}_{ij} \\
q_{i} = \Phi_i \\
q_{i+1} = \Phi_j \\
q_{i+2} = \Phi_j + \nabla\Phi_j\cdot \boldsymbol{r}_{ij}\\
\end{cases},
\end{equation}
where $\Phi_i$ and $\Phi_j$ represent the primitive values, 
i.e., $\rho$, $P$ and $\boldsymbol{v}\cdot \bold{e}_{ij}$, 
at particle $i$ and $j$ respectively. 
In this 4-point stencil, 
two visual particles, namely $i-1$ and $i+1$, 
are constructed along the interacting line $\mathbf r_{ij}$ with the gradients calculated from the SPH approximation 
of Eq. \eqref{eq:sph-grad}. 
Similar with Ref. \cite{zhang2019weakly}, 
Wang et al. \cite{wang2021new} introduced 5-point stencil by constructing more visual particles along the interacting line $\mathbf r_{ij}$ 
and whose values are calculated as
\begin{equation}\label{eq:weno-stencil}
q_{m} = \Phi_c + (i - m)\nabla\Phi_c \cdot \boldsymbol{r}_{ij}  \quad m \in \left\lbrace i-2, i-1, i+2 \right\rbrace 
\end{equation}
where $\Phi_c$ denotes the variable of particle $c$ 
which is the closed particle to the visual particle located at $\mathbf r_i + (i - m) \cdot \mathbf r_{ij}$.
Compared with the 4-point stencil construction, 
the construction proposed by Wang et al. \cite{wang2021new} provides the possibility of implementing 
the classical $5$-order WENO scheme, 
while requires extra computational efforts for nearest particle search of each visual particles. 
This searching procedure increases the complexity of neighbor searching from $O(N)$ to $5O(N)$ \cite{wang2021new}. 

Compared with the multi-dimension WENO reconstruction \cite{avesani2014new}, 
the one-dimensional reconstruction \cite{zhang2019smoothed, wang2021new} 
can no longer main the higher-order data reconstruction \cite{avesani2014new}. 
However, 
it is reasonable as 
the main objective of applying the WENO reconstruction 
aims to increase accuracy by decreasing the numerical dissipation 
other than increasing the formal approximation order of the SPH method \cite{avesani2014new},
which depends on many factors and is quite difficult to achieve in practice. 
It is shown that a general SPH method applying Gaussian-like kernel achieves 
only 2nd-order convergence even when the integration error is sufficiently small \cite{monaghan1992smoothed, litvinov2015towards}.

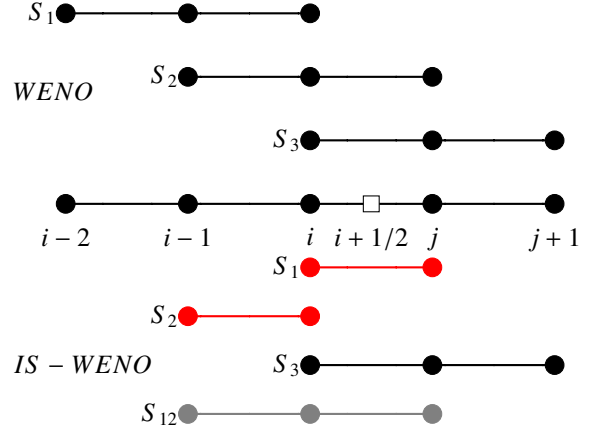
\begin{figure}[tbh!]
	\begin{center}
		\begin{tikzpicture}[
			dot/.style 2 args={circle,draw=#1,fill=#2,inner sep=2.5pt},
			square/.style 2 args={draw=#1,fill=#2,inner sep=3pt},
			mystar/.style 2 args={star,draw=#1,fill=#2,inner sep=1.5pt},
			mydiamond/.style 2 args={diamond,draw=#1,fill=#2,inner sep=1.5pt},
			scale=0.65
			]
			
			\draw[black,thick]  (1.25,-1) grid (11.25,-1);
			\draw[red,thick,yshift=-0.3cm]  (6.25,-2) grid (8.75,-2);
			\draw[red,thick,yshift=-0.3cm]  (3.75,-3) grid (6.25,-3);
			\draw[black,thick,yshift=-0.3cm]  (6.25,-4) grid (11.25,-4);
			\draw[gray,thick,yshift=-0.3cm]  (3.75,-5) grid (8.75,-5);

			\foreach \Fila in {1.25, 3.75,6.25,8.75,11.25}{\node[dot={black}{black}] at (\Fila,-1) {};}  
			\foreach \Fila in {7.5}{\node[square={black}{white}] at (\Fila,-1) {};}  
			\foreach \Fila in {6.25,8.75}{\node[dot={red}{red}] at (\Fila,-2.3) {};}  
			\foreach \Fila in {3.75,6.25}{\node[dot={red}{red}] at (\Fila,-3.3) {};}  
			\foreach \Fila in {6.25,8.75,11.25}{\node[dot={black}{black}] at (\Fila,-4.3) {};}   
			\foreach \Fila in {3.75,6.25,8.75}{\node[dot={gray}{gray}] at (\Fila,-5.3) {};}  
			
			\node[below] at (1.25,-1.3) {$i-2$};
			\node[below] at (3.75,-1.3) {$i-1$};
			\node[below] at (6.25,-1.3) {$i$};
			\node[below] at (8.75,-1.3) {$j$};
			\node[below] at (11.25,-1.3) {$j+1$};
			\node[below] at (7.5,-1.3) {$i + 1/2$};
			
			\node[left]  at (6.25,-2.3) {$S_{1}$};
			\node[left]  at (3.75,-3.3) {$S_{2}$};
			\node[left]  at (6.25,-4.3) {$S_{3}$};
			\node[left]  at (3.75,-5.3) {$S_{12}$};
			\draw[black,thick,yshift=0.3cm]  (6.25,0) grid (11.25,0);
			\foreach \Fila in {6.25,8.75,11.25}{\node[dot={black}{black}] at (\Fila,0.3) {};} 
			\node[left]  at (6.25,0.3) {$S_{3}$};
			\draw[black,thick,yshift=0.6cm]  (3.75,1.0) grid (8.75,1.0);
			\foreach \Fila in {3.75, 6.25,8.75}{\node[dot={black}{black}] at (\Fila,1.6) {};} 
			\node[left]  at (3.75,1.6) {$S_{2}$};
			\draw[black,thick,yshift=0.9cm]  (1.25,2.0) grid (6.25,2.0);
			\foreach \Fila in {1.25, 3.75, 6.25}{\node[dot={black}{black}] at (\Fila,2.9) {};} 
			\node[left]  at (1.25,2.9) {$S_{1}$};	
			
			\node[left]  at (2,1.3) {$WENO$};	
			\node[left]  at (3.2,-4.3) {$IS-WENO$};	
			
		\end{tikzpicture}
	\end{center}		
	\caption{Full stencil and candidate stencils for classic 5-point WENO \cite{jiang1996efficient} (upper panel) 
		and 4-point IS-WENO \cite{wang2018wenois,zhang2019weakly} (bottom panel) for the data reconstruction at $\mathbf r_{i + 1/2}$.}
	\label{figs:weno-stencils}
\end{figure}
Following the WENO reconstruction, the mid-point value, 
i.e., $ q_{1/2} $ as shown in Figure \ref{figs:weno-stencils}, 
is predicted by the non-linear weighted average
\begin{equation}\label{weno-1}
	q_{1/2} = \sum_k w_k  q^{(k)}_{1/2} ,
\end{equation}
where $q^{(k)}_{1/2}$ and $w_k$, $k=1,2,3$, 
are the reconstructed values from the candidate stencils and their non-linear weights. 

For classic 5-point stencil WENO scheme, 
the reconstructed values are defined as
\begin{equation}\label{eq:weno-5-stencil}
\begin{cases}
q^{(1)}_{1/2} = \frac{1}{6}\left(2 q_{i-2} - 7 q_{i-1} + 11q_i \right)  \\
q^{(2)}_{1/2} = \frac{1}{6}\left(- q_{i-1} + 5 q_i + 2 q_{i + 1} \right) \\
q^{(3)}_{1/2} = \frac{1}{6}\left(2 q_i + 5 q_{i+1} - q_{i+2} \right)  
\end{cases}, 
\end{equation}
with the renomalized nonlinear weights given by 
\begin{equation}\label{eq:weno-5-weights}
w_k = \frac{\alpha_k}{\sum_{s=1}^3 \alpha_s}, \quad \alpha_k = \frac{d_k}{\left(\beta_k + \epsilon \right) }, 
\end{equation}
Also, 
the smoothing indicator is calculated from 
\begin{equation}\label{eq:weno-5-smoothness}
\footnotesize
\begin{cases}
\beta_1 = \frac{1}{4}\left(q_{i-2} -4q_{i-1}+3q_i \right)^2 +  \frac{13}{12}\left(q_{i-2} -2q_{i-1}+q_i \right)^2  \\
\beta_2 = \frac{1}{4}\left(q_{i-1} -q_{i+1} \right)^2 +  \frac{13}{12}\left(q_{i-1} -2q_i + q_{i+1} \right)^2 \\
\beta_3 = \frac{1}{4}\left(q_i - 4q_{i+1} + q_{i+2} \right)^2  + \frac{13}{12}\left(q_i - 2q_{i+1} + q_{i+2}\right)^2 
\end{cases}.
\end{equation}

As for the 4-point stencile IS-WENO scheme, 
these reconstructed values are defined as \cite{zhang2019smoothed, wang2018wenois}
\begin{equation}\label{weno-2}
	\begin{cases}
		q^{(1)}_{1/2} = \frac{1}{2} q_{0} + \frac{1}{2} q_{1} \\
		q^{(2)}_{1/2} = -\frac{1}{2} q_{-1} + \frac{3}{2} q_{0} \\
		q^{(3)}_{1/2} = \frac{1}{3} q_{0} + \frac{5}{6} q_{1} - \frac{1}{6} q_{2}  \\
	\end{cases}.
\end{equation}
Also,  
the non-linear weights are 
\begin{equation}\label{weno-3}
	w_k = \frac{\alpha_k}{\sum_{s=1}^3 \alpha_s}, 
	\begin{cases} 
		\alpha_1 = d_1 \left( 1 + \frac{\tau_4}{\beta_1 + \varepsilon} \cdot \frac{\tau_4}{\beta_{12} + \varepsilon} \right)  \\
		\alpha_2 = d_2 \left( 1 + \frac{\tau_4}{\beta_2 + \varepsilon} \cdot \frac{\tau_4}{\beta_{12} + \varepsilon} \right) \\
		\alpha_3 = d_3 \left( 1 + \frac{\tau_4}{\beta_3 + \varepsilon}\right) 
	\end{cases},
\end{equation}
where  the linear weights are determined as  $d_1 = 1/3$, $d_2 = 1/6$ and $d_3 = 1/2$.
$\beta_k$, $k=1,2,3,$ and $\beta_{12}$ are the smoothness indicators for the candidate stencils, 
\begin{equation}\label{weno-4}
\footnotesize
	\begin{cases}
		\beta_1 = \left( q_{i+1} - q_i\right)^2 \\
		\beta_2 = \left( q_i - q_{i-1}\right)^2 \\
		\beta_{12} = \frac{1}{4} \left(q_{i-1} - q_i\right) ^2 + \frac{13}{12} \left(3q_{i-1} - 2q_i + q_{1+1}\right)^2 \\
		\beta_3 = \frac{13}{12} \left( q_i - 2q_{i+1} + q_{i+2}\right)^2 + \frac{1}{4}(3q_i - 4q_{i+1} + q_{i+2})^2 \\
	\end{cases},
\end{equation}
and $\tau_4$ is a global reference smoothness indicator \cite{fu2016family} given as
\begin{equation}\label{weno-tau}
\footnotesize
	\begin{split}
		\tau_4 =  \left[ q_{i-1} \left( 547  q_{i-1} - 2522  q_i + 1922 q_{i+1} - 494 q_{i+2}\right)  \right.  \\
		+  q_i \left(3423 q_i - 5966 q_{i+1} + 1602 q_{i+2}\right)  \\
		+ q_{i+1} \left(  2843 q_{i+1} - 1642 q_{i+2}\right)  \\
		\left. + 267 q_{i+2} \right] / 240.
	\end{split}
\end{equation}

Both the WENO and IS-WENO reconstruction can be further improved by the TENO scheme \cite{fu2016family} 
with the non-linear weights are reformulated as 
\begin{equation}\label{eq:teno-weights}
w_k = \frac{d_k \delta_k}{\sum_{r=0}^2 d_k \delta_k}. 
\end{equation}
Here, 
$d_k$ are optimal weights with respect to their dispersion and dissipation properties \cite{hu2010adaptive, fu2016family, wang2018wenois}
and $\delta$ is a sharp cutoff function to determine 
the contribution of each sub-stencil. 
Following Fu et al. \cite{fu2016family}, the cutoff function is defined as 
\begin{equation}\label{eq:teno-cutoff}
\delta_k = 
\begin{cases}
	0, \quad \mathrm{if} \chi_r < C_T \\
	1, \quad \mathrm{otherwise} 
\end{cases},
\end{equation}
with threshold $C_T = 10^5$ and the parameter $\chi_r$ is a normalized smoothness measure. 
For each sub-stencil, $\chi_r$ is defined as 
\begin{equation}\label{eq:teno-smoothness}
\chi_r = \frac{\gamma_r }{\sum_{r=0}^2 \gamma_r } ,
\end{equation}
where $\gamma$ is a scale separation to distinguish the discontinuity form the smoothe region. 
For $5$-point WENO scheme, 
\begin{equation}\label{eq:teno-smoothness-gammma-5}
\gamma_r = \left(1 + \frac{\tau_5}{\beta_r + \epsilon} \right)^6, \tau_5 =\left| \beta_0 - \beta_2\right| ,
\end{equation}
and for $4$-point IS-WENO
\begin{equation}\label{eq:teno-smoothness-gammma-4}
\gamma_r = \left(1 + \frac{\tau_4}{\beta_r + \epsilon} \right)^6 .
\end{equation}

\begin{figure*}[htb!]
	\centering
	\includegraphics[width=.75\textwidth]{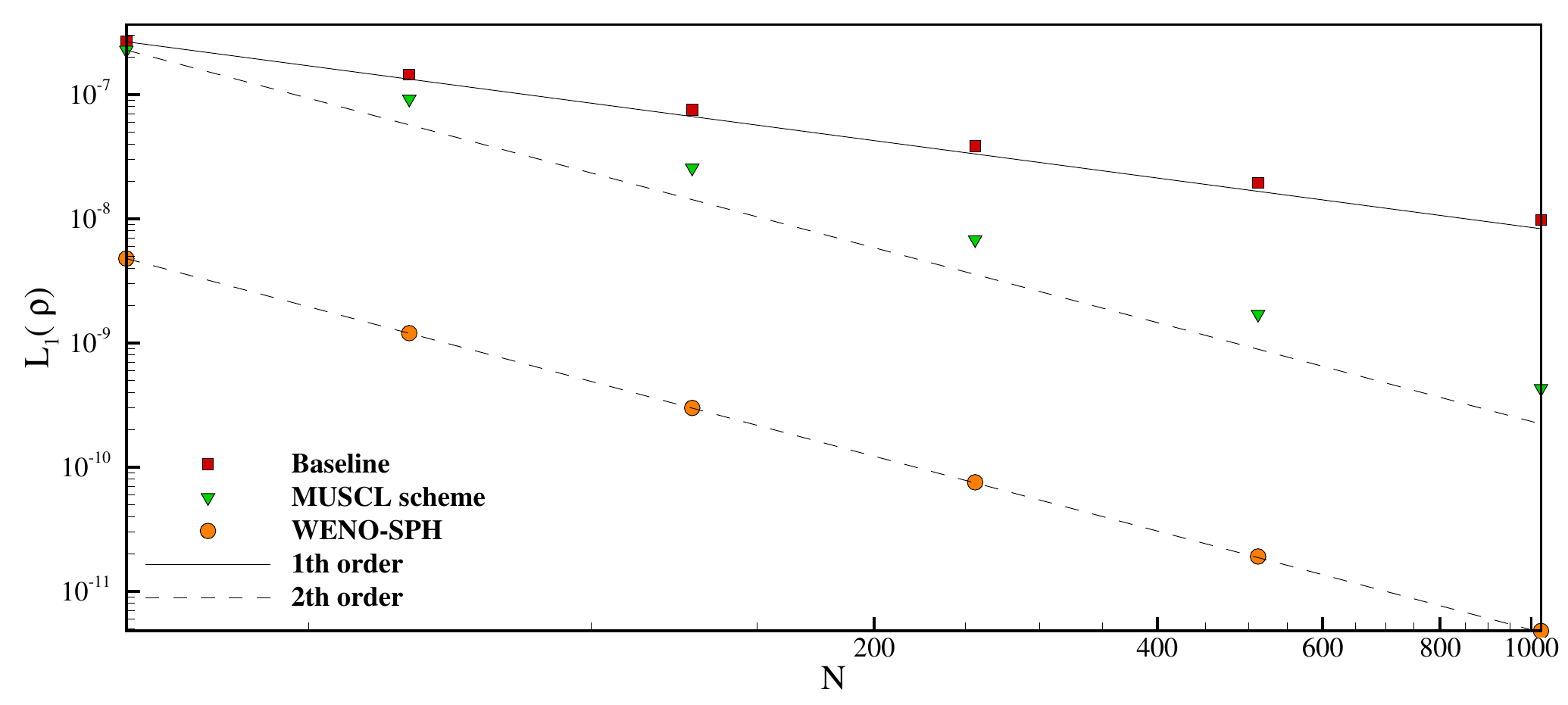}
	\caption{Numerical study of one-dimensional acoustic wave: the convergence of the density error 
		as a function of particle resolution by using the Baseline scheme, 
	MUSCL with the Sweby limiter and IS-WENO reconstructions.  }
	\label{figs:acoustic}
\end{figure*}

The convergence rate of different data reconstruction, 
i.e., 
piece-wise constant reconstruction termed as "Baseline", 
MUSCL and IS-WENO schemes, 
is presented in Figure \ref{figs:acoustic} 
which gives the density error with increasing particle resolution 
for one-dimensional acoustic wave propagation \cite{zhang2019weakly}. 
Both MUSCL and IS-WENO reconstructions achieve second-order convergence,
which is the formal accuracy of a general SPH approximation with Gaussian-like smoothing kernels 
when the particle integration error is negligible \cite{litvinov2015towards}.
As expected, 
the Baseline achieves first-order convergence only, 
and MUSCL exhibits considerably larger errors due to numerical dissipation.

\paragraph{MOOD scheme} \label{paragraph:mood}
The aforementioned MUSCL and WENO scheme provide a \textit{priori} limitation procedure, 
which is performed with data at time $t^n$ to eliminate the spurious numerical oscillation in the vicinity of discontinuity at time $t^{n+1}$ \cite{avesani2014new}.
Recently, 
Clain et al. \cite{clain2011high} proposed a \textit{posteriori} limiting paradigm, 
multi-dimensional optimal order detection (MOOD), within the finite volume Eulerian framework on unstructured mesh. 
The MOOD paradigm consists of detecting problematic situations after each time update of the solution and of reducing the local
polynomial degree before recomputing the solution.
Nogueira et al. \cite{nogueira2016high} implemented the MOOD paradigm in the MLS-WENO-SPH method \cite{avesani2014new,avesani2021alternative} 
to determine, 
a \textit{posteriori}, 
the optimal order of the polynomial reconstruction of MLS interpolation for each particle that provides the best compromise between accuracy and stability. 
Then,
Antona et al. \cite{antona2021towards} extended this method to the simulation of weakly-compressible viscous flow. 

\begin{figure}[htb]
	\centering
	\includegraphics[trim = 3cm 12cm 3cm 6cm, clip, width=.49\textwidth]{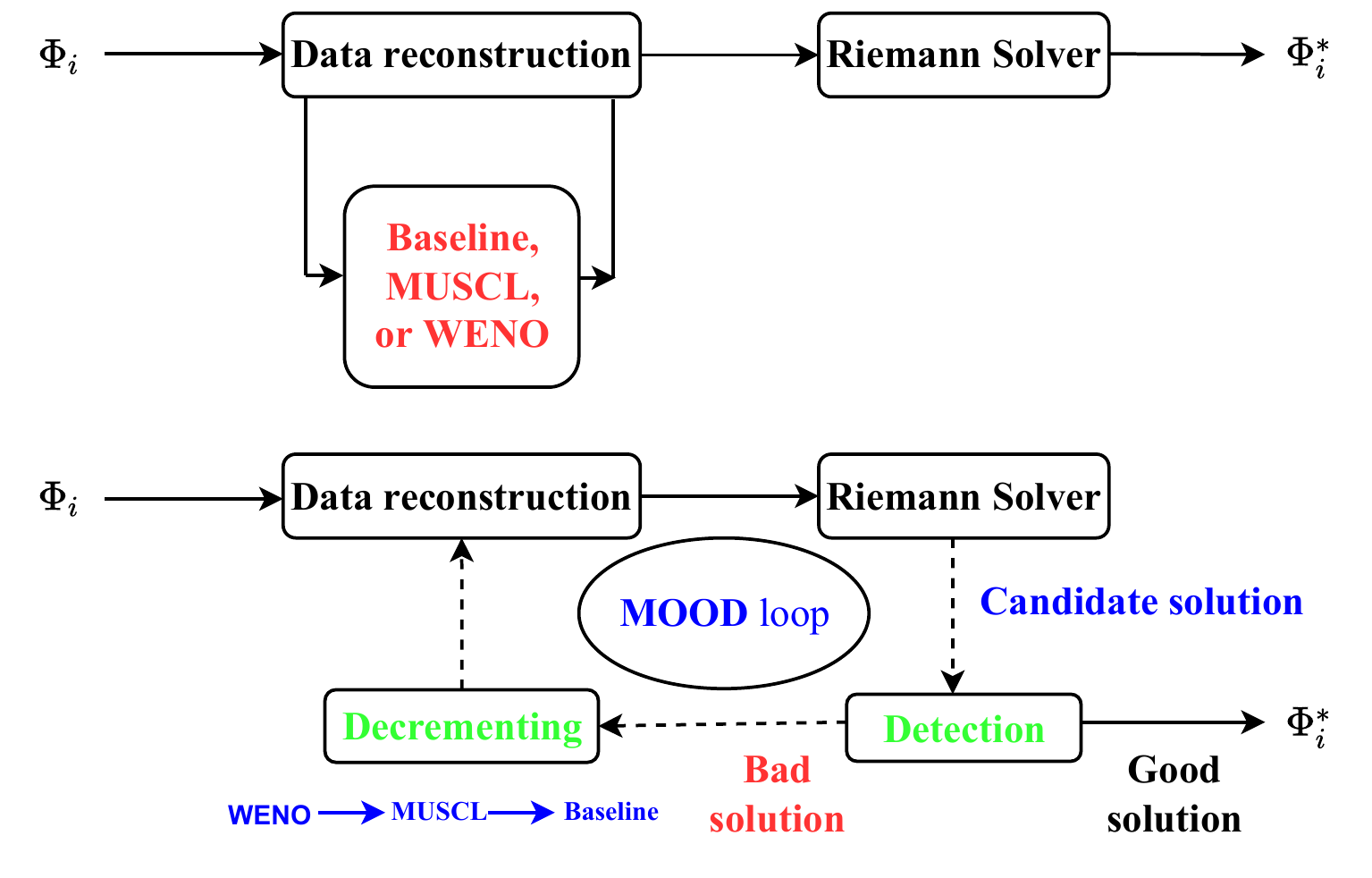}
	\caption{Sketch of the Riemann-based SPH (top panel) and the one with MOOD \textit{posteriori} limiting paradigm (bottom panel). }
	\label{figs:mood}
\end{figure}

Different with Refs. \cite{nogueira2016high, antona2021towards}, 
where the \textit{posteriori} limiting procedure is performed at the MLS interpolation process,
we derive herein the exploitation of MOOD paradigm to determine the optimal data reconstruction of the Left and Right states 
in the Riemann-based SPH method. 
The key idea is to introduce a Data Reconstruction Degree decrementing process to replace the counterpart based on Particle Polynomial Degree (PPD) 
applied in the original MOOD paradigm \cite{nogueira2016high, clain2011high}. 
More precisely, 
the present MOOD paradigm consists of two ingredients, a DRD and a detector. 
The DRD indicates the optimal data reconstruction, 
Baseline, MUSCL or WENO scheme, 
for the Riemann problem to obtain the candidate Riemann solution of $\Phi^\ast$, 
i.e., $U^\ast$ and $P^\ast$. 
The detector controls the admissibility of the resulting Riemann solution 
and the particle DRD will be decremented 
from the WENO scheme to MUSCL scheme and further to Godunov scheme when a detector is activated. 
The Riemann solution with first-order, robust Godunov scheme, i.e., the reconstruction with piecewise constant assumption, 
is assumed to be always valid as the original MOOD paradigm in Refs.\cite{clain2011high, nogueira2016high}.
Figure \ref{figs:mood} sketches the Riemann-based SPH method with Godunov in the top panel, 
MUSCL or WENO reconstruction, 
while displays in the bottom panel the present \textit{posteriori} MOOD procedure.
Concerning the detector, 
the physical admissibility detection
\begin{equation}
	0 <	P^\ast \leq P_{threshold}, 
\end{equation}
or the discrete maximum principle 
\begin{equation}
\min\left( \rho_i, \rho_j \right) \leq \rho^\ast \max \left( \rho_i, \rho_j\right), 
\end{equation}
proposed by Ref. \cite{nogueira2016high} can be applied. 

%
\subsection{Particle-interaction configuration} \label{sec:dualcriteria}
In the particle-base methods, 
the pairwise interaction between neighboring particles is determined through a Gaussian-like kernel function 
which has radial-symmetric compact support. 
Therefore, 
implementing the particle-interaction configuration, 
i.e., 
searching of neighbor particles and computing corresponding kernel weights and gradients, 
is a critical aspect of the high-performance particle-based solver. 
Concerning the searching of neighbor particles, 
two different approaches, 
i.e.,
cell-linked list (CLL) \cite{mattson1999near} and Verlet list (VL) \cite{verlet1967computer}, 
are widely used in the particle-method community \cite{dominguez2011neighbour, dominguez2021dualsphysics, viccione2008defining}. 
With different neighboring search technique, 
the particle-interaction configuration can be updated accordingly. 
For the CLL approach, 
the neighbouring search procedure must be performed at each numerical iteration, 
indicating that the particle-interaction configuration is also updated accordingly. 
As for the VL approach, 
a VL containing all potential neighboring particles is created and stored for each particle. 
Therefore, 
a VL may be used for multiple times without executing neighbor search 
if the particle-interaction configuration can be obtained \cite{dominguez2011neighbour, viccione2008defining}.
Notwithstanding the wide implementation of the CLL and VL approach in the particle-based methods, 
they may become not sufficiently efficient when adaptive particle resolution with variable smoothing lengths is applied 
\cite{springel2005cosmological, springel2010smoothed}, 
where tree-based neighboring search technique can be applied 
for address this issue \cite{arge2008priority,fu2019isotropic, khorasanizade2019improving}. 
With the CLL and VL approaches in hand, 
several schemes, 
data sorting with space filling curve \cite{dominguez2011neighbour,dominguez2013optimization, winkler2019gpusphase, fair2019particle}, 
dual-criteria time stepping \cite{zhang2020dual} and multi-cell linked lists \cite{zhao2021high}, 
are developed for further improvement of the computational efficiency. 
\subsubsection{Cell-linked and Verlet lists}
\begin{figure*}[htb!]
	\centering
	\includegraphics[trim = 1mm 1mm 1mm 1cm, clip, width=.45\textwidth]{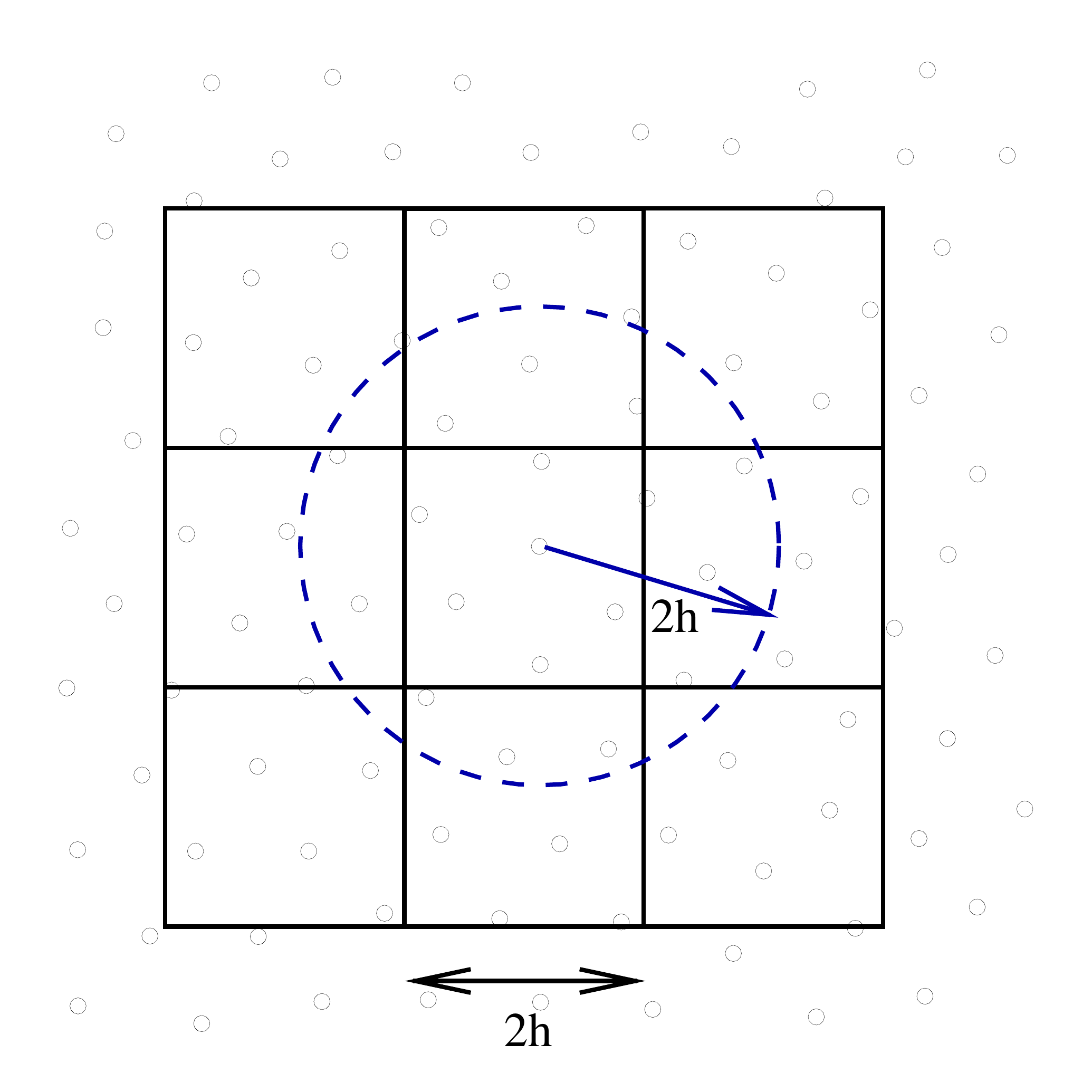}
	\includegraphics[trim = 1mm 1mm 1mm 1cm, clip, width=.45\textwidth]{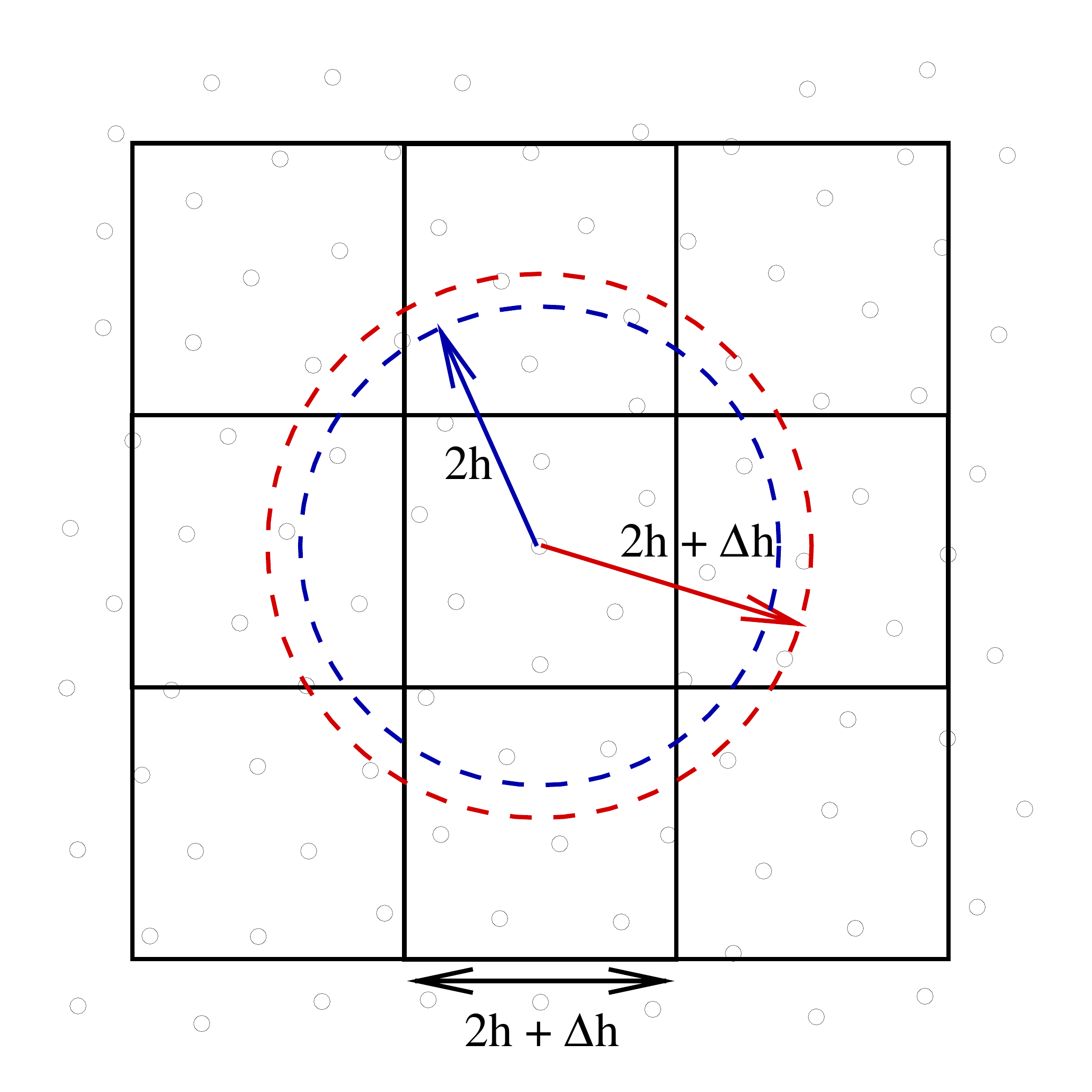}
	\caption{Sketch of the different approaches used to create the neighbour particle list: 
		The cell-linked list (left panel) and the Verlet list (right panel) approaches.
	}
	\label{figs:list}
\end{figure*}

In the linked list approach, 
i.e., CLL and VL, 
the whole computational domain is partitioned into equisized cells 
and each cell creates a list consisting of the references of all the particles located within it, 
as shown in Figure \ref{figs:list}.

For the CLL approach, 
the cell size is equal to that of the support or cut-off radius of the kernel function, 
i.e., $2h$, 
and the searching of neighbor particles is restricted to the nearest neighboring cells,
9 cells in two dimensions as shown in Figure \ref{figs:list}.
After the neighbor-searching operation, 
the kernel function values and interaction forces can be calculated with respect to 
the particles belong to these neighboring cells only if they are found within the cut-off radius.
As the CLL approach does not store the neighboring-particle identities 
and the corresponding kernel function values, 
the particle-interaction configuration must be updated multiple times 
during a single time step in the time integration, 
e.g. the particle-interaction configuration is updated twice 
when the kick-drift-kick time integration scheme is applied \cite{zhang2017weakly, monaghan2005smoothed, adami2013transport}. 

Different with the CLL approach, 
the VL increases the cell size to $2h + \Delta h$, 
creates and stores a Verlet list 
which contains the references to all potential neighboring particles 
for each particle by checking all particles within the adjacent cells \cite{dominguez2011neighbour}, 
as shown in Figure \ref{figs:list}.
Without conducting neighbor particle search, 
a VL can be used for multiple times, 
twice for one single time step with the kick-drift-kick scheme, 
with non-vanishing kernel function values \cite{dominguez2011neighbour}. 

Dominguez et al. \cite{dominguez2011neighbour} conducted a comprehensive study of the 
CLL and VL approaches, 
and concluded that the VL approach has to use cells with considerably larger size than the cut-off radius 
to increase the reuse of the Verlet lists. 
They demonstrated that with a 50\% increase of the cell size, only a slight performance gain of 
$8\%$ is achieved when the Verlet lists are reused for $13$ times in $7$ time steps. 
Winkler et al. \cite{winkler2018neighbour} also found that in general this approach is not able to substitute the CLL due to its poor performance. 
More recently, 
Fraga Filho et al. \cite{fraga2020investigation} evaluated the performance of 
the CLL and the VL approaches, 
and concluded that the VL approach is an optimisation proposal in which the neighbour list is not update at each numerical iteration 
through an appropriate choice of the cutoff radius ensuring no accuracy loss in the location of neighbor particles.
\subsubsection{Particle sorting with space filling curve}
Traditionally, 
particles are generated and stored following a given order, 
row or column order for Lattice distributed particles, 
and the corresponding memory allocation of particle data is unaltered during the simulation. 
This results poor temporal and spatial data access and insufficient usage of memory hierarchy 
\cite{springel2005cosmological, dominguez2011neighbour} due to the full Lagrangian feature of 
the particle-based methods. 
Therefore, 
sorting particle data to change their memory location for better data locality, 
which has positive effects on hardware caching \cite{hockney2021computer},
can improve the memory access and decrease the computational time, 
achieving scalability and efficiency for large scale particle simulations. 
To that end, 
particle date arrays including particle index and other physical variables 
are rearranged so that the neighbouring particles are close in computer memory space. 
This procedure can be realized by implementing sorting algorithm with proper space filling curve (SFC), 
for example the Morton SFC \cite{morton1966computer} and the Hilbert SFC \cite{hilbert1891aijber}, 
which traverses higher dimensional space in a continuous fashion \cite{moon2001}. 
Note that particle sorting does not change the particle-interaction configuration. 

Springel \cite{springel2005cosmological} implemented an efficient Hilbert SFC in a 
cosmological N-body/SPH code to domain decomposition and particle sorting within each processor, 
exhibiting approximated speedup of $2$ compared with random sorting.  
For pure SPH simulation, 
the CLL approach is introduced to avoid the naive $O(n^2)$ neighbor searching and
particle sorting can be conducted with respect to the mapped cell index \cite{dominguez2011neighbour, winkler2019gpusphase}. 
Dominguez et al. \cite{dominguez2011neighbour} showed that implementing particle sorting with 
the Morton SFC \cite{morton1966computer} can increase the computational performance about $20\%$ for weakly-compressible SPH simulations. 
Since then, 
particle sorting has been implemented in particle-based code with the coupling 
of MPI \cite{hofmann2010parallel}, 
GPU-acceleration \cite{dominguez2013optimization, ihmsen2011parallel, winkler2018neighbour} 
and share-memory high-performance computing strategy \cite{winkler2019gpusphase}. 

Following Refs. \cite{dominguez2011neighbour, winkler2019gpusphase}, 
the cell index in the CLL will be defined by a SFC function \cite{morton1966computer, hilbert1891aijber} 
$H\left(x_{cell}, y_{cell}, z_{cell}\right) = n_{cell}$
which produces one dimensional indices $n_{cell}$ 
where two cells $i$ and $j$ that are geometrically close will be ordinally close. \
As pointed out by Dominguez et al. \cite{dominguez2011neighbour} particle sorting provides 
a way to improve the data access pattern and neighboring particle search, 
while it increases the memory requirements. 
\subsubsection{Dual-criteria time stepping}
Zhang et al. \cite{zhang2020dual} proposed a dual-criteria time stepping scheme 
to optimize the computational efficiency of the WCSPH method by introducing two time-step criteria 
characterized by the particle advection and the acoustic speeds, respectively.
In this scheme, 
the advection criterion determines the updating frequency of the particle-interaction configuration, 
i.e., the simplest VL approach with a cell size of $2h$ and the corresponding kernel weights and gradients,
and the acoustic criterion controls the frequency of
the pressure relaxation process, 
i.e., 
the time integration of the particle density, position 
and velocity due to the action of pressure gradient.

The time-step size determined by the advection criterion, 
termed $\Delta t_{ad}$, 
has the following form
\begin{equation}\label{eq:dt-advection}
	\Delta t_{ad}   =  {CFL}_{ad} \min\left(\frac{h}{|\mathbf{v}|_{max}}, \frac{h^2}{\nu}\right),
\end{equation}
where $CFL_{ad} = 0.25$, 
$|\mathbf{v}|_{max}$ is the maximum particle advection speed in the flow 
and $\nu$ the kinematic viscosity. 
The time-step size according to the acoustic criterion, 
termed $\Delta t_{ac}$, 
has the form
\begin{equation}\label{eq:dt-relax}
	\Delta t_{ac}   = {CFL}_{ac} \frac{h}{c + |\mathbf{v}|_{max}},
\end{equation}
where ${CFL}_{ac} = 0.6$. 
Therefore,  
the pressure relaxation process is carried out approximately $k \approx \frac{\Delta t_{ad}}{\Delta t_{ac}} $ times, 
for example $k$ is about $4$ to $5$ when considering inviscid flow \cite{zhang2020dual}, 
during one advection step. 
Also,  
the particle-interaction configuration is not altered in one advection time step, 
a large $CFL_{ac} = 0.6$ value typically for a Eulerian method is allowable without introducing numerical instability. 

As reported in Ref. \cite{zhang2020dual}, 
the dual-criteria time stepping scheme can achieve an speedup up to $2.80$ with good robustness and accuracy, 
in comparison to the traditional counterpart where the CLL approach is applied.
%
%
\section{Solid mechanics}\label{sec:solid}
In the SPH method, 
there generally two types of formulations, 
namely update Lagrangian (UL) and total Lagrangian (TL) formulations, 
have been developed for solid dynamics. 
The UL formulation, 
where the current configuration is used as the reference, 
suffers from several shortcomings, 
e.g. the presence of tensile instability \cite{gray2001sph, zhang2017generalized} exhibiting non-physics fracture, 
the appearance of zero-energy modes due to the rank-deficiency inherent to the use of under-integrated particle integration \cite{puso2008meshfree} 
and the reduced order of convergence for derived variable \cite{bonet1998simple}.
To address these problems, 
many modifications by correcting the kernel function \cite{liu1995reproducing, bonet2000correction, ganzenmuller2015hourglass}, 
improving the interpolation integral \cite{randles1996smoothed, zhang2017smoothed, libersky1991smooth, vignjevic2000treatment, bonet2002simplified} or 
introducing transport-velocity formulation\cite{zhang2017generalized} have been proposed.
Compared with the UL formulation, 
the TL formulation shows promising potential in the simulation of finite deformation 
due to its attractive advantages in
being free from tensile instability and ensuring $1$st-order consistency when computing deformation gradient by introducing 
the kernel gradient correction. 
Since its inception, 
it has been applied for the problems of necking and fracture in  
thermomechanical deformations \cite{ba2018thermomechanical}, 
fluid-structure interaction (FSI) \cite{antoci2007numerical, khayyer2018enhanced, liu2019smoothed, zhang2020multi} 
and biomechanics \cite{zhang2020sphinxsys, zhang2020integrative}, 
among many others.  
In this paper, 
we focus on the TL formulation with highlights on stablized term, 
the steady state solution and the hourglass control scheme. 
\subsection{Governing equations} \label{sec:solidgoverning}
The kinematics of the finite deformations can be characterized by introducing a deformation map $\varphi$, 
where a material point $\mathbf{\mathbf{r}^0}$ can thus be mapped from
the initial reference configuration $\Omega^0 \subset \mathbb{R}^d $ 
to the point $\mathbf{r} = \mathbf{\varphi}\left(\mathbf{r}^0, t\right)$ 
in the deformed configuration $\Omega = \mathbf{\varphi} \left(\Omega^0\right)$. 
Here, the superscript $\left( {\bullet} \right)^0$ denotes the quantities in the initial reference configuration. 
Accordingly, 
the deformation tensor $\mathbb{F}$ can be defined by its derivative with respect to the initial reference configuration as 
\begin{equation} \label{eq:deformationtensor}
	\mathbb{F} = \nabla^{0} {\varphi} =  \frac{\partial \varphi}{\partial \mathbf{r}^0}  = \frac{\partial \mathbf{r}}{\partial \mathbf{r}^0} .
\end{equation}
With the definition of the displacement $\mathbf{u} = \mathbf{r} - \mathbf{r}^0$, 
the deformation tensor $\mathbb{F}$ can also be calculated through
\begin{equation} \label{eq:deformationtensor-displacement}
	\mathbb{F} = \nabla^{0} {\mathbf{u}}  + \mathbb{I},
\end{equation}
where $\mathbb{I}$ represents the unit matrix. 

In total Lagrangian framework, 
the conservation of mass and the linear momentum corresponding to the solid mechanics can be expressed as
\begin{equation}\label{eq:mechanical-mom}
	\begin{cases}
		\rho =  {\rho^0} \frac{1}{J} \quad \\
		\rho^0 \frac{\text{d} \mathbf{v}}{\text{d} t}  =  \nabla^{0} \cdot \mathbb{P}^T  + \rho^0 \mathbf{g} \quad  
	\end{cases} \Omega^0 \times \left[0, T \right],
\end{equation}
where $\rho$ is the density, $J = \det(\mathbb{F})$ and $\mathbb{P}$ the first Piola-Kirchhoff stress tensor 
and $\mathbb{P} =  \mathbb{F} \mathbb{S}$ with $\mathbb{S}$ denoting the second Piola-Kirchhoff stress tensor. 
In particular, when the material is linear elastic and isotropic, the constitutive equation can be simply given by
\begin{eqnarray}\label{isotropic-linear-elasticity}
	\mathbb{S} & = & K \tr\left(\mathbb{E}\right)  \mathbb{I} + 2 G \left(\mathbb{E} - \frac{1}{3}\tr\left(\mathbb{E}\right)  \mathbb{I} \right) \nonumber \\
	& = & \lambda \tr\left(\mathbb{E}\right) \mathbb{I} + 2 \mu \mathbb{E} ,
\end{eqnarray}
where $\lambda$ and $\mu$ are the Lamé parameters \cite{sokolnikoff1956mathematical}, 
$K = \lambda + (2\mu/3)$ the bulk modulus and $G = \mu$ the shear modulus. 
The relation between the two modulus reads
\begin{equation}\label{relation-modulus}
	E = 2G \left(1 + \nu\right) = 3K\left(1 - 2\nu\right),
\end{equation}
with $E$ denoting the Young's modulus and $\nu$ the Poisson's ratio. 
Note that the sound speed of solid structure is defined as $c^{S} = \sqrt{K/\rho}$. 
The Neo-Hookean material model can be defined in a general form with the introduction of the strain-energy density function 
\begin{eqnarray}\label{Neo-Hookean-energy}
	\mathfrak W  =  \mu \tr \left(\mathbb{E}\right) - \mu \ln J + \frac{\lambda}{2}(\ln J)^{2} .
\end{eqnarray}
Then, the second Piola-Kirchhoff stress $\mathbb{S}$ is derived as 
\begin{equation}\label{2rd-PK}
	\mathbb{S} = \frac{\partial \mathfrak W}{\partial \mathbb{E}}.
\end{equation}
%
\subsection{Total Lagrangian formulation}\label{sec:solid-totallagrangian}
In the TL formulation, 
the correction matrix of Eq. \eqref{eq:kernel-matrix-correction} of the kernel gradient correction 
is calculated form the initial reference configuration as \cite{vignjevic2006sph}
\begin{equation} \label{eq:sph-correctmatrix}
	\mathbb{B}^0_a = \left( \sum_b V_b \left( \mathbf{r}^0_b - \mathbf{r}^0_a \right) \otimes \nabla^0_a W_{ab} \right) ^{-1} ,
\end{equation}
where 
\begin{equation}\label{strongkernel}
	\nabla^0_a W_{ab} = \frac{\partial W\left( |\mathbf{r}^0_{ab}|, h \right)}  {\partial |\mathbf{r}^0_{ab}|} \mathbf{e}^0_{ab} ,
\end{equation}
denotes the gradient of the kernel function.
Here, 
the subscript $a$ and $b$ are introduced to denote the solid particles. 
It is worth noting that the correction matrix is only calculated once before the simulation as it is evaluated at the initial reference configuration.
Then,
the discretization form of the mass and momentum conservation equations, Eq.\eqref {eq:mechanical-mom}, yields
\begin{equation}\label{eq:sph-mechanical-mom}
	\begin{cases}
		\rho_a =  {\rho^0} \frac{1}{\text{det}\left(\mathbb{F}\right) } \quad \\
		\frac{\text d\mathbf v_a}{\text d t} = \frac{2}{m_i} \sum_b V_a V_b \tilde{\mathbb{P}}_{ab} \nabla^0_a W_{ab} + \mathbf{g} + \mathbf f^f
	\end{cases}, 
\end{equation} 
where $\mathbf f^f$ denotes the force exerting on the solid particles due the existence of the fluid particles and 
$\tilde{\mathbb{P}}$ denotes the inter-particle averaged first Piola-Kirchhoff stress and is defined by
\begin{equation}
	\tilde{\mathbb{P}}_{ab} = \frac{1}{2} \left( \mathbb{P}_a \mathbb{B}^0_a + \mathbb{P}_b \mathbb{B}^0_b \right). 
\end{equation}
Note that the first Piola-Kirchhoff stress tensor is computed from the constitutive law with the deformation tensor $\mathbb{F}$ given by
\begin{equation}
	\mathbb{F} = \left( \sum_b V_b \left( \mathbf{r}_b - \mathbf{r}_a\right) \otimes \nabla^0_a W_{ab}  \right) \mathbb{B}^0_a + \mathbb{I} .
\end{equation}
%
\subsection{Stablized scheme}\label{sec:solid-stablization}
Without appropriate stabilization technique, 
the original TL formulation may exhibit spurious fluctuations especially in the vicinity of sharp spatial gradients. 
This deficiency can result in numerical instability and lead to wrongly predicted deformation for problems involving large strain. 
To rectify this deficiency, 
Lee et al. \cite{lee2016new} proposed a Jameson-Schmidt-Turkel SPH (JST-SPH) method, 
which shows good performance of eliminating spurious pressure oscillations 
in the simulation of nearly incompressible solid. 
In JST-SPH methodology, 
the nodally conservative JST stabilization is additively decomposed into harmonic operator (2nd-order) and biharmonic operator (4th-order) 
which require excessive computational efforts \cite{lee2017variationally}. 
In a more recent work, 
Lee et al. \cite{lee2019total} further proposed a total Lagrangian upwind SPH (TLU-SPH) method 
by introducing a characteristic-based Riemann solver in conjunction 
with a linear reconstruction procedure to guarantee the consistency and conservation of the overall algorithm. 
This method also shows good performance in the simulation of nearly and truly incompressible explicit fast solid dynamics with large deformations. 
 
More recently,
Zhang et al. \cite{zhangAsimple} proposed an efficient artificial damping method by introducing a Kelvin-Voigt (KV) type damper 
for total Lagrangian formulation by introducing appropriate damping terms into the constitutive equation.
Besides, many stabilization strategies for update Lagrangian formulation, 
e.g. artificial viscous fluxes \cite{monaghan1992smoothed, randles1996smoothed, randles2000normalized}, 
conservative strain smoothing regularization \cite{bonet1998simple, chen2001stabilized} and Riemann-based scheme \cite{parshikov2000improvements}, 
have also been proposed.
In KV model,
a viscous damper and a purely elastic spring connected in parallel are involved.
Following the same idea,
an elastic solid undergoing large strains can also be modeled with the mechanical components of springs and dashpots.
Thus,
the total stress $\sigma_{total}$ can be decomposed into two parts, 
i.e., the elastic stress $\sigma_{S}$ and the damper stress $\sigma_{D}$ as
\begin{equation}\label{eq:kv-stress}
	\sigma_{total} = \sigma_{S} + \sigma_{D}, 
\end{equation}
Where the damper stress is defined by
\begin{equation}\label{eq:kv-mdoel}
	\sigma_{D} =  \eta \frac{\text{d}\epsilon(t)}{\text{d}t} .
\end{equation}
Here, 
$\text{d}\epsilon(t)/\text{d}t$ denotes the strain rate 
and $\eta$ the physical viscosity. 
Applying the KV model to TL-SPH formulation, 
the second Piola-Kirchhoff stress $\mathbb{S}$  can be rewritten as 
\begin{equation}
	\mathbb{S} = \mathbb{S}_{S} + \mathbb{S}_D ,
\end{equation}
where $\mathbb{S}_{S}$ is given by the constitutive equation of Eq. \eqref{isotropic-linear-elasticity} or Eq. \eqref{Neo-Hookean-energy}, 
and the damper $\mathbb{S}_D$ is defined as 
\begin{equation}\label{eq:s-d}
	\mathbb{S}_D =  \pi \frac{\text{d}\epsilon(t)}{\text{d}t} 
	= \frac{\pi}{2}\left[\left( \frac{\text{d}\mathbb{F}}{\text{d}t} \right)^T \mathbb{F} + \mathbb{F}^T  \left( \frac{\text{d}\mathbb{F}}{\text{d}t}\right)  \right].
\end{equation}
By introducing a von Neumann-Richtmyer type scaling factor with the speed of sound $c$,
the artificial viscosity $\pi$ in Eq. \eqref{eq:s-d} is defined by
\begin{equation}\label{eq:artificial-v}
	\pi = \alpha \rho c h, 
\end{equation}
where $\alpha=0.5$ is a constant parameter and $h$ denotes the smoothing length.

\begin{figure*}
	\centering
	\begin{subfigure}{\textwidth}
		\includegraphics[trim = 2mm 2mm 1mm 2mm, clip, width=0.485\textwidth]{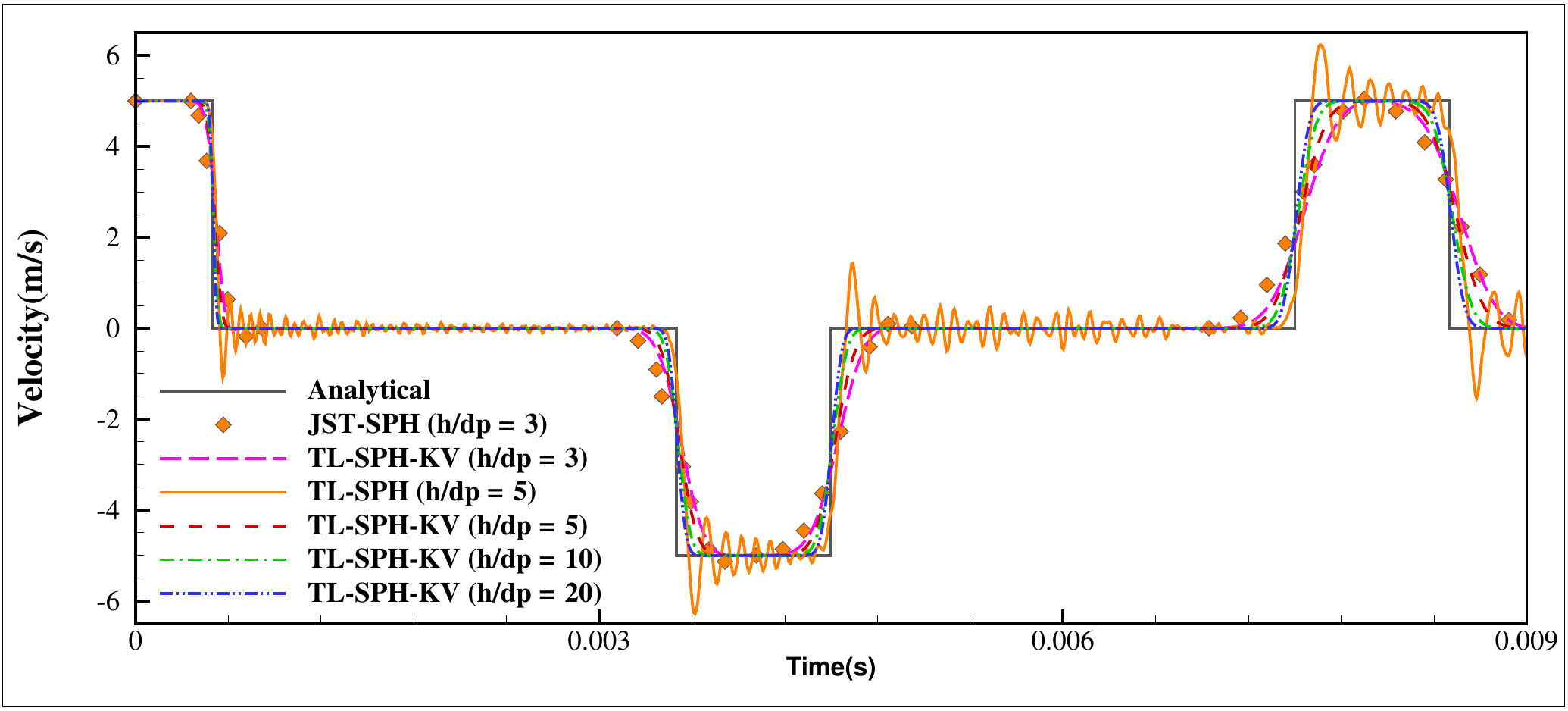} 
		\includegraphics[trim = 2mm 2mm 1mm 2mm,clip, width=0.485\textwidth]{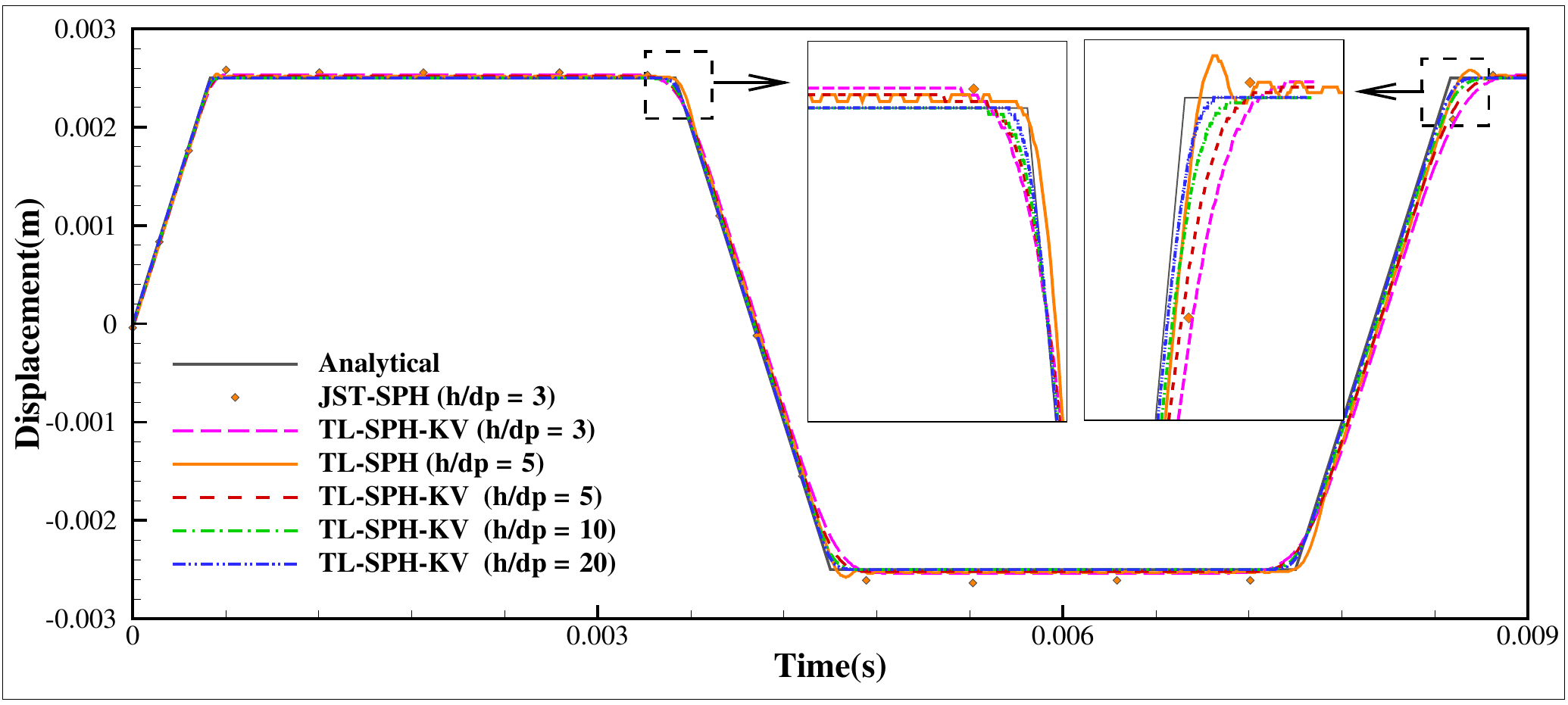} 
		\caption{Wave propagation in an elastic cable.}
		\label{figs:solid-waveincable}
	\end{subfigure}
	\newline
	\begin{subfigure}{\textwidth}
		\includegraphics[trim = 1.75cm  1mm 12cm  1mm, clip, height=0.18\textwidth]{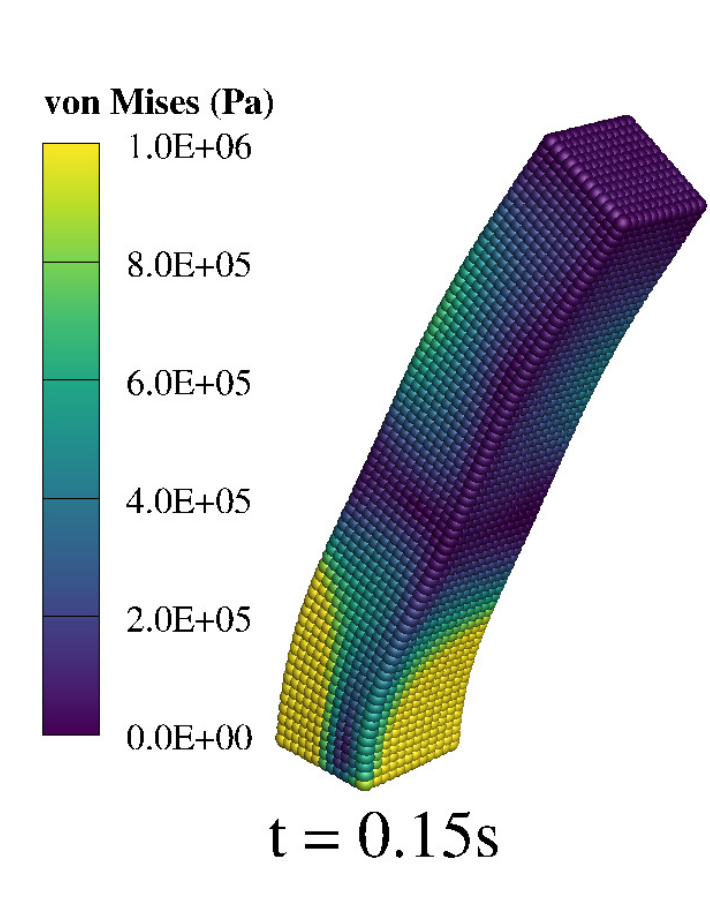}
		\includegraphics[trim = 4.5cm  1mm 10.5cm 1mm, clip, height=0.18\textwidth]{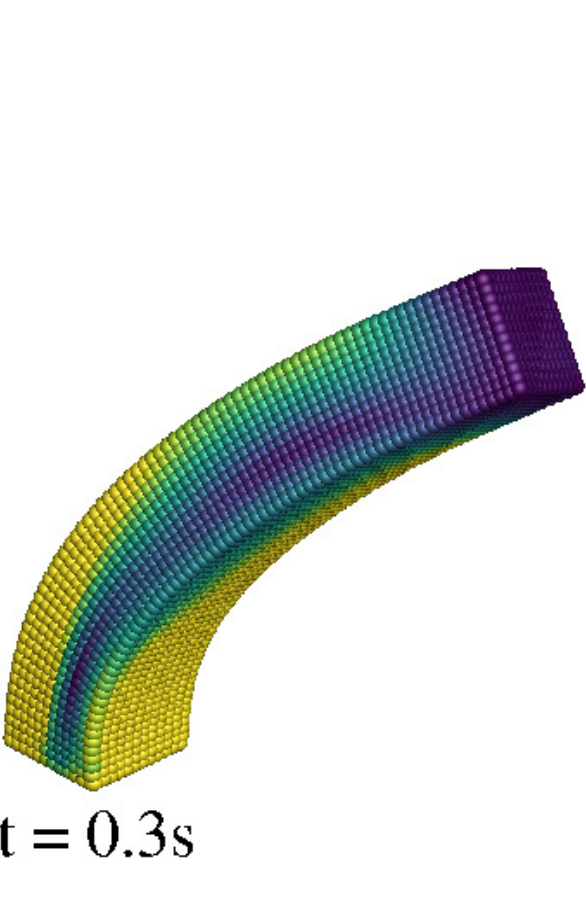}
		\includegraphics[trim = 4.5cm  1mm 9.5cm  1mm, clip, height=0.18\textwidth]{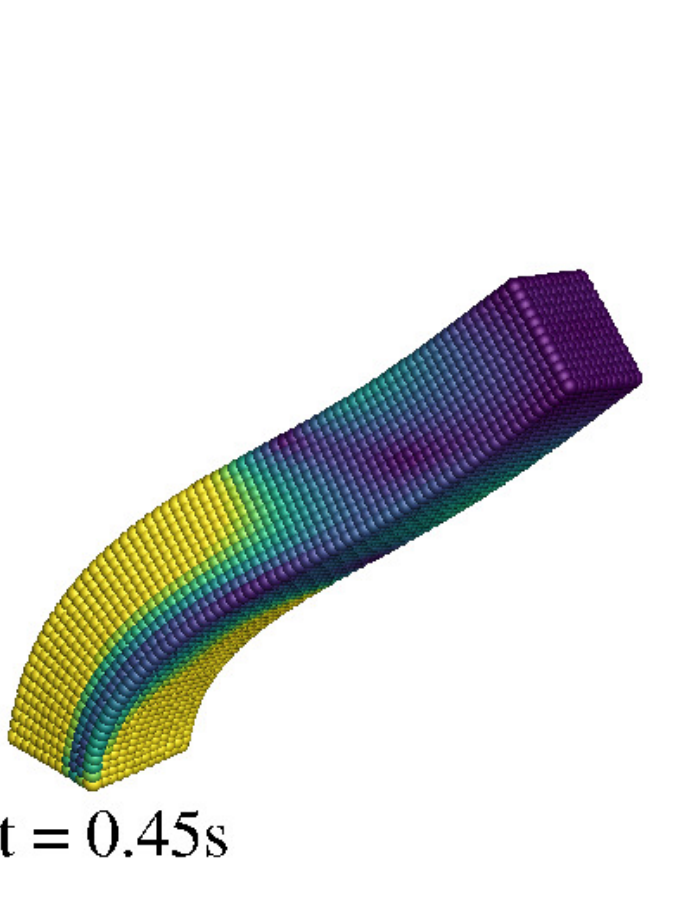}
		\includegraphics[trim = 4.5cm  1mm 9.5cm  1mm, clip, height=0.18\textwidth]{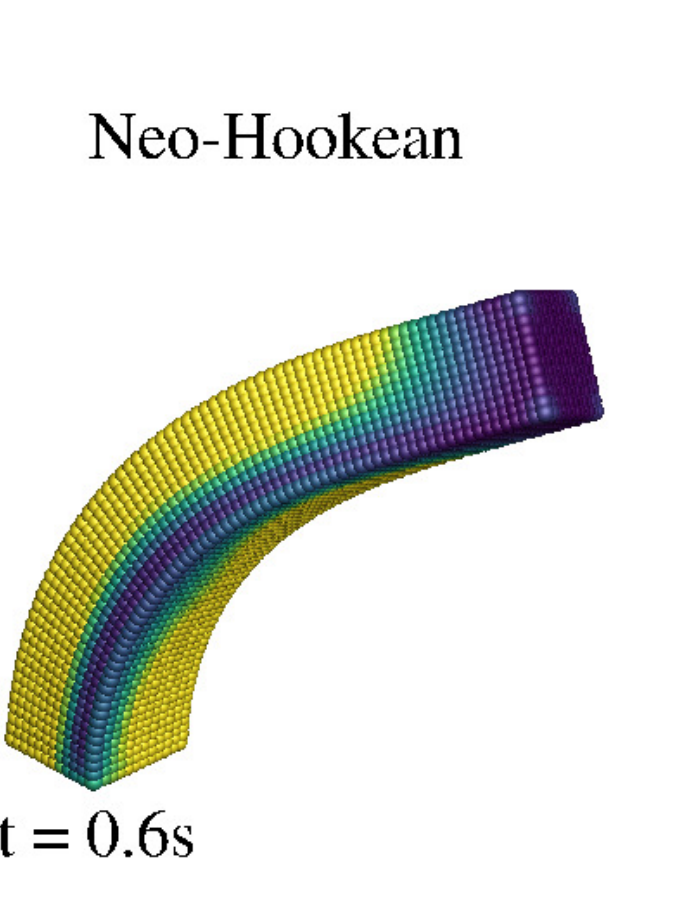} 
		\includegraphics[trim = 2mm 1.5mm 2mm 2mm, clip, width=.42\textwidth]{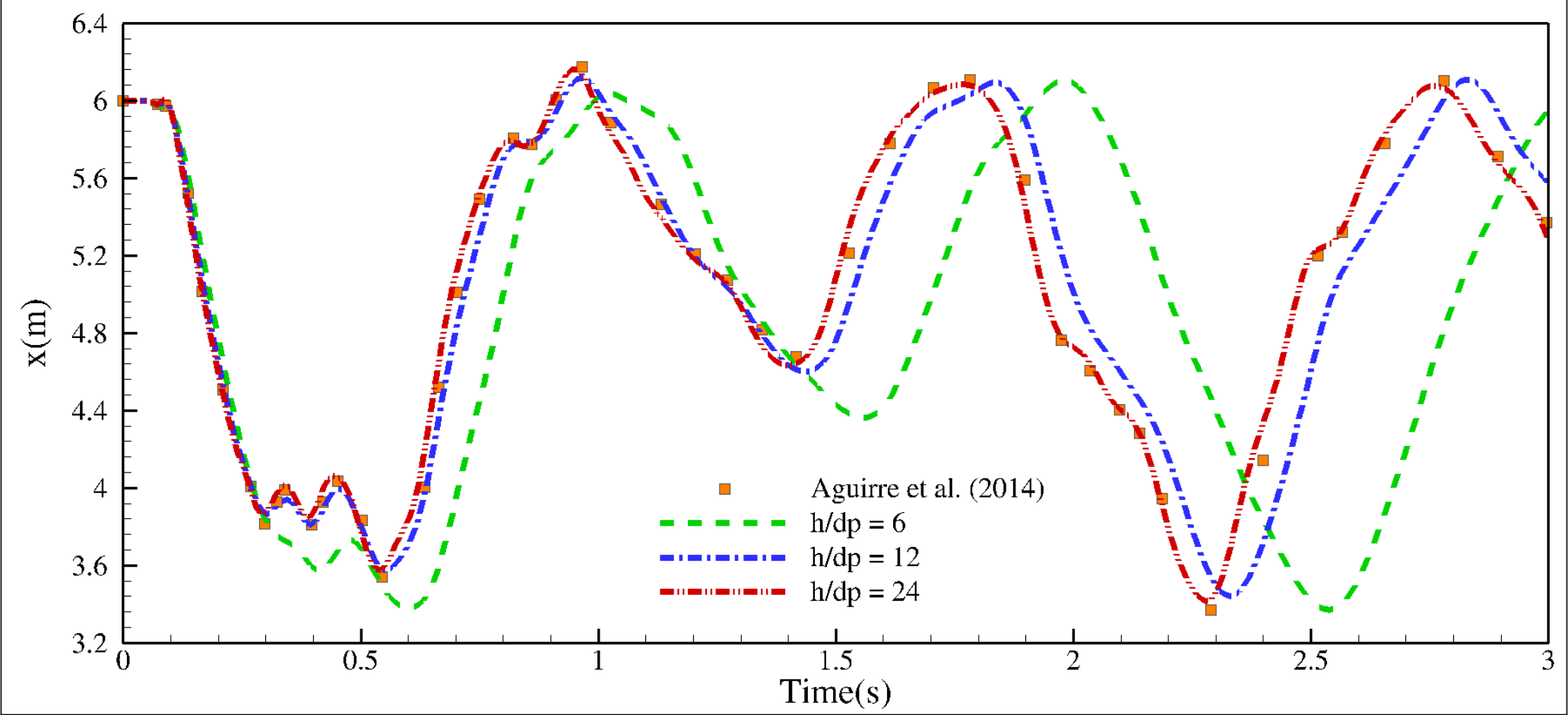}
		\caption{Bending cantilever.}
		\label{figs:solid-bending}
	\end{subfigure}
	\newline
	\begin{subfigure}{\textwidth}
		\centering
		\includegraphics[trim = 1mm 1mm 1mm 1mm, clip,width=0.65\textwidth]{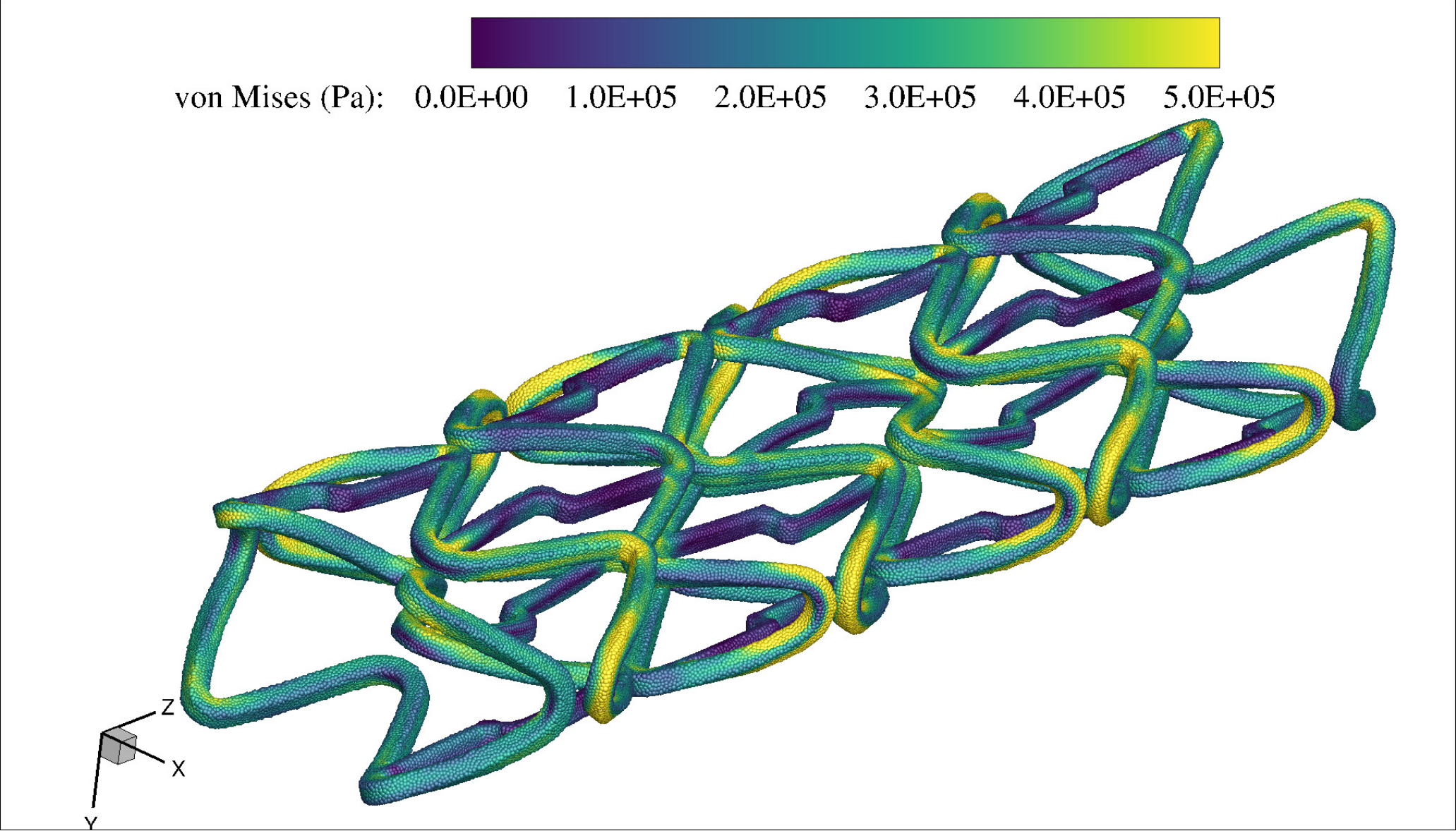}
		\caption{C-shaped stent deformation}
		\label{figs:solid-stent}
	\end{subfigure}
	\caption{Total Lagrangian SPH formulation for solid mechanics: 
		(a) Velocity (left panel) and displacement (right panel) profiles for wave propagation in an elastic cable with different stablized scheme.   
		(b) Deformed configurations with von Mises stress contour (left panel) and the displacement profile at the free end (right panel)
		for bending rubber-like cantilever. 
		(c) The deformed configuration with von Mises stress contour of C-shaped stent.}
	\label{figs:solid-validation}
\end{figure*}

Figure \ref{figs:solid-validation} shows the validations of the total Lagrangian formulation and 
the KV-type damper
for solid mechanics and its applications in bio-mechanics. 
Figure \ref{figs:solid-waveincable} presents 
the time histories of velocity and displacement  
in the length direction at the right tip end of an elastic cable 
which experiences a wave propagation initialized by imposing a velocity $v = 5 ~\text{m} / \text{s}$ 
along the length direction on the right quarter \cite{zhang2021simple}. 
For the original TL formulation without stablized schemes, 
excessive oscillation and similar overshoots in the velocity and displacement profiles are exhibited. 
With both JST  \cite{lee2016new} and KV-type \cite{zhang2021simple} stablized scheme, 
correct velocity and displacement are predicted as expected, 
whereas small overshoots are observed with the JST scheme \cite{lee2016new}. 
Figure \ref{figs:solid-bending} reports the deformed configuration with von Mises stress contour and the displacement of the free end 
for three-dimensional bending rubber-like cantilever whose bottom face is clamped to the ground and its
body is allowed to bend freely by imposing an initial uniform velocity \cite{zhang2021integrative}. 
Compared with the numerical data in literature obtained by mesh-based method \cite{aguirre2014vertex}, 
the nonlinear deformation of the structure is accurately predicted by the TL formulation with the KV-type damper. 
The robustness and versatility in biomedical applications of the TL formulation is 
portrayed in Figure \ref{figs:solid-stent} 
where the C-shaped stent is considered by imposing initial velocity at its the top and bottom \cite{zhang2021simple}.  
To the best knowledge of the authors, 
this is first time that an SPH-based method is successfully extended to the simulation of realistic cardiovascular stent and this will open up interesting possibilities for modeling bio-mechanical applications.
%
\subsection{Steady state solution}\label{sec:steady-state}
Coupling a mechanical system to converge to a static equilibrium state plays a key role in static and dynamic analyses. 
For SPH-based simulations, 
where an explicit time integration scheme is mainly applied, 
fast achieving the static equilibrium state is a very critical numerical challenging. 
Also, 
traditional implicit methods solving the entire system with iterative solvers
are not applicable for large scale applications especially in nonlinear cases.
Instead, dynamic relaxation,
which was originally proposed for finite element method (FEM)   
\cite{otter1966dynamic,belytschko2013nonlinear,jung2018dynamic},
is attractive because its explicit iterative algorithm is simpler and more efficient. 

Generally,
there are two groups of dynamic relaxation technique,
i.e., 
viscous dynamic relaxation (VDR) and kinetic dynamic relaxation (KDR) \cite{rodriguez2011numerical,alamatian2012new,rezaiee2014fictitious}, 
with respect to the manner of applying damping. 
In VDR, an artificial viscous damping term is added into 
the equation of motion to reduce the number of iterations.
While VDR is able to obtain system equilibrium without the loss of momentum conservation, 
it has one difficulty that the solution for nonlinear problems may be path dependent 
and vary with different damping ratio as noted in Ref. \cite{lin2014efficient}.
Also, the relaxation process is slow if insufficient damping ratio is applied, 
on the other hand, 
excessive damping not only can lead to numerical stability issue for explicit damping scheme but also can hinder the system 
from achieving the correct final steady state as the damped velocity can be very small.
The inefficiency of the damping leads low efficiency to achieve the final solution,
which is the main drawback of VDR \cite{lee2011simple}. 
As an alternative approach,
the KDR was first proposed by Cundall \cite{cundall1976explicit} for static structure analysis.
In the KDR,
the damping is introduced to a dynamic system by resetting all current velocities to zero when kinetic energy peak is detected \cite{jung2018dynamic}.
After performing this relaxation procedure several times for successive local peak kinetic energy,
the final state can be achieved and the static equilibrium solution is obtained.
The procedure is simple with no damping coefficient required and has been applied to a wide range of engineering problems \cite{lee2011simple, douthe2009design}.
However,
it also exhibits two obvious drawbacks.
One is that it violates the momentum conservation. 
Therefore, it cannot be applied to moving systems for their equilibrium state analysis due to the variation of the entire mechanical energy.  
The other is that the additional process for kinetic energy peak detection brings extra computational efforts \cite{alamatian2012new,zardi2020new}.
Concerning the development of dynamic relaxation in SPH method, 
the first effort is credited to Lin et al. \cite{lin2014efficient} where 
a dynamic relaxation scheme is developed for shell-based SPH 
by exploiting the formulations in finite element analysis. 
However,
the damping matrix as well as the damping ratio have to be chosen suitably, which is not an easy task.

Recently,
Zhu et al. \cite{zhuAsplitting} 
proposed a VDR-type dynamic relaxation method for SPH by introducing efficient momentum-conservative damping.
Specifically,
an artificial viscous force is first introduced into the momentum conservation equation which is rewritten as
\begin{equation}
\frac{\text{d} \mathbf v}{\text{d} t} =\frac{1}{\rho^0 } \nabla^{0} \cdot \mathbb{P}^T  +  \mathbf{g} + \mathbf f^f  + \mathbf f^{\nu},
\label{momentum-conservation-m}
\end{equation}
where $\mathbf f^{\nu} $ denotes the added damping term. 
Following the TL formulation,
this viscous damping term can be discretized as
\begin{equation}
\mathbf f^{\nu}_a =  \frac{\eta}{\rho_a}  (\nabla^0)_a^2 \mathbf v = \frac{2\eta}{m_a}\sum_{b} V_a V_b \frac{\mathbf v_{ab}}{r_{ab}} \left.  \frac{\partial W_{ab}}{\partial r_{ab}} \right|^0  ,
\label{damping-term}
\end{equation}
where 
$\eta$ is the dynamic viscosity.

Note that along with the acoustic and body-force time-step size criteria, the time-step size during the simulation would be also constrained by
\begin{equation}
\Delta t \leq 0.5\frac{h^2}{\nu D},
\label{viscous-criteria-o}
\end{equation}
where $D= \left\lbrace 1, 2, 3 \right\rbrace$ for one-, two- or three-dimensional cases, respectively. 
This limitation may lead excessive computational efforts especially when large damping ratio and high resolution ratio are applied.
To release this limitation,
an operator splitting scheme \cite{wang2019split, zhang2020integrative} is first applied to 
decouple the momentum conservation equation Eq. \eqref{momentum-conservation-m} into the original momentum part and the damping part.
Then, two operators $S_m$ and $S_d$, which yield
\begin{equation}
S_m \quad :\quad \mathbf v (t+\Delta t) = \mathbf v (t) + (\frac{1}{\rho^0 } \nabla^{0} \cdot \mathbb{P}^T  +  \mathbf g) \Delta t ,
\label{fist-step-m}
\end{equation}
and
\begin{equation}
S_d \quad :\quad \mathbf v (t+\Delta t) = \mathbf v (t) +  \mathbf f^{\nu} \Delta t, 
\label{second-step-m}
\end{equation}
are introduced as forward Euler scheme is adopted for time integration.
Subsequently,
the first order Lie-Trotter splitting scheme \cite{mclachlan2002splitting} is applied to approximate the solution from time $t$ to $t+\Delta t$ by
\begin{equation}
\mathbf v \left( t+\Delta t \right)  =	S_d^{(\Delta t)} \circ	S_m^{(\Delta t)}\mathbf v (t),
\label{operator-splitting}
\end{equation}
where the symbol $\circ$ denotes the separation of each operator and
indicates that $S_d^{(\Delta t)}$ is applied after $S_m^{(\Delta t)}$.
As demonstrated in Refs. \cite{litvinov2010splitting,monaghan2019integration},
a larger time-step size is allowed when suitable implicit formulations are constructed to solve the viscous term.

To avoid large scale matrix operations for traditional implicit formulations,
the entire-domain-related damping step is sequentially split into particle-by-particle operators, e.g. by second-order Strang splitting \cite{strang1968}, as
\begin{equation}
\footnotesize
S_d^{(\Delta t)}  = D_{1}^{(\frac{\Delta t}{2})}  \circ	D_{2}^{(\frac{\Delta t}{2})} \circ \dots \circ D_{N_t-1}^{(\frac{\Delta t}{2})} \circ D_{N_t}^{(\frac{\Delta t}{2})} \circ D_{N_t}^{(\frac{\Delta t}{2})}  \circ	D_{N_t-1}^{(\frac{\Delta t}{2})} \circ \dots \circ D_{2}^{(\frac{\Delta t}{2})} \circ D_{1}^{(\frac{\Delta t}{2})},
\label{particle-by-particle splitting}
\end{equation}
where $N_t$ denotes the total number of particles and $D_a$ the split damping operator corresponding to particle $a$.
Two efficient schemes, the particle-by-particle splitting scheme and the pairwise splitting scheme, are then proposed for the local damping operator \cite{zhuAsplitting}.
The new time-step velocity updating for the entire field can be thus achieved 
by carrying out the local split operator to all particles for half a time step and then performing the operator to these particles in a reverse sequence for another half time step \cite{nguyen2009mass} as shown in Eq.\eqref{particle-by-particle splitting}.
\subsubsection{Particle-by-particle splitting scheme} \label{sec:PP-splitting}
In an implicit formulation, 
the local damping term in Eq. \eqref{damping-term} can be rewritten as 
\begin{equation}
\mathbf f^{\nu} = \left( \frac{\text{d} \mathbf v_a}{\text{d} t}\right) ^{\nu}  = \frac{2\eta}{m_a}\sum_{b} V_a V_b \frac{\mathbf v^{n+1}_{ab} }{r_{ab}} \left.  \frac{\partial W_{ab}}{\partial r_{ab}} \right|^0,
\label{damping-term-implicit}
\end{equation}
where $\mathbf v^{n+1}_{ij}  = \mathbf v^{n}_{ij} + \text d \mathbf v_i - \text d \mathbf v_j$
with $\text{d} \mathbf v_i$ and $\text{d} \mathbf v_j$ representing the incremental change of velocity for particle $a$ and its neighboring particles $b$ induced by viscous acceleration. 
After denoting
\begin{equation}
B_b = 2 \eta  V_a V_b \frac{1}{r_{ab}}  \left.  \frac{\partial W_{ab}}{\partial r_{ab}} \right|^0 \text{d} t ,
\label{parameters1}
\end{equation}
and 
\begin{equation}
\mathbf E_a = - 2\eta \sum_{b} V_a V_b \frac{ \mathbf v^{n}_{ab} }{r_{ab}}  \left.  \frac{\partial W_{ab}}{\partial r_{ab}} \right|^0 \text{d} t= - \sum_{b} B_b \mathbf v^{n}_{ab},
\label{parameters2}
\end{equation}
the implicit formulation Eq.\eqref{damping-term-implicit} can be simplified to  
\begin{equation}
\mathbf E_a = \left( \sum_{b} B_b - m_a\right) \text{d} \mathbf v_a - \sum_{b} B_b \text{d} \mathbf v_b.
\label{implicit-equation}
\end{equation}
A gradient descent method \cite{nielsen2015neural} is then adopted to evaluate $\text{d} \mathbf v_b$ and $\text{d} \mathbf v_b$. 
In Eq. \eqref{implicit-equation}, 
the gradient $\nabla \mathbf E_a$ with respect to variables $\left( \text{d} \mathbf v_a, \text{d} \mathbf v_1, \text{d} \mathbf v_2, \cdots, \text{d} \mathbf v_N \right)^T$ gives
\begin{equation}
\nabla \mathbf E_a = \left( \sum_{j} B_b - m_a, -B_1, -B_2, \cdots, -B_N \right)^T.
\label{gradient-E}
\end{equation}
Let 
\begin{equation}
\left( \text{d} \mathbf v_a, \text{d} \mathbf v_1, \text{d} \mathbf v_2, \cdots, \text{d} \mathbf v_N \right)^T = \mathbf k \nabla \mathbf E_a,
\label{gradient-learning-rate}
\end{equation}
where $\mathbf k$ is known as the learning rate \cite{nielsen2015neural}.
By substituting Eqs. \eqref{gradient-E} and \eqref{gradient-learning-rate} into Eq. \eqref{implicit-equation},
the learning rate can be obtained, i.e. 
\begin{equation}
\mathbf k=\left(\left( \sum_{b} B_b - m_a\right)^2+ \sum_{b} \left( B_b\right)^2 \right)^{-1} \mathbf E_a. 
\label{learning-rate-solve}
\end{equation}
According to Eqs. \eqref{gradient-E} and \eqref{gradient-learning-rate}, 
the incremental change of velocity by viscous damping can be thus achieved.
In order to ensure momentum conservation,
the velocities of neighboring particles are then modified by the above predicted incremental change.
In summary,
the local update of velocities includes two steps as follows.
The first step calculates the incremental change for velocity by gradient descent method, i.e.,
\begin{equation}
\begin{cases}
\mathbf v_a^{n+1}&=\mathbf v_a^{n} + \text{d} \mathbf v_a = \mathbf v_a^{n} + \left( \sum_{b} B_b - m_a\right) \mathbf k\\
\mathbf v_1^{p}&=\mathbf v_1^{n} + \text{d} \mathbf v_1 = \mathbf v_1^{n}  -B_1 \mathbf k   \\
\mathbf v_2^{p}&=\mathbf v_2^{n} + \text{d} \mathbf v_2 = \mathbf v_2^{n}  -B_2 \mathbf k   \\
&\cdots \\
\mathbf v_N^{p}&=\mathbf v_N^{n} + \text{d} \mathbf v_N = \mathbf v_N^{n}  -B_N \mathbf k   \\
\end{cases},
\label{first-step}
\end{equation} 
where the superscript $p$ denotes the predicted value.
The second step ensures momentum conservation, which yields
\begin{equation}
\begin{cases}
\mathbf v_1^{n+1}&= \mathbf v_1^{n} -B_1 \left( \mathbf v_i^{n+1} -\mathbf v_1^{p} \right) / m_1   \\
\mathbf v_2^{n+1}&= \mathbf v_2^{n} -B_2 \left( \mathbf v_i^{n+1} -\mathbf v_2^{p} \right) / m_2   \\
&\cdots \\
\mathbf v_N^{n+1}&= \mathbf v_N^{n} -B_N \left( \mathbf v_i^{n+1} -\mathbf v_N^{p} \right) / m_N   \\
\end{cases}.
\label{second-step}
\end{equation} 

As the velocities are updated implicitly,
much larger time-step size is allowed and the following viscous criterion
\begin{equation}
\Delta t \leq 50\frac{h^2}{\nu D},
\label{viscous-criteria}
\end{equation}
which is about 100 times larger than the corresponding explicit method as presented in Eq. \eqref{viscous-criteria-o}, is adopted.
For solid, the artificial dynamic viscosity  $\eta = \rho \nu$ as shown in Eq. \eqref{momentum-conservation-m}
is defined by
\begin{equation}
\eta = \frac{1}{4} \beta \rho \sqrt{\frac{E}{\rho}}L=\frac{\beta}{4} \sqrt{\rho E}L,
\label{viscous-parameter}
\end{equation}
where $E$ is the Young's modulus, $L$ the characteristic length scale of the problem and $\beta$ denotes a parameter relating to the body shape. Note that choosing different value for the parameter $\beta$ 
may alter, though not much, the speed to final state. 
For fluid,
the viscosity is defined by
\begin{equation}
\eta =  \rho U_{\text{max}} L .
\label{viscous-parameter-fluids}
\end{equation}
\subsubsection{Pairwise splitting scheme} \label{sec:PS-splitting}
The pairwise splitting scheme is inspired by the work of Ref. \cite{litvinov2010splitting}, 
where particle velocity is updated implicitly and locally in a pairwise fashion. 
By adopting the second-order Strang splitting \cite{strang1968},
the damping operator corresponding to each particle $i$ as given in Eq. \eqref{particle-by-particle splitting} 
is further split based on its neighbors, i.e.,
\begin{equation}
\footnotesize
D_a^{(\Delta t)}  = D_{a,1}^{(\frac{\Delta t}{2})}  \circ	D_{a,2}^{(\frac{\Delta t}{2})}  \circ \dots \circ  D_{a,N-1}^{(\frac{\Delta t}{2})} \circ  D_{a,N}^{(\frac{\Delta t}{2})} \circ D_{a,N}^{(\frac{\Delta t}{2})}  \circ	D_{a,N-1}^{(\frac{\Delta t}{2})}  \circ \dots \circ  D_{a,2}^{(\frac{\Delta t}{2})} \circ  D_{a,1}^{(\frac{\Delta t}{2})},
\label{pairwise-splitting}
\end{equation}
where $D_{a,b}^{(\frac{\Delta t}{2})}$ denotes the interaction between particle $a$ and its neighbors.
Specifically,
the incremental changes for velocity of a specific particle pair induced by viscosity can be written in implicit form as
\begin{equation}
\begin{cases}
m_a \text{d} \mathbf v_a =  B_{b} \left( \mathbf v_{ab} + \text d \mathbf v_a - \text d \mathbf v_b\right) \\
m_b \text{d} \mathbf v_b = -B_{b} \left( \mathbf v_{ab} + \text d \mathbf v_a - \text d \mathbf v_b\right) \\
\end{cases}.
\label{pairwise-eq}
\end{equation} 
Here, 
$B_b$ is defined in Eq. \eqref{parameters1} and it is obvious that this process does not change the conservation of momentum.
Then, $\text{d} \mathbf v_a$ and $\text{d} \mathbf v_b$ can be obtained straightforwardly by solving Eq. \eqref{pairwise-eq},
which yields
\begin{equation}
\begin{cases}
\text{d} \mathbf v_a = m_b\frac{  B_{b} \mathbf v_{ab}}{ m_a m_b - \left( m_a + m_b\right) B_b } \\
\text{d} \mathbf v_b = - m_a\frac{ B_{b} \mathbf v_{ab}}{ m_a m_b - \left( m_a + m_b\right) B_b } \\
\end{cases}.
\label{pairwise-solve}
\end{equation}
By sweeping over all neighboring particle pairs for half a time step and then over these particles in a reverse sequence for another half time step,
the incremental changes for velocity of particle $a$ and all its neighbors can be thus achieved.
Compared to the particle-by-particle splitting method,
this scheme leads more errors in solving viscosity due to the further splitting in pairwise fashion.
However, it is unconditional stable and thus more suitable for problems with high spatial resolution and high damping ratio. 
\subsubsection{Random-choice strategy} \label{sec:random-choice}
It is worth noting that
the added viscous force would hinder the system achieving correct steady state 
especially when the damping radio is large,
which may lead to the different solutions of the nonlinear problems with different damping ratio \cite{lin2014efficient}.
Thus, a suitable damping ratio has to be selected \cite{lin2014efficient,crisfield1997non}
for faster reaching to final state of the system with the aforementioned viscous damping methods.

To avoid this damping-ratio-related problem and relax the limitation on the choice of large damping ratio,
a random-choice strategy is presented, 
in which the viscosity term is imposed randomly rather than at every time step.
To achieve this, the artificial dynamic viscosity $\eta$ is modified as
\begin{equation}
	\tilde{\eta} = 
	\begin{cases}
	\eta / \alpha  \quad \quad & {\rm if} \quad \alpha > \phi\\
	0      & {\rm otherswise}   \\
	\end{cases},
	\label{viscosity-reset}
\end{equation}
where $\phi$ is a random number uniformly distributed between $0$ to $1$, 
and $\alpha=0.2$ a parameter determining the probability.
Therefore, the resistance on displacement 
induced by the large artificial viscosity can be released randomly,
which eliminates the damping-ratio-related issue and accelerates the achievement to the final state.
Note that this strategy also helps to save much computational cost 
since the computation of damping is only carried out 
at a small fraction of time steps.

\begin{figure*}[htb!]
	\centering
	\includegraphics[trim = 8cm 7cm 8cm 8cm, clip,width=0.5\textwidth]{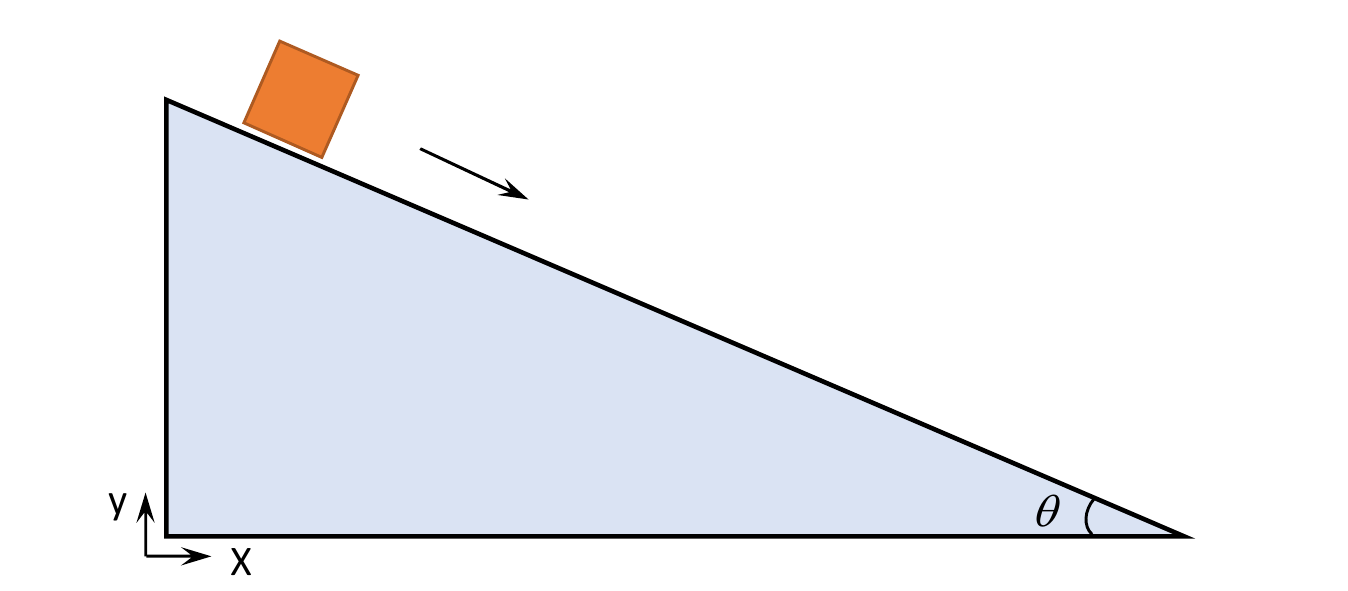} \\
	\includegraphics[trim = 0.5cm 0.5cm 0.5cm 0.5cm, clip,width=0.4\textwidth]{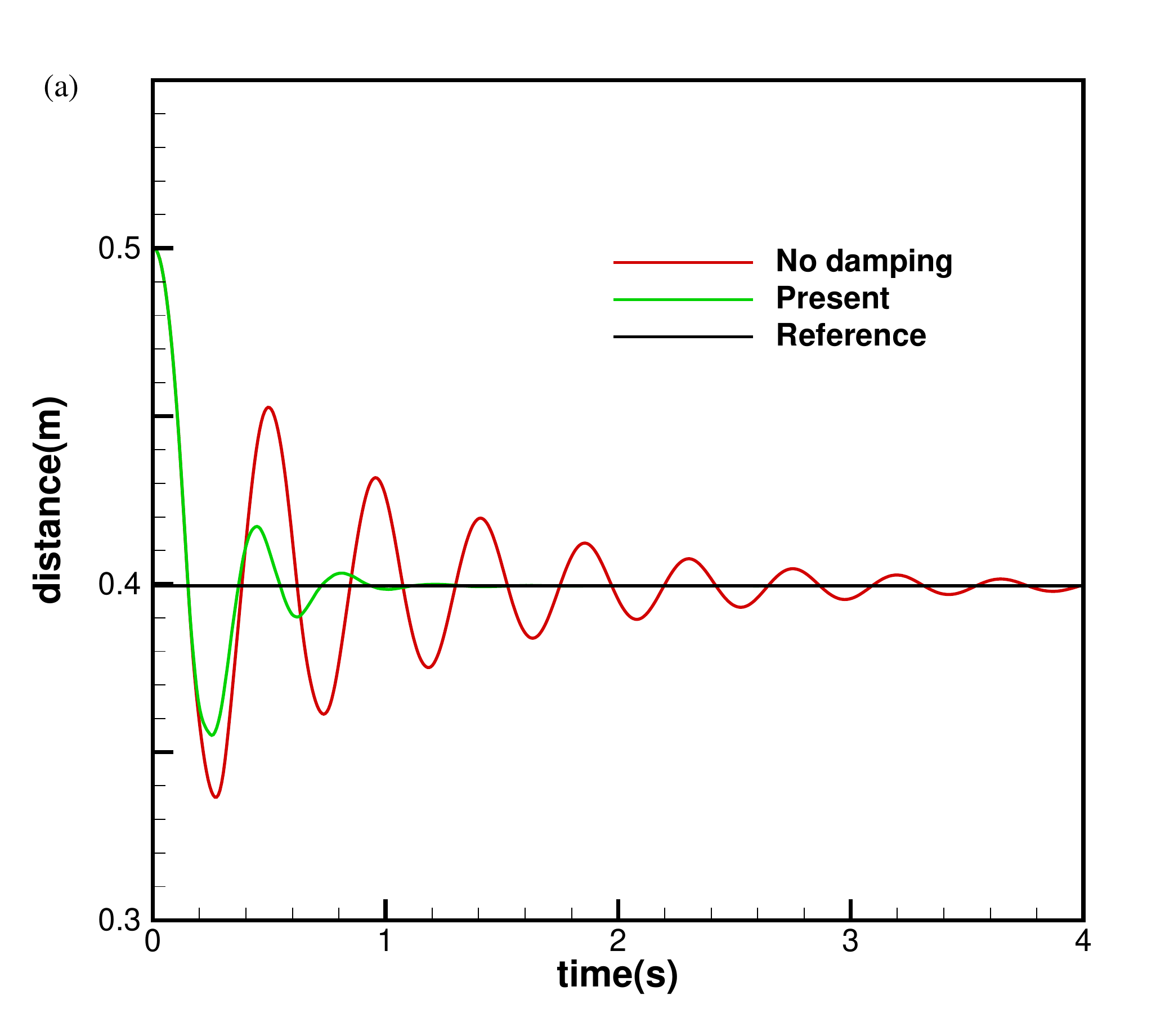}
	\includegraphics[trim = 0.5cm 0.5cm 0.5cm 0.5cm, clip,width=0.4\textwidth]{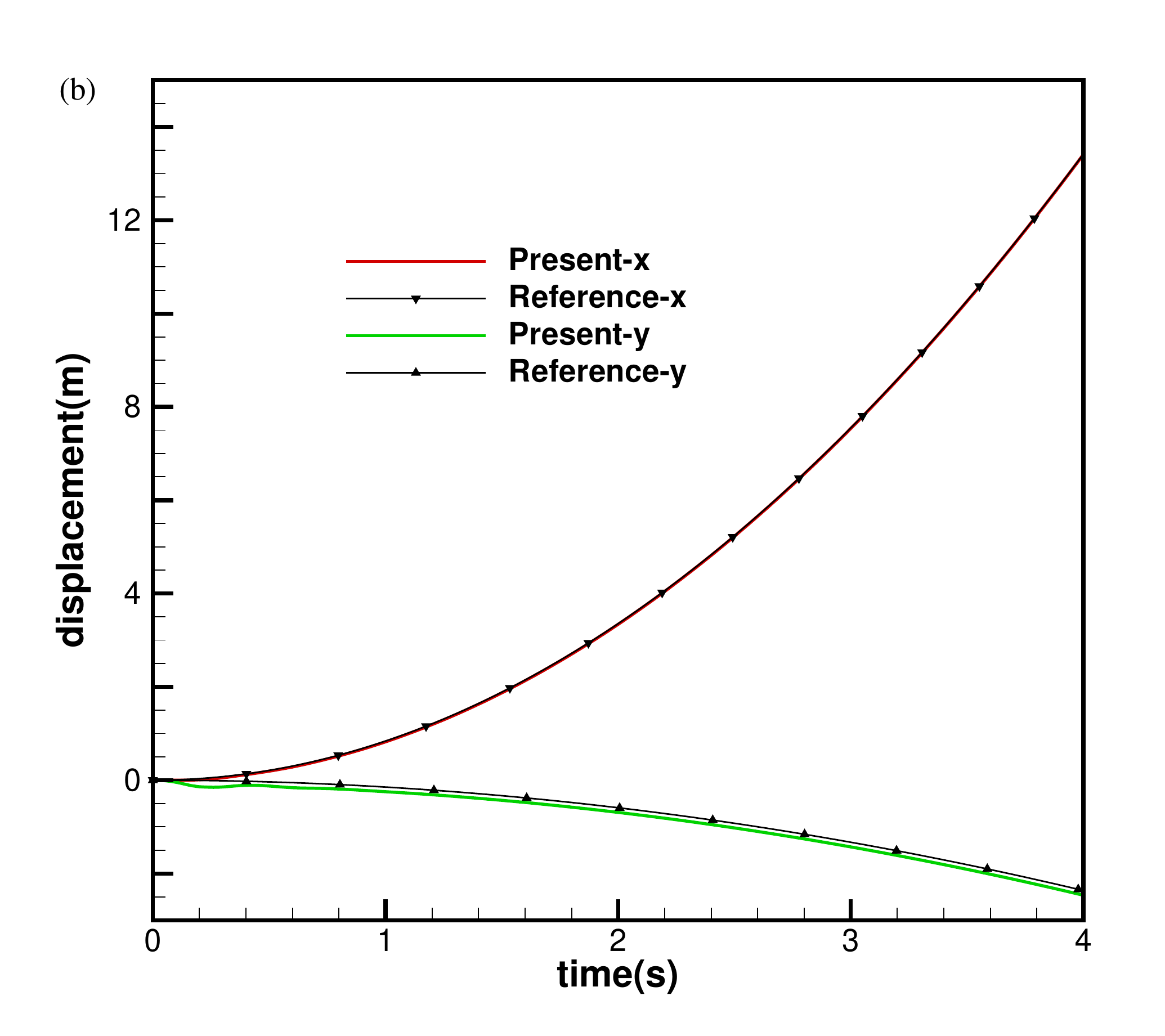}
	\caption{Numerical investigation of the block sliding along slope. The configuration (upper panel) 
		and the time histories of the distance between the block center and the slope (a) and displacement of the block center in $x-$ and $y-$ directions (b). }
	\label{figs:block-sliding}
\end{figure*}

The performance of the VDR-type dynamics relaxation scheme is validated by considering 
an elastic block sliding along a smooth slope accelerated by the gravity \cite{zhuAsplitting}. 
Figure \ref{figs:block-sliding} presents the time histories of the distance between the block center and the slope, 
and the displacements of the block center in $x-$ and $y-$ direction. 
With the dynamics relaxation scheme, 
the steady state is quickly achieved by surpassing oscillations. 
%
\subsection{Hourglass control scheme}\label{sec:hourglass}
In the FEM method, 
hourglass modes represent zero-energy modes in the sense that 
the element deforms without an associated increase of the elastic energy 
when reduced-integration elements are employed. 
Insufficient integral points lead to rank deficiency of FEM stiffness matrix, 
which further causes the non-uniqueness of governing equations. 
Similarly, 
the SPH method is also susceptible to hourglass modes as all field variables and their derivatives are evaluated at the same position \cite{vignjevic2000treatment, dyka1997stress}.
These modes cannot be detected and can be developing over time \cite{belytschko2000unified} if the number of integration points is reduced, 
and therefore the solution is polluted with arbitrary amounts of strain energy and entirely dominated by these modes \cite{ganzenmuller2015hourglass}.

To address the hourglass modes in SPH method,
Swegle et al. \cite{swegle1994analysis} proposed to replace the strain measure by a non-local approximation based on gradient approach.
Beissel and Belytschko \cite{beissel1996nodal} treated these singular modes by the addition to the potential energy functional of a stabilization term which contains the square of the residual of the equilibrium equation.
A more straightforward idea is to introduce additional integral points to calculate derivatives away from particles with zero derivatives of the kernel function \cite{randles2000normalized, randles1999neighbors}. 
Two sets of points, velocity points and stress points,
are applied to discretize the calculation domain, 
one carrying the velocity and the other carrying the stress.
While the velocity gradient and stress are computed on stress points, 
the divergence of stress is calculated using stress points as neighbor particles and then sampled at velocity points.
The enhanced stability of additional stress points was confirmed\cite{vignjevic2006sph, rabczuk2004stable, xiao2005material, maurel2008sph, ming2015smoothed}.
However, 
estimating the stresses induces extra computational efforts and 
how to place the stress points is till not fully addressed. 

When the FEM using the one-point reduced finite element,
a mean deformation gradient is obtained, 
resulting mean strain and stress over a single element.
The SPH method also evaluates a mean strain and stress at the center of a particle through the weighted averaging over the neighboring particles.
Recently, Ganzenm{\"u}ller \cite{ganzenmuller2015hourglass} recognized the analogy between SPH collocation and FEM using  the one-point reduced element.
Inspired by Flanagan and Belytschko \cite{flanagan1981uniform}, 
Ganzenm{\"u}ller pointed out 
that the mean stress-strain description can only represent a fully linear velocity field,
which implies node or particle displacement should be exactly described by the deformation gradient, 
and node or particle displacement incompatible with the linear deformation field are identified as the hourglass modes.
The distance of particle $a$ and $b$ can be estimated by the deformation gradient $\mathbb{F}_a$ as
\begin{equation}
	\left< \mathbf{r}_{ab}\right>^a =  \mathbb{F}_a \mathbf{r}_{ab}^0.
	\label{estimated_distance}
\end{equation}
When the hourglass mode exists, the estimated distance between the two particles is inconsistent with the actual distance. 
This difference defines the error vector 
\begin{equation}
	\Phi_{ab}^a =  \left< \mathbf{r}_{ab}\right>^a  - \mathbf{r}_{ab}.
	\label{error_vector}
\end{equation}
A correction force, proportional to the error vector, is introduced,
i.e., an artificial stiffness is added to counteract hourglass modes.
Then, the scalar $\delta_{ab}^a$ can be defined as
\begin{equation}
	\delta_{ab}^a =  \frac{\Phi_{ab}^a \mathbf{r}_{ab}}
	{\left| \mathbf{r}_{ab} \right|}.
	\label{error_scalar}
\end{equation}
$\delta_{ab}^a$ is the projection of error vector onto current particle distance vector.
The hourglass correction force per unit volume is derived as 
\begin{equation}
	\hat{\mathbf{f}}_{ab}^a = - \frac{E}{\left| \mathbf{r}_{ab}^0 \right|} 
	\frac{\delta_{ab}^a}{\left| \mathbf{r}_{ab}^0 \right|}
	\frac{\mathbf{r}_{ab}}{\left| \mathbf{r}_{ab} \right|}.
	\label{correction_force_i}
\end{equation}
The stiffness is linear in $\delta_{ab}^a$, and described using the Young's modulus $E$ of the material.
A normalized smoothing kernel is used to be consistent with SPH collocation,
and a explicit systematization via the arithmetic mean is applied as the hourglass forces,
$\hat{\mathbf{f}}_{ab}^a $ and $\hat{\mathbf{f}}_{ba}^b $, 
are unsymmetrical.
Therefore,
the smoothed and symmetric correction force between particle $a$ and $b$ can be expressed as
\begin{equation}
	\hat{\mathbf{f}}_{ab}^{HG} = - \frac{1}{2} \alpha W_{ab}
	\left( \hat{\mathbf{f}}_{ab}^a + \hat{\mathbf{f}}_{ba}^b \right),
	\label{correction_force_ij}
\end{equation}
where $\alpha$ is a dimensionless constant which determines the amplitude of the hourglass control. 
Finally, the total hourglass correction force of particle $a$ over all neighboring particles is 
\begin{equation}
	\begin{split}
		\mathbf{f}_{ab}^{HG} &= - V_a \sum_b V_b W_{ab}  \hat{\mathbf{f}}_{ab}^{HG} \\
		& = \sum_b - \frac{1}{2} \alpha 
		\frac{V_a V_b W_{ab}}{{\left| \mathbf{r}_{ab}^0 \right|}^2}
		\left(E_a \delta_{ab}^a + E_b \delta_{ba}^b\right)
		\frac{ \mathbf{r}_{ab}}{\left| \mathbf{r}_{ab} \right|} ,
	\end{split}
	\label{correction_force}
\end{equation}
This hourglass control algorithm has successfully applied in geomaterials \cite{islam2019stabilized}, FSI with GPU acceleration \cite{o2021fluid, zhan2019stabilized}, etc.
%
%
\section{Fluid-structure interaction}\label{sec:fsi}
Fluid-structure interaction, 
where the structure represents either movable rigid or flexible structures
is ubiquitous in natural phenomena, e.g. 
aerial animal flying, aquatic animal swimming and blood circulation, 
and also plays a crucial role in the design of many engineering systems, 
e.g. automobile, aircraft, spacecraft, engines and energy harvesting device. 
This phenomenon is characterized by the multiphysics coupling between the laws that describe fluid dynamics and structural mechanics. 
Due to the intrinsic complexity of the interaction between a movable or flexible structure and a surrounding or internal fluid flow, 
computational study of this type multiphysics problems is highly challenging. 

The conventional FSI algorithms are based on 
mesh-based methods,  
i.e., the finite difference method (FDM) \cite{forsythe1960finite}, 
the FEM \cite{tezduyar1992new} 
and the finite volume method (FVM) \cite{versteeg2007introduction}, 
by implementing monolithic or partitioned approach. 
The monolithic approach treats the coupled problem as a whole with proper combination of the sub-system, 
applying a single solver to simultaneously solve the governing equations of the fluid and solid dynamics 
in FSI.
One typical example is the arbitrary Lagrangian-Eulerian (ALE) description of the FEM method \cite{souli2000ale} 
where moving mesh is introduced to flud discretization 
for addressing the issues of unacceptable mesh distortion near the structure undergoing large deformations. 
This approach encounters difficulties of the convective terms treatment 
and the challenging of complex mesh regeneration, 
in particular when large structure deformation is evolved \cite{onate1996finite}. 
The partitioned approach strives to solve each sub-problem separately, 
applying computational fluid and computational solid solves for the fluid and solid, respectively, 
with communication of FSI interface data. 
Typical example are the immersed-boundary method (IBM) \cite{peskin2002immersed} 
which utilizes two overlapped Lagrangian and Eulerian meshes. 
In IBM method, 
the fluid equation is solved on the Eulerian mesh 
and the effects of solid structure are taken into account 
by distributing the forces computed on the deformed Lagrangian mesh to the Eulerian counterpart using proper kernel function.  
Compared with the monolithic one, 
the partitioned approach suffers from Lagrangian-Eulerian mismatches on the kinematics and the distribution of solid structure forces due to the fairly weak coupling formulation.

As an alternative for tackling FSI problems, 
the meshless methods, 
i.e., 
the SPH \cite{lucy1977numerical, gingold1977smoothed, zhang2017weakly}, 
the MPS \cite{koshizuka1996moving} and the DEM \cite{mishra1992discrete}
provide unified monolithic approach with pure Lagrangian discretization of both the fluid and solid equations.
In recent years, 
the meshless methods have attracted significant attention in studying FSI problems, 
owing to their peculiar advantages in handling material interfaces \cite{rezavand2020weakly, zhang2017generalized} 
and the capability of capturing violent events such as wave impact and breaking \cite{zhang2019weakly}. 
Promising results have been obtained by the weakly-compressible SPH method 
\cite{antoci2007numerical, liu2019smoothed, zhang2019smoothed, zhang2019weakly, zhan2019stabilized, oger2009simulations, liu2013numerical, wang2019sph, han2018sph, sun2021accurate}, 
incompressilbe SPH method \cite{khayyer2018enhanced, rafiee2009sph, sun2019study}, 
MPS method \cite{khayyer2019multi}, 
SPH-DEM or MPS-DEM methods \cite{ren2013sph, xie2021numerical} 
and their combinations with FEM by using partition approach 
\cite{yang2012free, zhang2019mps, chen2019numerical, hermange20193d, zhang2021partitioned}.  

In this part, 
we focus on the recent developments of unified SPH method for FSI problems and special attention are devoted to the FSI interface treatment, 
multi-resolution discritzation and time stepping schemes, 
which are key components of the accurate and efficient FSI algorithm. 
Concerning the applications of FSI algorithm in engineering, 
comprehensive reviews can be found in Refs. 
\cite{luo2021particle, liu2019smoothed, zhang2017smoothed, ye2019smoothed, gotoh2021entirely, lyu2022review}.
\subsection{Treatments of FSI interface}\label{sec:fsi-interface}
In unified SPH-FSI computation, 
the fluid-strucutre coupling is resolved by treating the surrounding movable or flexible structure 
as moving solid boundary for fluid 
with imposing free- or no-slip boundary condition at the fluid-structure interface.

Concerning the treatment of movable or flexible boundary,
several methods have been proposed and they are generally categorized into three schemes, 
i.e., boundary force particle, dummy particle and one-sided Riemann scheme. 
In the first scheme, 
the structure particle is behaving as the moving boundary force particle \cite{monaghan2009sph} 
and a repulsive force is introduced to prevent particle penetration. 
This scheme was first proposed by Monaghan and Kajtar \cite{monaghan2009sph}, 
where one layer boundary particles provide Lennard-Jones potential repulsive force for fluid particle, 
and was improved by Liu et al. \cite{liu2012treatment} and further by Zhang et al. \cite{zhang2018meshfree} 
with introducing a new numerical approximation scheme for estimating field functions of solid particles.
In the second scheme,  
the structure is presented by the dummy particle whose velocity and pressure are interpolated from fluid particles 
\cite{marrone2011delta, adami2012generalized} 
to solve governing equations.
Notwithstanding its wide application, 
this approach exhibits excessive computational efforts which are inherently expensive in three-dimensional simulations 
due to the data interpolation.
In the third scheme, 
the one-sided Riemann problem is constructed along the structure norm and 
solved to determine the FSI coupling \cite{zhang2017weakly}.  
This scheme has demonstrated its robustness, accuracy and efficiency \cite{zhang2020dual, zhang2017weakly, zhang2020multi}. 

Note that there are other schemes, 
for example ghost particles \cite{liu2003smoothed} and semi-analytical approach \cite{ferrand2013unified, mayrhofer2015unified,chiron2019fast}, 
have been applied for handing the FSI interface in the coupling of particle-based method and FEM \cite{yang2012free, hermange20193d}, 
which is not included in this survey.  
\subsubsection{Boundary force particle}
\begin{figure*}[htb!]
	\centering
	\includegraphics[trim = 1mm 1mm 1mm 1mm, clip, width=.495\textwidth]{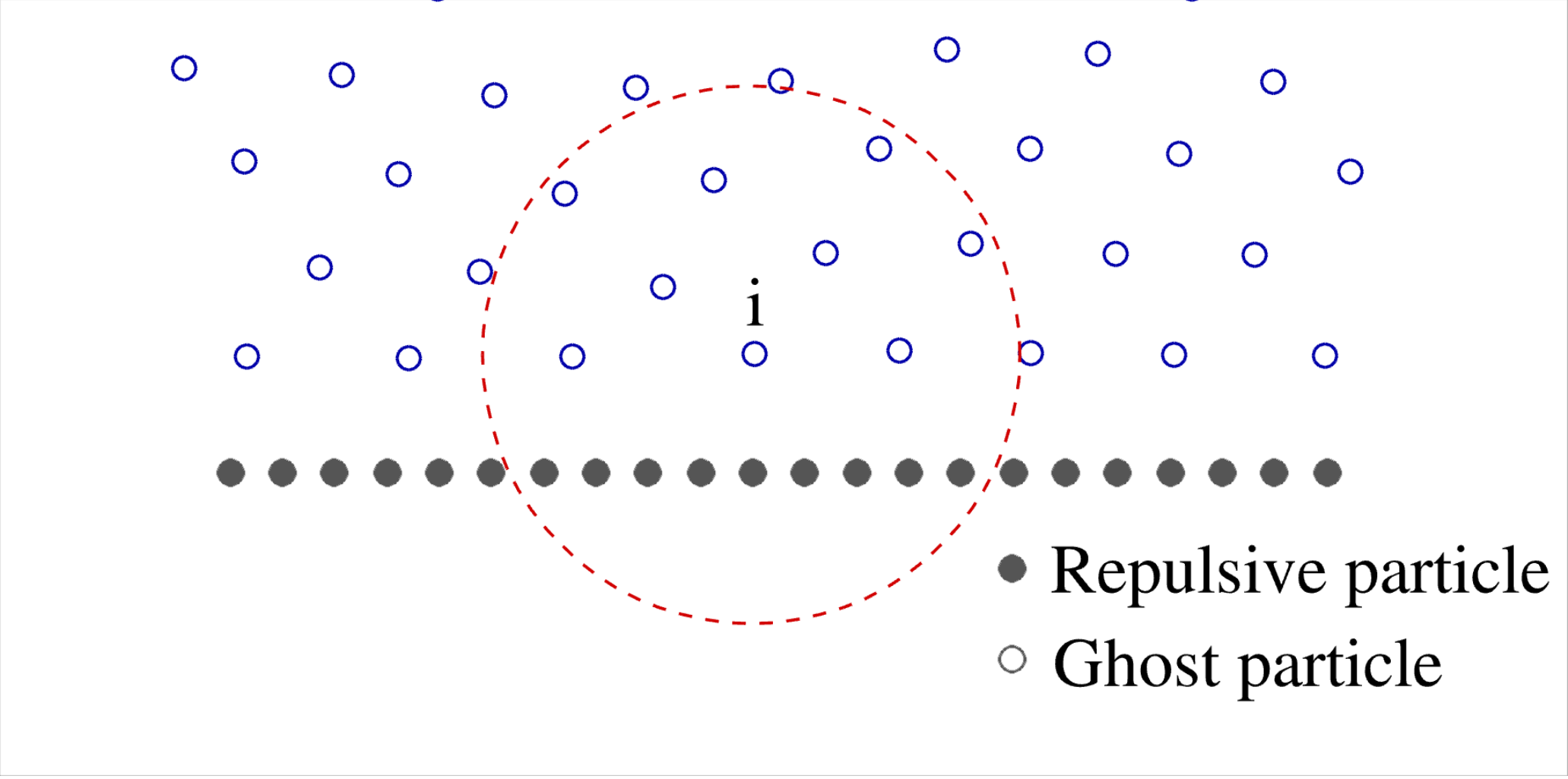}
	\includegraphics[trim = 2cm 1mm 1mm 1mm, clip, width=.45\textwidth]{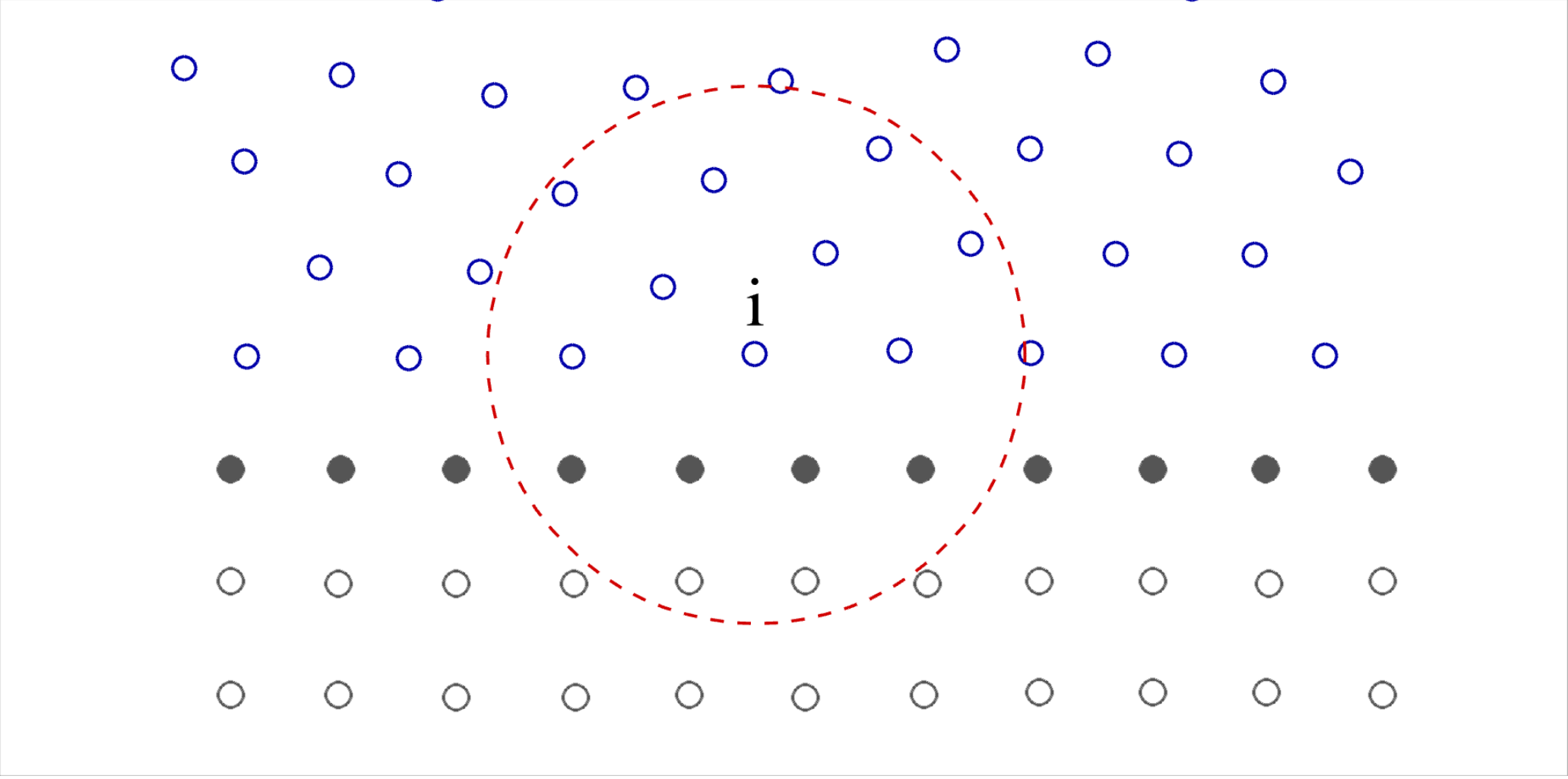}
	\caption{Sketch of multi-phase particles interacting with solid particles along the normal vector 
		through the one-side Riemann problem.}
	\label{figs:fsi-bfparticle}
\end{figure*}
The boundary force particle scheme is firstly proposed by Monaghan and Kajtar \cite{monaghan2009sph} 
for approximating arbitrarily shaped boundaries. 
The movable boundary is modeled by one layer boundary particles with particle spacing as a factor of $3$ 
less than the fluid particle spacing, 
as shown in the left panel of Figure \ref{figs:fsi-bfparticle}, 
and these particles interact with the fluid particles by forces depending on the separation of the particles 
and pre-determiend parameters. 
Then, 
the force $\mathbf{f}_i^{s}$ in Eq. \eqref{eq:fluid-governing} 
acting on a fluid particle $i$, due to the presence of the neighboring solid particle $a$, 
is given by \cite{monaghan2009sph, valizadeh2015study} 
\begin{equation}
	\mathbf f_{i}^s =  \beta \sum_a \phi_{ia} \frac{\mathbf r_{ia}}{r_{ia}( r_{ia} - \Delta)},
\end{equation}
where
\begin{equation}
\footnotesize
\phi_{ai} =
\begin{cases}
	\frac{1.8}{32}(1+\frac{5}{2}q + 2q^2)(1.5-q)^5 \quad q < 1.5\\
	0 \quad q \leq 1.5
\end{cases}; q = r_{ia} /h. 
\end{equation}
Here, $\beta = \nu_{max} / (\rho_0 dp_0^2)$ with $\nu = \frac{1}{8}\alpha h c$ 
and $\Delta = dp_0 / 3$ is the separation of boundary particles. 

This solid boundary treatment is widely applied in SPH simulations 
for dealing with FSI interface treatment involving complex solid shape 
in single- and multi-phase flows \cite{monaghan2009sph, monaghan2013simple}, 
and also extended for SPH-FEM coupling scheme \cite{yang2012free}.
However, 
using the artificial repulsive forces 
violates the kernel truncation in the immediate vicinity 
of the solid boundaries as only a single layer of particles are required to mimic the boundaries. 
To address this issue,
Liu et al. \cite{liu2012treatment} developed a coupled dynamic SBT scheme, 
termed as CD-SBT, 
by introducing ghost particles along with the repulsive force particles, 
as shown in the right panel of Figure \ref{figs:fsi-bfparticle}. 
In the CD-SBT scheme, 
the repulsive particles are similar to that of Ref. \cite{monaghan2009sph} 
with identical particle spacing of fluid particle. 
Ghost particles are located outside the repulsive ones and initially generated 
in a regular or irregular distribution \cite{liu2012treatment}. 
An improved repulsive force in the form of 
\begin{equation}
\mathbf f^s_{ia} = 0.01 c^2 \sum_a \chi f(\eta) \frac{\mathbf r_{ia}}{r^2_{ia}}, 
\end{equation}
with
\begin{equation}
\begin{split}
	\eta & = r_{ia} / 0.75h, \\
	\chi & = 1 - \frac{r_{ia}}{dp_0}, \quad 0 < r_{ia} < dp_0,\\
	f_{\eta} & =
	\begin{cases}
		2/3 & 0 < \eta < 2/3\\
		(2\eta - 1.5\eta^2) & 2/3 < \eta < 1 \\
		0.5(2 - \eta)^2& 1 < \eta < 2  \\
		0 & otherwise
	\end{cases},
\end{split}
\end{equation}
is proposed. 
Also, both the repulsive particles and ghost particles are dynamically evolved in the SPH approximation 
of the governing equations, and their density and velocity can be interpolated from the fluid particles by 
\begin{equation}
\begin{cases}
	\rho_a = \sum_i m_i W^{new}_{ai}  \\
	\mathbf v_a = - \sum_i \mathbf v_i W^{new}_{ai} V_i
\end{cases}, 
\end{equation}
where $W^{new}$ represents the corrected kernel function with Shepard filter or MLS method. 
Its straightforward to note that the repulsive force particle provides a penetration force 
and its combination with ghost particle restores consistency by extending the full support domain of fluid particles. 
The CD-SBT has shown its accuracy and robustness in the simulation of multi-phase flow \cite{chen2015sph}, 
free-surface flows interacting with movable rigid objects\cite {liu2014sph} and  hydro-elastic FSI \cite{liu2013numerical}.
\subsubsection{Dummy particle}
\begin{figure*}[htb!]
	\centering
	\includegraphics[trim = 1mm 1mm 1mm 1mm, clip, width=.495\textwidth]{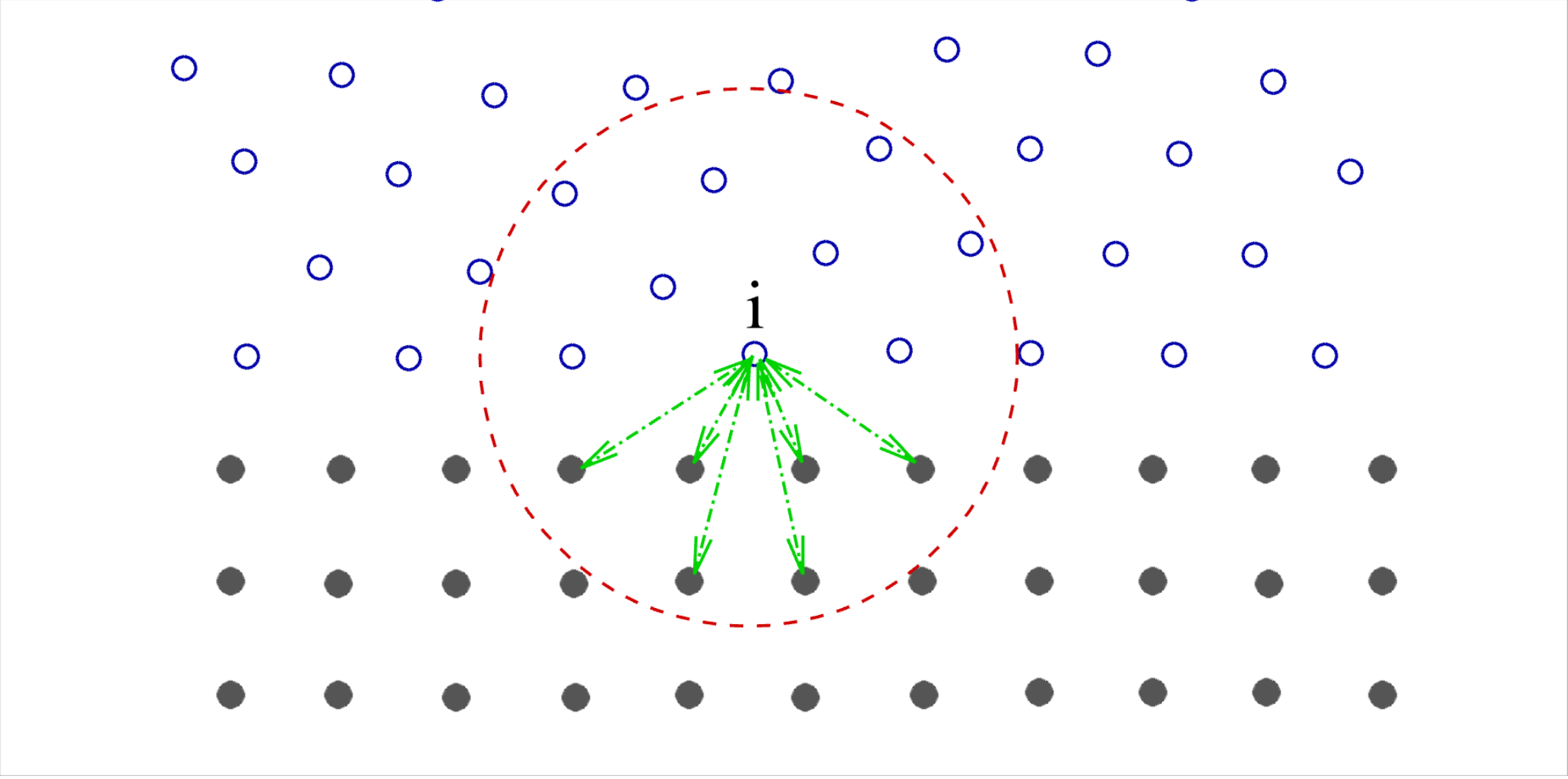}
	\includegraphics[trim = 1mm 1mm 1mm 1mm, clip, width=.495\textwidth]{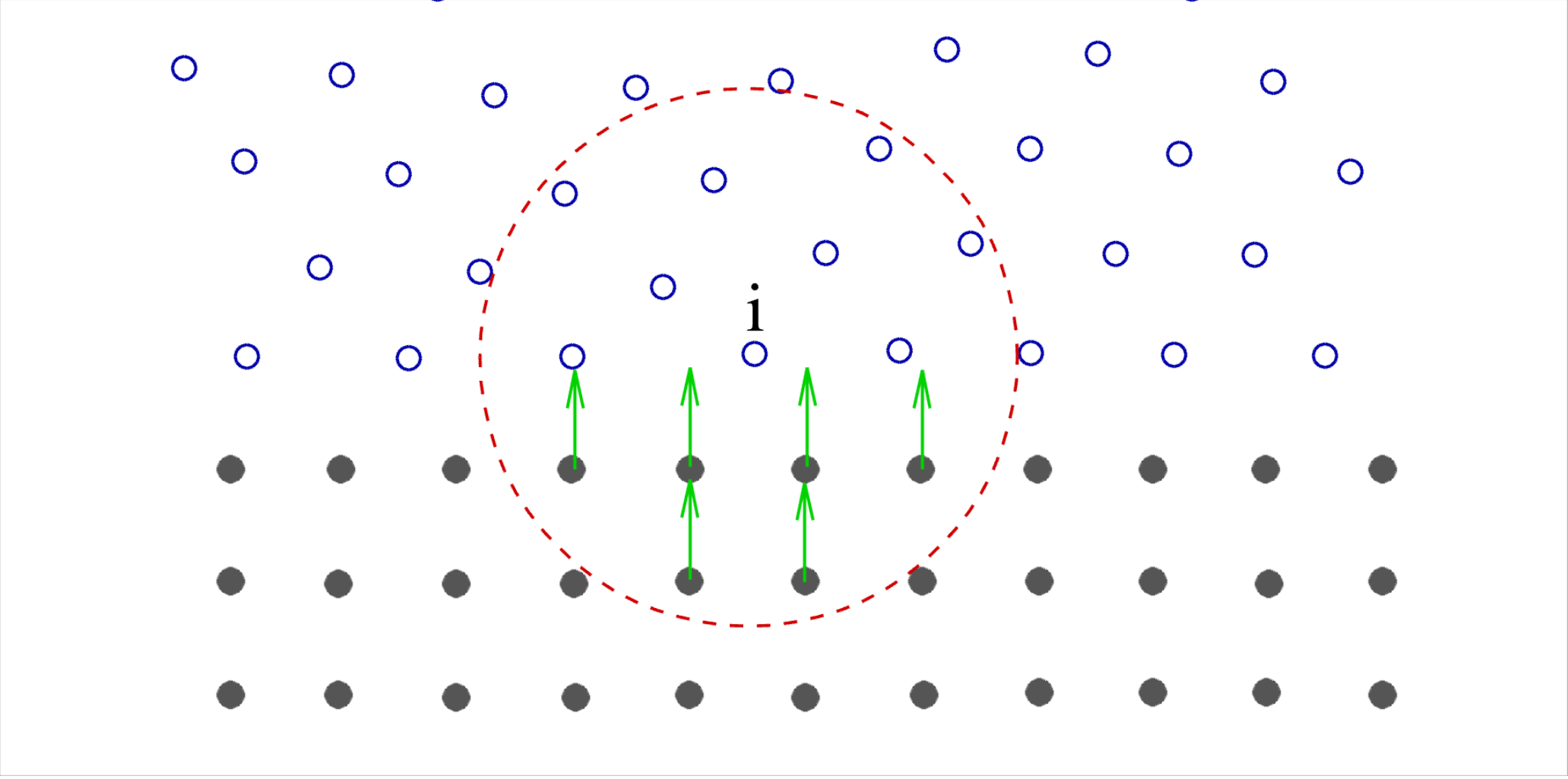}
	\caption{Sketch of multi-phase particles interacting with solid particles along the normal vector 
		through the one-side Riemann problem.}
	\label{figs:fsi-dummyparticle}
\end{figure*}
In dummy particle scheme, 
the solid is discretized with dummy particles in a layer of width $r_c$, 
the cutoff radius of the kernel function, 
along the interface 
or the particles of the flexible structure to represent dummy particles when interacting with fluid particles. 
As shown in the left panel of Figure \ref{figs:fsi-dummyparticle}, 
the fluid particles (in blue) near the wall interact with 
the dummy particles (in black) which lie within the support radius of the smoothing kernel function. 
Then, 
the force $\mathbf f^s$ acting on the fluid is decomposed into 
the pressure force $\mathbf f^{s:p}$ and viscous force $\mathbf f^{s:\nu}$ which are defined as 
\begin{equation} \label{eq:fluid-structure-forces}
\begin{cases}
\mathbf f_i^{s:p} = - \frac{2}{m_i}\sum_a V_i V_a \widetilde p_{ia} \nabla_i W_{ia} \\
\mathbf f_i^{s:\nu} = \frac{2}{m_i}\sum_a \eta V_i V_a \frac{\mathbf v_i - \widetilde{\mathbf v}_a}{r_{ia}} \frac{\partial W_{ia}}{\partial r_{ia}}
\end{cases},
\end{equation}
where the density-weighted inter-particle averaged pressure $\widetilde p_{ia}$ is defined as 
\begin{equation}\label{eq:fsi-dummy-pressure}
\widetilde p_{ia} = \frac{\rho_i p_a + \rho_a p_i}{\rho_i + \rho_a},
\end{equation}
and the dummy particle velocity $\widetilde{\mathbf v}_a$ is extrapolated from the fluid phase as 
\begin{equation} \label{eq:fsi-dummy-velocity}
\widetilde{\mathbf v}_a = \mathbf v_a - \frac{\sum_i \mathbf v_i W_{ai}}{\sum_i \mathbf W_{ai}}, 
\end{equation}
to impose the no-slip boundary condition. 
Note that the $\mathbf v_a$ of Eq. \eqref{eq:fsi-dummy-velocity} represents the velocity of the movable solid or flexible structure. 
And the dummy particle pressure $p_a$ of Eq. \eqref{eq:fsi-dummy-pressure} can be calculated with a summation over all 
contribution of the neighboring fluid particles by  \cite{marrone2011delta}
\begin{equation}\label{eq:fsi-wall-pressure-mls}
p_a = \sum_i p_i W^{MLS}\left(\mathbf r_i \right) dV_i + 2d\rho \mathbf n_a \cdot \mathbf g ,
\end{equation}
or \cite{adami2012generalized}
\begin{equation}\label{eq:fsi-wall-pressure-adami} 
p_a = \frac{\sum_{i} p_i W_{ai} + (\mathbf{g}-\mathbf{a}_{a}) \cdot \sum_{i} \mathbf{r}_{ai} 
	W_{ai}}{\sum_{i} W_{ai}} .
\end{equation} 
Here, 
$\mathbf{a}_{a}$ is solid acceleration. 
With the interpolated pressure $p_a$, the dummy particle density $\rho_a$ can be derived from the EoS of Eq. \eqref{eq:fluid-eos}. 
Compared with the interpolation of Eq. \eqref{eq:fsi-wall-pressure-mls}, 
Eq. \eqref{eq:fsi-wall-pressure-adami} is written in the more general formulation with moving solid present, 
making it a suitable choice for FSI involving the moving or flexible structure. 

With the dummy particle, 
the kernel truncation of fluid particles close to the structure is avoided, 
ensuring approximation accuracy and consistency. 
The dummy particle has been widely applied in the simulation of free-surface flow \cite{adami2012generalized, rezavand2020weakly}, 
multi-phase flow \cite{rezavand2020weakly, rezavand2021generalised} or multi-phase FSI \cite{sun2019study}.
Notwithstanding its wide application, 
the dummy particle results excessive computational efforts which are inherently expensive in three-dimensional simulations 
due to the introduction of physical variable interpolation.
\subsubsection{One-sided Riemann-based scheme}
Similar to the dummy particle scheme, 
the one-sided Riemann scheme also represents the movable or flexible structure with dummy particles, 
while a one-sided Riemann problem is constructed along the solid normal and solved to realize the FSI coupling \cite{zhang2017weakly}
as shown in the right panel of Figure \ref{figs:fsi-dummyparticle}. 

In the one-sided Riemann scheme \cite{zhang2017weakly}, 
the pressure forces $\mathbf{f}_i^{s:p}$ is rewritten as 
\begin{equation} \label{eq:fluid-structure-forces-rie}
\mathbf f_i^{s:p}  = - \frac{2}{m_i}\sum_a V_i V_a p^\ast \nabla_i W_{ia},
\end{equation}
where $p^\ast$ is the Riemann solution of the one-sided Riemann problem 
whose left and right states are defined as 
\begin{equation}\label{eq:onesided-rie}
\begin{cases}
\left( \rho_L, U_L, p_L\right) = \left( \rho_f,- \mathbf n_a \cdot \mathbf v_i, p_i \right)  \\
\left( \rho_R, U_R, p_R\right) = \left(\rho_a, - \mathbf n_a \cdot \left(2 \mathbf v_i  - \mathbf v_a \right) , p_a \right) 
\end{cases}. 
\end{equation}
Here, 
$\mathbf n_a$ is the local normal vector pointing from solid to fluid, 
and 
the pressure $p_a = p_i + \rho_i \mathbf g \cdot \mathbf r_{ia}$. 
With the dummy particle pressure, 
its density $\rho_a$ is also calculated through the EoS presented in Eq. \eqref{eq:fluid-eos}. 
Compared with the aforementioned dummy particle scheme,
the present one is more simple and efficient due to the fact that 
the one-sided Riemann problem is solved in a particle-by-particle fashion and no interpolation of
states for the solid particles is required. 

For each solid particles, 
the normal vector in the reference configuration can be calculated by \cite{zhang2017weakly, randles1996smoothed}
\begin{equation}\label{eq:rienorm}
\mathbf n_a^0 = \frac{\mathbf \Phi \left( \mathbf r^0_a \right) }{\lvert \mathbf \Phi \left(\mathbf r^0_a \right) \lvert}, \quad
\mathbf \Phi \left( \mathbf r^0_a \right)  = - \sum_{b \in{s}}V^0_b \nabla W^0_{ab},
\end{equation}
where the summation is over wall particles only. 
For static solid structure, 
the normal vector is not altered during the computation. 
For flexible structure, 
the normal vector should be updated accordingly when deformation occurs. 
To void of frequent summation computation in Eq. \eqref{eq:rienorm}, 
the updated normal vector for flexible structure can be obtained by 
\begin{equation}
\mathbf n_a = \mathbb Q \mathbf n_a^0 , 
\end{equation}
where $\mathbb Q$ is the orthogonal matrix of the deformation tensor $\mathbb F$ 
and can be calculated with polar decomposition. 

\begin{figure*}
	\centering
	\includegraphics[trim = 2mm 2mm 2mm 2mm, clip, width=0.495\textwidth]{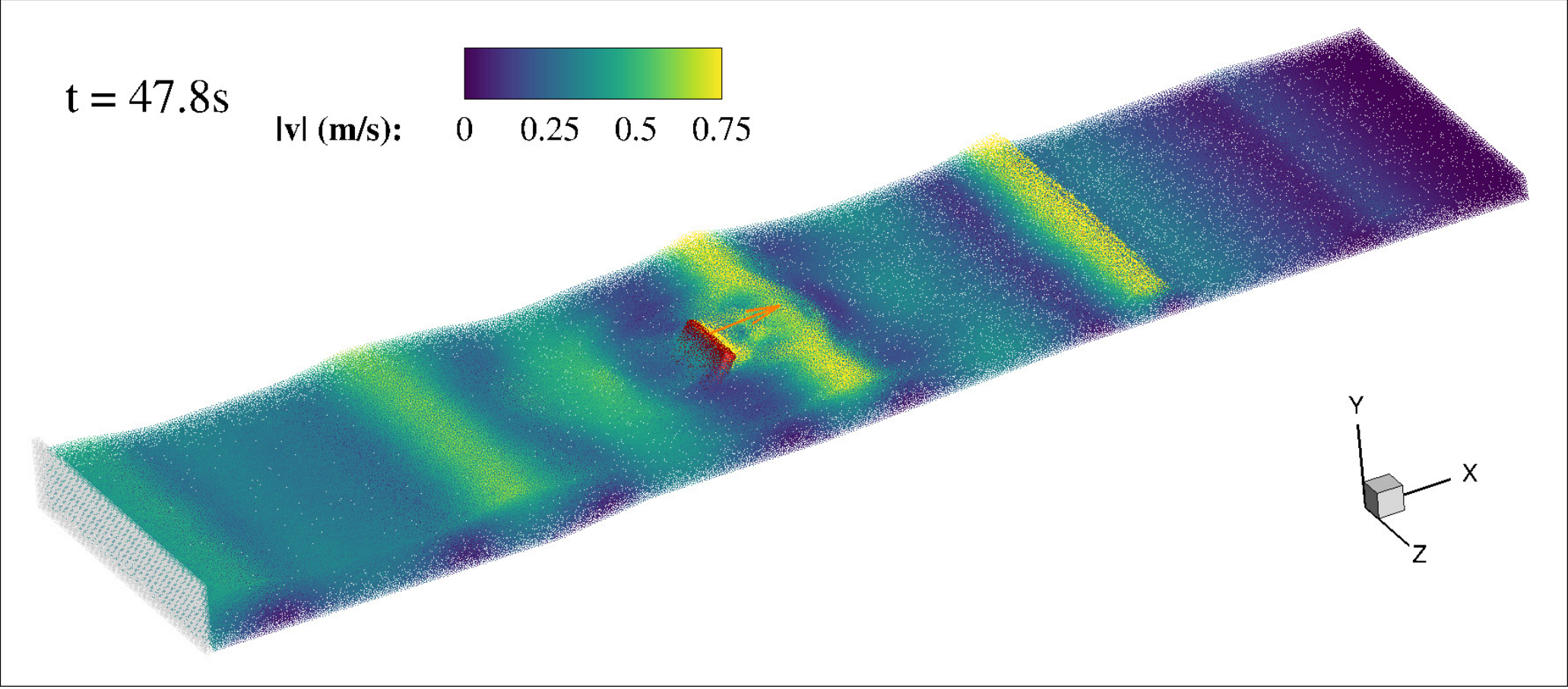}\\
	\includegraphics[trim = 2mm 2mm 2mm 2mm, clip, width=0.495\textwidth]{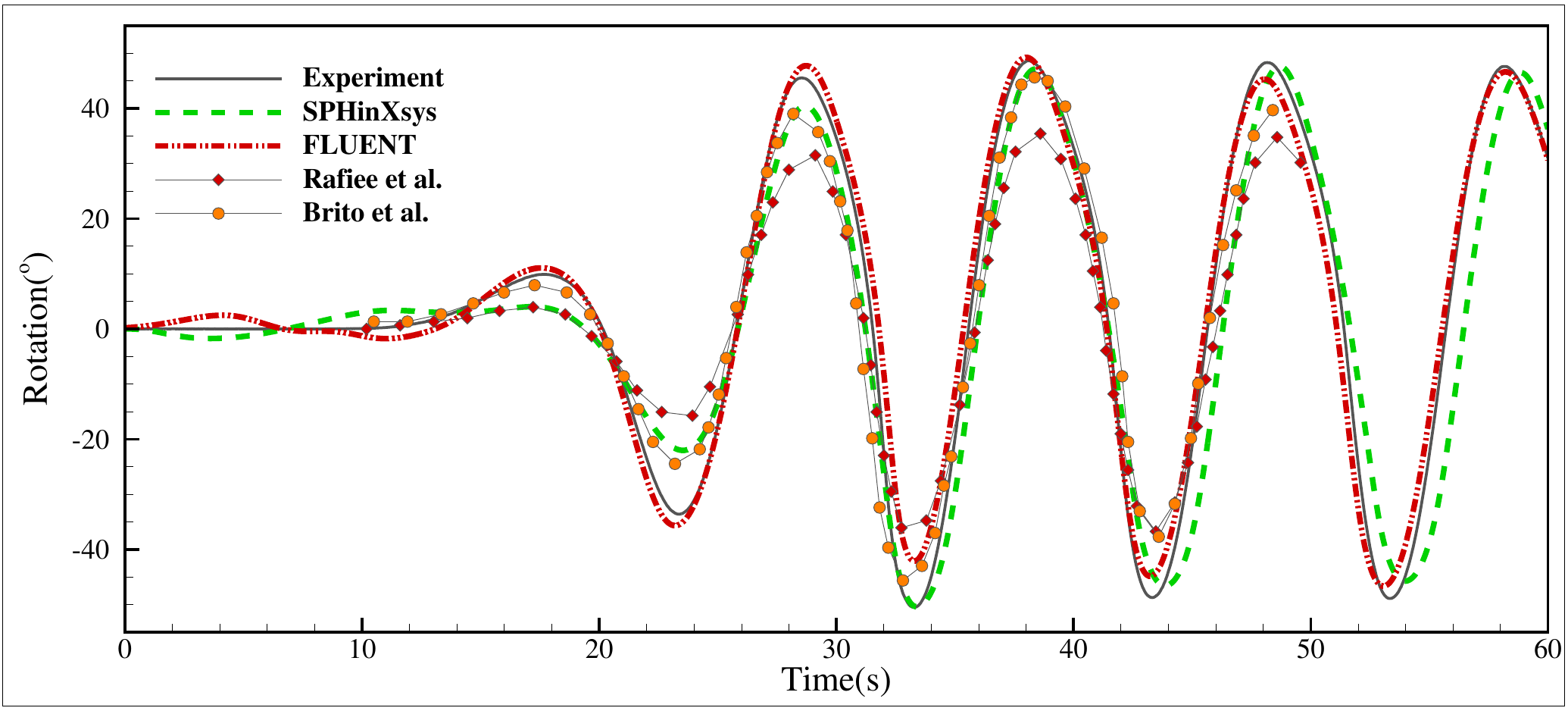}
	\includegraphics[trim = 2mm 2mm 2mm 2mm, clip, width=0.495\textwidth]{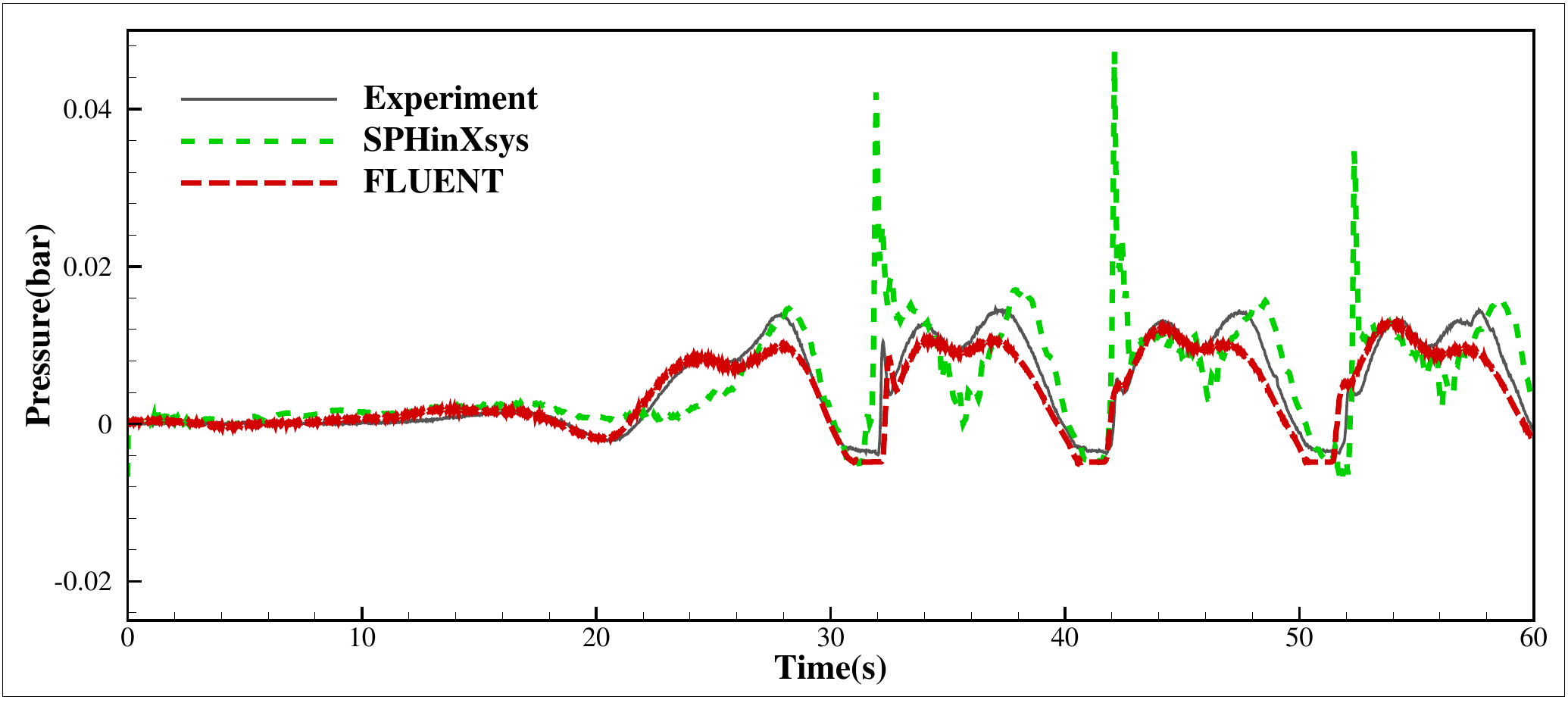}
	\caption{Wave interaction with an OWSC investigated by SPHinXsys library:  
		(a) Free surface and flap motion with fluid particles colored by the velocity magnitude (upper panel). 
		(b) Validations of the flap rotation (bottom left panel) and wave load on the flap (bottom right panel).}
	\label{figs:owsc}
\end{figure*}

Figure \ref{figs:owsc} presents the numerical investigation of 
wave interaction with an oscillating wave surge converter (OWSC) \cite{zhang2021efficient} 
with SPHinXsys library \cite{zhang2021sphinxsys} by using Riemann-based WCSPH method 
with one-sided Riemann-based fluid-solid interface treatment. 
It is observed that 
smooth velocity fields are produced even when complex interactions between the wave and the flap are involved. 
Also, 
wave-structure interaction and wave loading are well predicted 
in comparison with the experimental data \cite{wei2015wave}, 
numerical results obtained with commercial software FLUENT \cite{wei2015wave} 
and SPH results in the literature \cite{rafiee2013numerical, brito2020numerical}.
%
\subsection{Multi-resolution scheme}\label{sec:fsi-multi-discretization}
When applying the particle-based solver for modeling FSI problems, 
single spatial-temporal resolution, 
where a uniform particle spacing is used for discretizing the entire computational domain and a single time step 
being the smallest one of these required by the fluid and solid structure is used for time integration \cite{antoci2007numerical}, 
is commonly employed. 
In such a case, 
the single-resolution approach is computationally expensive and high memory consumption 
in the application where the structure requires locally refined resolution or the global computational domain is quite larger 
with respect to the critical sub-domain.  
Therefore, 
developing a multi-resolution scheme for particle-based FSI solver is desirable in the computational efficiency point of view. 

For particle-based simulation, 
different accurate, stable and consistent multi-resolution schemes have been developed for discretizating the fluid equations
and they are generally classified into four classes:
adaptive particle refinement (APR) with or without particle splitting/merging 
\cite{springel2005cosmological,lastiwka2005adaptive, vacondio2016variable, khorasanizade2016dynamic, hu2019consistent}, 
non-spherical particle scheme \cite{liu2006adaptive, owen1998adaptive}, 
domain-decomposition based scheme \cite{bian2015multi, shibata2017overlapping} or the hybrid scheme \cite{barcarolo2014adaptive, tanaka2018multi,omidvar2012wave}. 
Notwithstanding these progress, 
the development of multi-resolution scheme for particle-base FSI solver has merged in the very recent years.
The pioneering work is credited to Khayyer et al. \cite{khayyer2019multi} 
where a multi-resolution scheme is developed for MPS-FSI solver. 
Then, 
Sun et al. \cite{sun2019study} extended the multi-resolution scheme developed by Barcarolo et al. \cite{barcarolo2014adaptive} 
to the simulation of multi-phase hydroelastic FSI problems. 
More recently, 
Zhang et al. \cite{zhang2021multi} proposed a multi-resolution SPH method for fluid-flexible structure interaction 
with special attention in dealing with enforcing the momentum conservation and force matching at the fluid-structure interface. 

It is worth noting that consistent particle resolution is applied through FSI interface in Ref. \cite{sun2019study} 
where the APR scheme is applied in predefined area interface located in the fluid domain, 
with proper particle splitting/merging \cite{barcarolo2014adaptive}. 
On the other hand, 
different particle resolution is adopted in the FSI interface of Refs. \cite{khayyer2019multi, zhang2021multi, khayyer2021multi} 
with proper handling of momentum conservation and force matching. 
In this work, 
we focus on the second approach as the first one involves splitting/merging scheme which is not the main objective of the survey. 
\subsubsection{Multi-resolution discretizaiton}
In the multi-resolution framework of Ref. \cite{zhang2021multi}, 
the fluid and solid equations are discretized by different spatial-temporal resolutions. 
In this case, 
the solid structure can be resolved at a higher spatial resolution, 
and the computational efficiency is enhanced when a lower resolution discretization for the fluid is sufficient. 
This strategy is suitable for applications where the structure is relatively thin, 
i.e. the structure has a considerable small spatial scale compared with fluid,  
or when the structure has a high Poisson ratio which results smaller time step size than that required by the fluid. 

With different spatial resolutions are applied across the FSI interface, 
the interaction pressure force $\mathbf{f}_i^{s:p}$ and viscous force $\mathbf{f}_i^{s:\nu}$ are rewritten as 
\begin{equation} \label{eq:mr-fsi-forces}
\begin{cases}
\mathbf f_i^{s:p} \left(h^f\right) = - \frac{2}{m_i}\sum_a V_i V_a \widetilde p_{ia} \nabla_i W_{ia} W(\mathbf{r}_{ia}, h^f ) \\
\mathbf f_i^{s:\nu} \left(h^f\right) = \frac{2}{m_i}\sum_a \eta V_i V_a \frac{\mathbf v_i - \widetilde{\mathbf v}_a}{r_{ia}} \frac{\partial W(\mathbf{r}_{ia}, h^f )}{\partial {{r}_{ia}}}
\end{cases},
\end{equation}
where $h^f$ denotes the smoothing length used for fluid
with the assumption of $h^f \geqslant h^s$, 
ensuring that a fluid particle $i$ can be searched and tagged as a neighboring particle 
of a solid particle $a$ which is located in the neighborhood of particle $i$. 
Having Eq. \eqref{eq:mr-fsi-forces} in hand, 
the fluid forces exerting on the solid structure can be derived straightforwardly. 
\subsubsection{Multi-time stepping with position-based Verlet scheme}
Following Ref. \cite{zhang2020integrative, zhang2020multi}, 
the time-step criterion for the solid integration is given as
\begin{equation}\label{dts-advection}
\Delta t^S   =  0.6 \min\left(\frac{h^S}{c^S + |\mathbf{v}|_{max}},
\sqrt{\frac{h^S}{|\frac{\text{d}\mathbf{v}}{\text{d}t}|_{max}}} \right) .
\end{equation}
With the advection criterion $\Delta t_{ad}^F$ of Eq. \eqref{eq:dt-advection} 
and the acoustic criterion $\Delta t_{ac}^F$ of Eq. \eqref{eq:dt-relax} in hand, 
three different time step sizes are introduced. 
Generally, 
$\Delta t^S < \Delta t_{ac}^F$, 
due to the fact that $c^S >  c^F$.
Other than choosing $\Delta t^S$ as the single time step for both fluid and structure, 
one can carry out the structure time integration $\kappa = [\frac{\Delta t_{ac}^F}{\Delta t^S}] + 1$ times, 
where $[\cdot]$ represents the integer operation, during one acoustic time step of fluid integration. 
As different time steps are applied in the integration of fluid and solid equations, 
the issue of force mismatch in the fluid-structure interaction may be encountered. 
That is, 
in the imaginary pressure and velocity calculation, 
the velocity and acceleration of solid particles in Eq. \eqref{eq:mr-fsi-forces} may present several different values updated after each $\Delta t^S$. 
Another issue is that the momentum conservation in the fluid and structure coupling may be violated. 
To address the force-calculation mismatch, 
Zhang et al. \cite{zhang2021multi} proposed to calculate the imaginary pressure $p_a^d$ and velocity $\mathbf{v}_a^d$ as
\begin{equation} \label{fs-coupling-mr }
\begin{cases}
p_a^d = p_i + \rho_i max(0, (\mathbf{g} - \widetilde{\frac{\text{d} \mathbf{v}_a}{\text{d}t}}) \cdot \mathbf{n}^S) (\mathbf{r}_{ia} \cdot \mathbf{n}^S) \\
\mathbf{v}_a^d = 2 \mathbf{v}_i  - \widetilde{\mathbf{v}}_a
\end{cases}, 
\end{equation}
where $\widetilde{\mathbf{v}}_a$ and $\widetilde{\frac{d\mathbf{v}_a}{dt}}$ 
represents the single averaged velocity and acceleration of solid particles during a fluid acoustic time step. 

\begin{figure*}
	\centering
	\includegraphics[trim = 1cm 0cm 2cm 0cm, clip,width=0.65\textwidth]{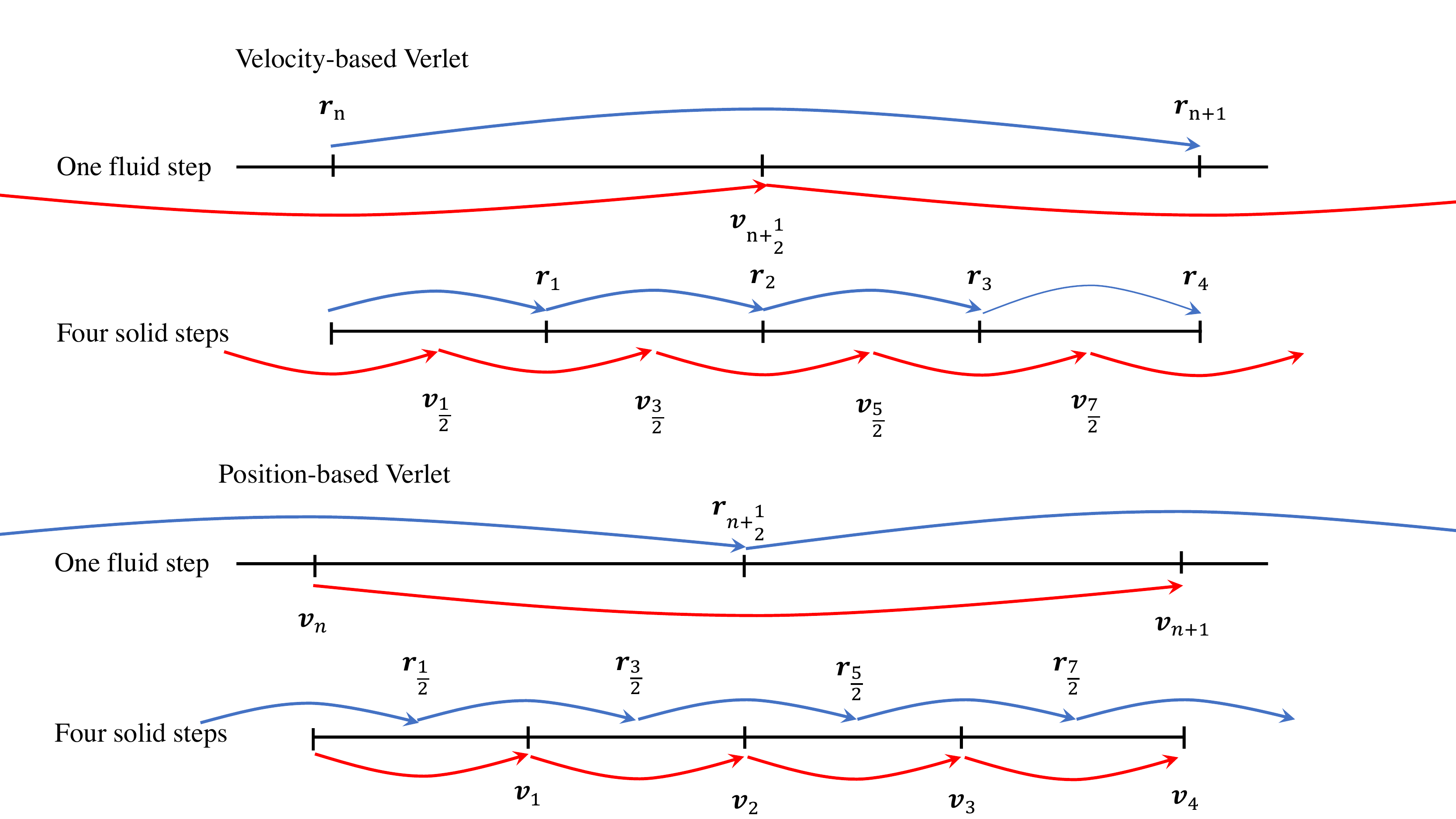}
	\caption{Sketch of velocity- and position-based Verlet schemes with assumption that $\kappa = 4$.}
	\label{figs:verletsetup}
\end{figure*}

Also, 
to address the momentum conservation issue, 
Zhang et al. \cite{zhang2021multi} developed a position-based Verlet scheme. 
Instead of starting with a half step for velocity  
followed by a full step for position and another half step for velocity 
as in the velocity-based Verlet scheme \cite{adami2012generalized}, 
the position-based Verlet does the opposite: 
a half step for position followed by a full step for velocity and another half step for position. 
Figure \ref{figs:verletsetup} depicts the velocity- and position-based Verlet schemes assuming that $\kappa = 4$ for the integration of fluid and solid equations. 
Note that since the position is updated twice and the velocity once with the acceleration at the half step, 
using the Taylor expansion
one can find that the position-based scheme has the same 2nd-order accuracy as the original one. 

In the position-based Verlet scheme, 
as the velocity field is updated only once in the current fluid acoustic time step criterion, 
time marching of the momentum equations for fluid and solid are exactly consistent as the velocity marching interval $ (\Delta t_{ac}^F)_n= \sum_{\varkappa = 0}^{\kappa - 1} (\Delta t^S)_{\varkappa}$, as shown Figure \ref{figs:verletsetup}.
Therefore, 
the position-based Verlet algorithm achieves strict momentum conservation in fluid-structure coupling, when multiple time steps is employed. 
In contrast, 
the velocity-based Verlet scheme does not guarantee momentum conservation as 
$ 0.5 \left[  (\Delta t_{ac}^F)_{n-1} +  (\Delta t_{ac}^F)_{n} \right] \neq  0.5 \left[ \sum_{\varkappa = 0}^{\kappa -1} \left( (\Delta t^S)_{\varkappa - 1} + (\Delta t^S)_{\varkappa} \right)  \right]$, 
as also shown in Figure \ref{figs:verletsetup}.

\begin{figure*}
	\centering
	\begin{subfigure}[b]{\textwidth}
		\includegraphics[trim = 1mm 1mm 1mm 2mm, clip,width=.5\textwidth]{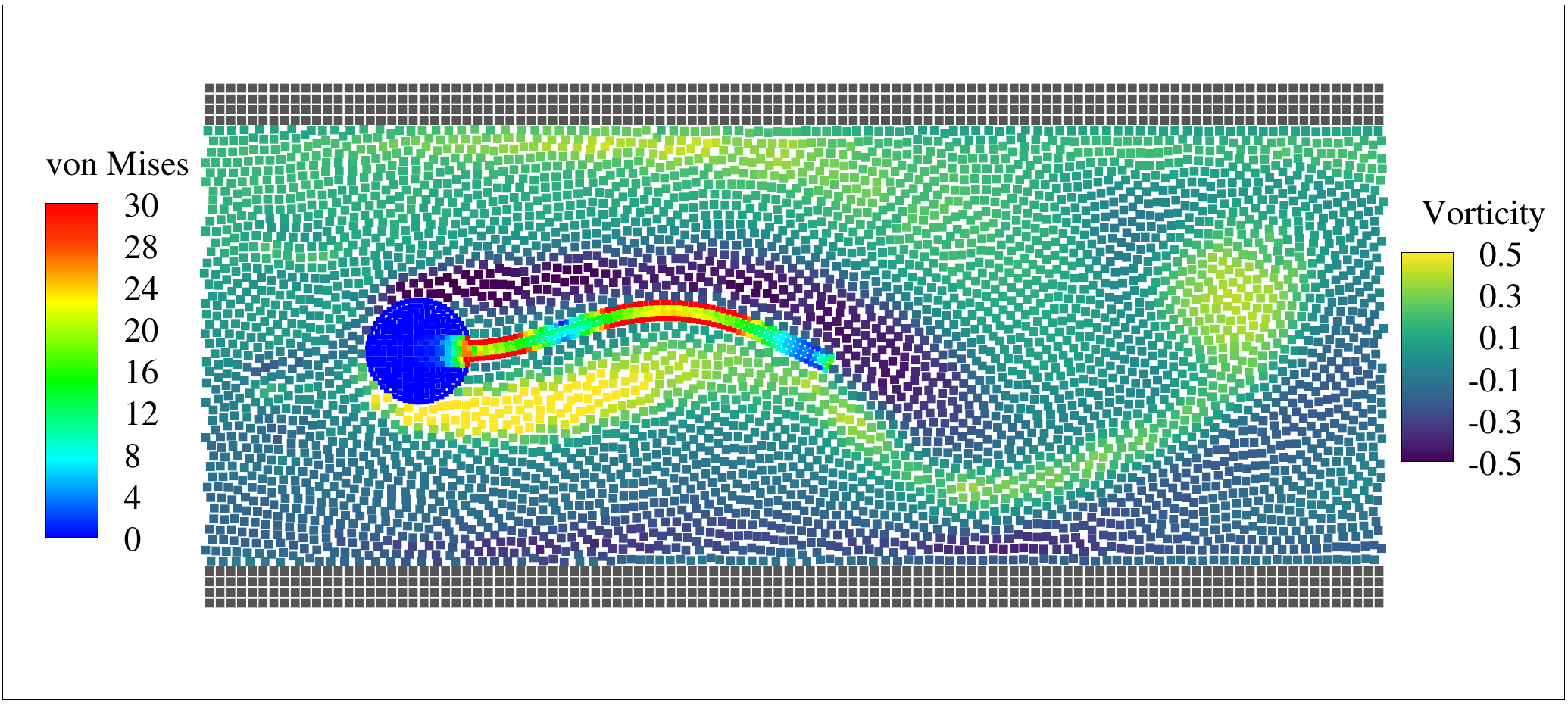}
			\begin{subfigure}{0.45\textwidth}
				\vspace{-3.5cm}
			\begin{tabular}{lcc}
				\hline
				Refs.    & \begin{tabular}{@{}c@{}} Amplitude \\ in $y$-axis\end{tabular}    & Frequency  \\ 
				\hline
				Turek and Hron \cite{turek2006proposal}   &$0.83$       & $0.19$      \\ 
				Bhardwaj and Mittal \cite{bhardwaj2012benchmarking}  &$0.92$       & $0.19$      \\
				Tian et al. \cite{tian2014fluid} &$0.784$ &$0.19$ \\
				SPHinXsys & $0.855$       & $0.189$     \\ 
				\hline
			\end{tabular}
		\end{subfigure}
		\caption{Flow-induced vibration of a elastic beam attached to a cylinder.}
		\label{figs:fsi-vibration}
	\end{subfigure}
	\newline 
	\begin{subfigure}[b]{0.95\textwidth}
		\centering
		\includegraphics[trim = 5mm 1mm 2mm 1cm, clip,width=0.49\textwidth]{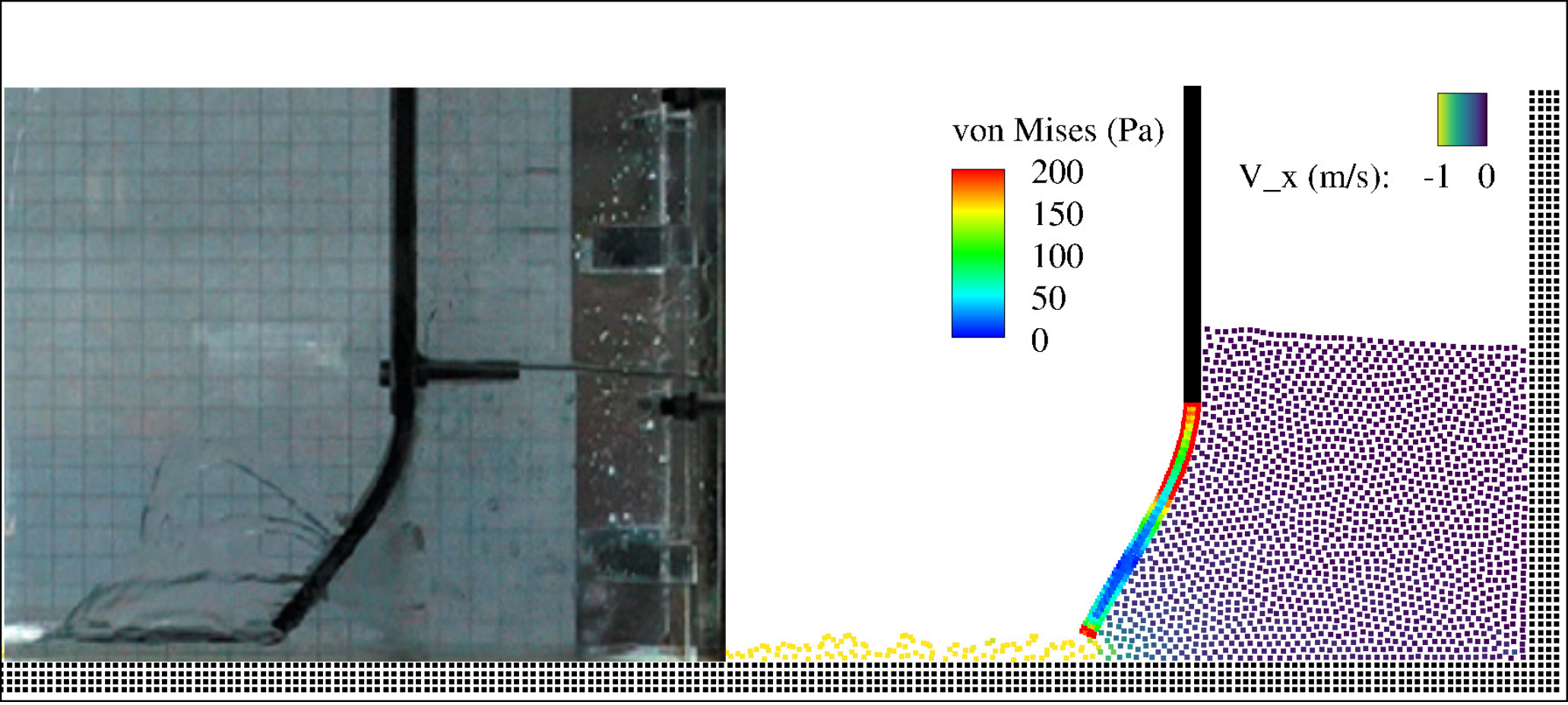}\\
		\includegraphics[trim = 1mm 1mm 1mm 1mm, clip,width=.485\textwidth]{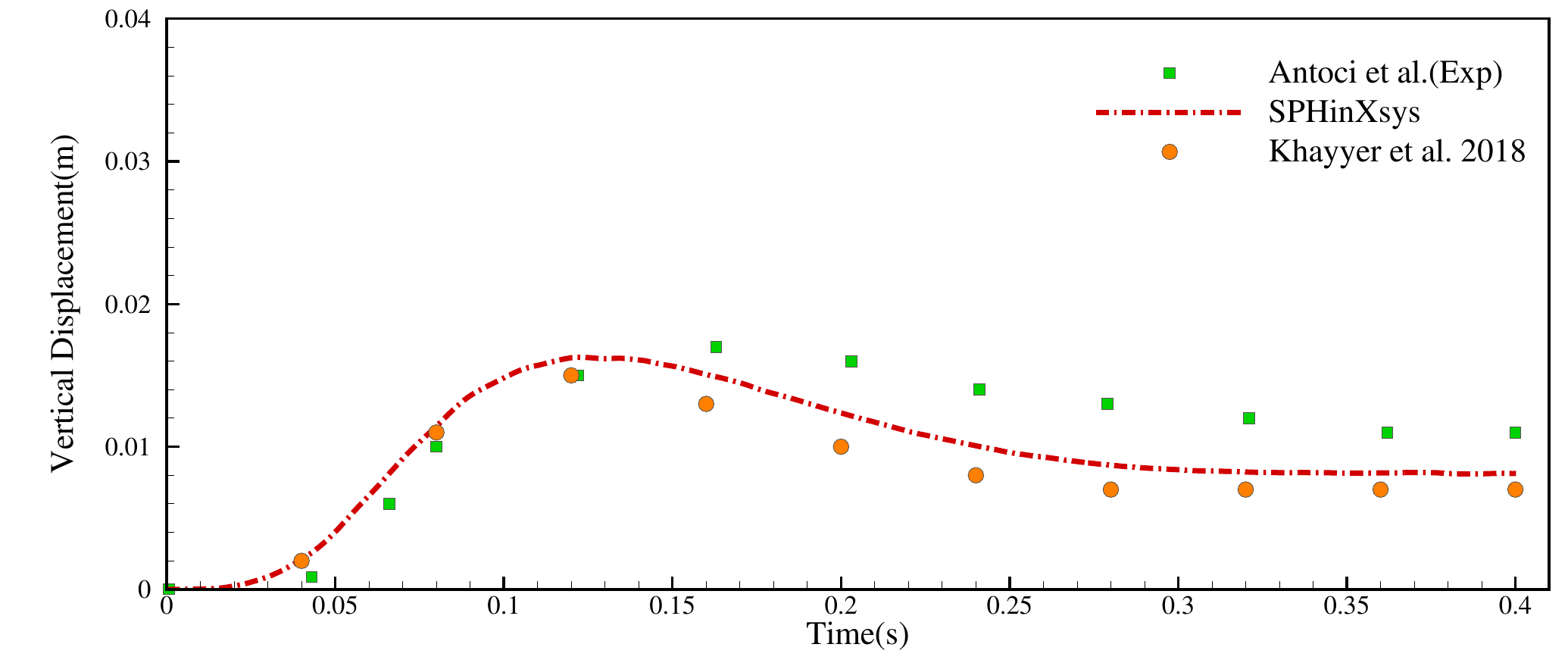}
		\includegraphics[trim = 1mm 1mm 1mm 1mm, clip,width=.485\textwidth]{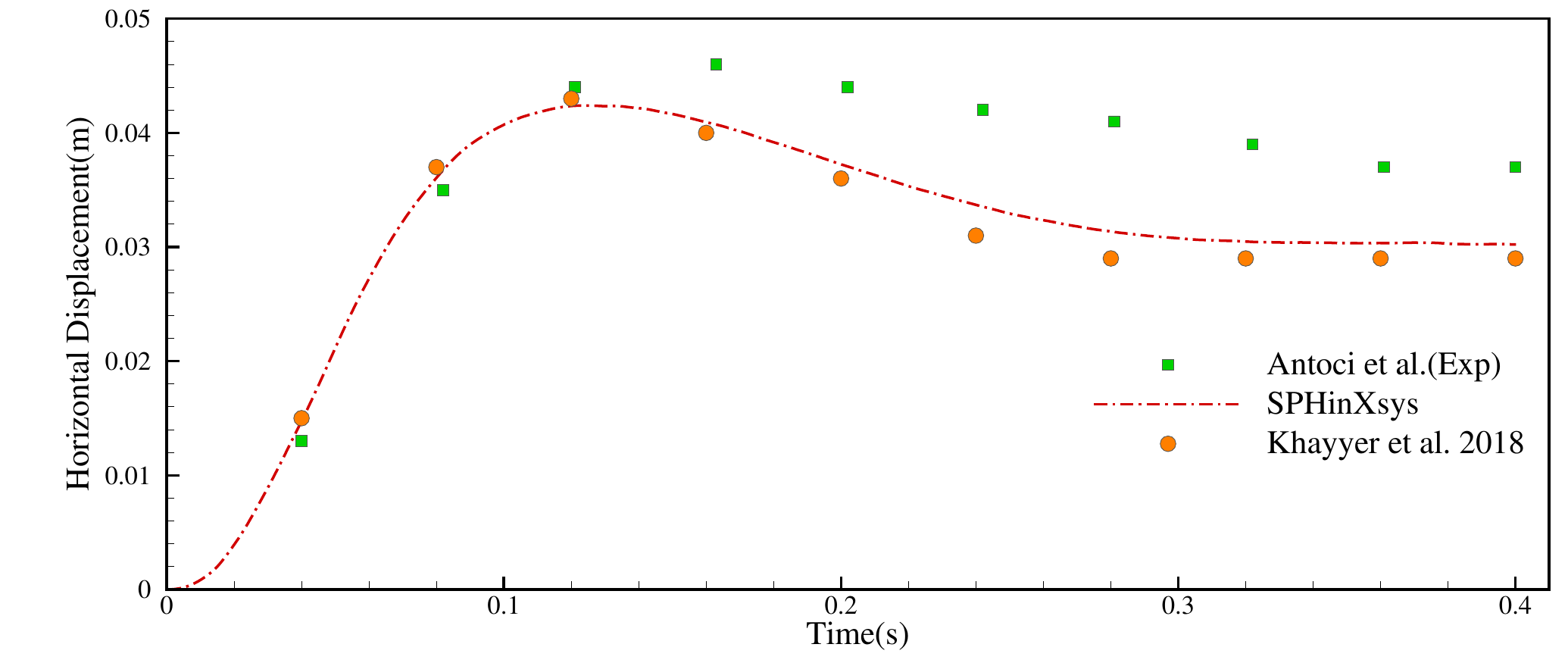}
		\caption{ Dam break flow through an elastic gate.}
		\label{figs:fsi-damgate}
	\end{subfigure}
	\caption{Numerical investigation of FSI problems with SPHinXsys library in multi-resolution scenario. 
		(a) Flow-induced vibration of a elastic beam attached to a cylinder: snapshots of vorticity field 
		and deformed bream configuration with von Mises contour (left panel) and the comparisons of the amplitude and the frequency of the beam oscillation with data in the literature.
		(b) Dam-break flow through an elastic gate : snapshots compared against experimental frames \cite{antoci2007numerical} (upper panel), 
		vertical (bottom left panel) and horizontal (bottom right panel) displacements of the free end of the plate and their comparisons with experimental \cite{antoci2007numerical} 
		and numerical \cite{khayyer2018enhanced} data.}
	\label{figs:fsi}
\end{figure*}

Figure \ref{figs:fsi} reports the validation of the multi-resolution scheme implemented in the open-source library SPHinXsys \cite{zhang2021sphinxsys, zhang2020sphinxsys} by simulating two FSI benchmark tests, 
i.e., flow-induced vibration of a beam attached to a cylinder and dam-break flow with elastic gate \cite{zhang2020multi}. 
For both tests, 
the deformation of the flexible structure induced by the fluid-structure interaction is accurately captured 
in comparison with experimental data \cite{antoci2007numerical} 
and numerical data in the literature \cite{antoci2007numerical,khayyer2018enhanced,han2018sph, turek2006proposal, bhardwaj2012benchmarking}. 
Concerning the computational efficiency, 
approximated seedup in the order of $10^2$ is achieved compared with the simulation in single-resolution scenario 
as reported in Ref. \cite{zhang2021multi}.   
%
%
%
\section{Particle and mesh generation}\label{sec:generator}
Generating high-quality unstructured mesh or particle distributions is essentially important 
to mesh-based or particle-based methods in scientific computing \cite{liu2003smoothed, 2016Improving}. 
However,
complex geometries are always involved in industrial applications bringing a critical challenge 
for generating high-quality particle distributions or mesh for arbitrarily complex geometry.

For particle-based methods,
there are generally two approaches to generate the initial particle distributions, 
i.e., 
(a) initiating particles on a lattice structure and (b) generating particles on a volume element mesh.
The first approach, positioning particles on a cubic lattice structure,
is widely used in particle-based methods community. 
For example, 
Dominguez et al. \cite{dominguez2011development} proposed a pre-processing tool for the DualSPHysics library 
where particles are generated on lattice structure and three-dimensional object is represented by particle model with excluding the outside particles. 
In this approach, the particles are equally distributed,
however, a very fine spatial resolution is needed to correctly portray the complex geometry.
The second approach generating particles at the center of tetra- or hexahedron volume elements has been widely used in application of bird strike \cite{vignjevic2013parametric, heimbs2011computational} and provided by state-of-art commercial pre-processing tools. 
This approach can accurately portray the complex surface, however, 
comprise drawbacks of non-uniform particle spacing and volume which may reduce the interpolation accuracy. 
Recently, 
the weighted Voronoi tessellation (WVT) method 
has been applied for generating initial particle distribution by Diehl et al. \cite{diehl2015generating} for SPH astrophysical simulation and 
by Siemann and Ritt \cite{siemann2019novel} for SPH modeling of bird-strike. 
Also, 
Vela et al. \cite{vela2018alaric} proposed an algorithm for constructing complex initial density distributions with low noise. 

Concerning high-quality mesh generation,
various mesh generation techniques have also been developed,
e.g. advancing front/layer methods \cite{1996Progress,19983D} initiating meshing from the boundary to domain interior, 
refinement based Delaunay triangulation \cite{Jonathan2002Delaunay,1989Guaranteed} inserting new Steiner points into a Delaunay mesh, 
centroidal Voronoi tessellations (CVT) \cite{1999Centroidal,2002Grid,2009OnCen,2008Generic} 
and particle-based method \cite{1994Using,2010Particle,2005Mesh,2013Particle,Meyer2005Robust}
in which a relaxation strategy basing on the physical analogy between a simple mesh and a truss structure is applied.
Among them,
the particle-based mesh generation method has been widely studied due to the efficiency and versatility feature.

More recently, 
Fu et al. \cite{fu2019isotropic} and Zhu et al. \cite{zhu2021cad} presented a novel application of the particle-relaxation in SPH methodology 
for high-quality unstructured mesh and body-fitted particle distribution, respectively, 
for arbitrarily complex geometry. 
Ji et al. \cite{ji2020consistent, ji2021feature} further improved the particle-relaxation for mesh generation by exploiting multi-phase algorithm 
and introducing feature boundary correction term. 
The general procedure consists of three steps. 
First,
the geometry surface is represented by zero level-set function by parsing corresponding computer-aided design (CAD) file \cite{zhu2021cad}. 
Second, 
several steps of particle-relaxation is conducted by solving a set of physically-motivated model equations in the SPH methodology \cite{fu2019isotropic}. 
In this step, 
proper surface bonding techniques were developed for achieving body-fitted feature of particle distribution. 
Third, 
a set of neighboring particles generates a locally valid Voronoi diagram at the interior of the domain by using Delaunay triangulation method \cite{fu2019isotropic}. 
Note that the third step is only for mesh generation. 
\subsection{Surface representation}\label{sec:geometry-definition}
By parsing a CAD file, 
the geometry surface can be constructed and represented by the zero level-set of the signed-distance function
\begin{equation}
\Gamma = \left\lbrace  \left(x, y, z \right)| \phi \left(x, y, z, t \right) = 0 \right\rbrace  .
\label{level-set}
\end{equation}
Then, 
the normal direction $\mathbf N= \left(  n_x, n_y, n_z \right) ^T$ of the surface can be evaluated from 
\begin{equation}
\mathbf N= \frac{\nabla \phi}{\left| \nabla \phi \right| }.
\label{normal}
\end{equation}
To discretize the level-set function,
a Cartesian background mesh is generated in the whole computational domain.
The level-set value $\phi$ is equal to the distance from the cell center to the geometry surface. 
Besides,  
the negative phase with $\phi < 0$ is defined if the cell center is inside the geometry and positive phase with $\phi > 0$ otherwise. 
\subsection{Particle relaxation}\label{sec:physics-driven relaxation}
Starting from a preconditioned Lattice or random particle distribution generated inside the domain of the geometry,
a physics-driven relaxation process is introduced to define particles evolution.
The relaxation is governed by the momentum conservation equation
\begin{equation} \label{eq:momentum} 
\begin{split}
	\frac{\text{d}\mathbf{v_i}}{\text{d}t} =  & -\frac{2}{m_i}\sum_j  V_i V_j \overline{p}_{ij}  \nabla_i W_{ij}  \\
	& + \frac{2}{m_i}\sum_j  V_i V_j \frac{\eta_i \eta_j}{\eta_i + \eta_j} \frac{\mathbf v_{ij}}{r_{ij}} \frac{\partial W_{ij}}{r_{ij}},
\end{split}
\end{equation}
where $\mathbf{v}$ is the advection velocity and $\eta = 0.2 h \rho |\mathbf v|$. 
In Fu et al. \cite{fu2019isotropic}, 
$\overline{p}_{ij} = \frac{1}{2}(p_i + p_j)$ and the particle pressure is defined by an equation of state 
\begin{equation} 
p = p_0 \frac{\rho^2}{\rho_t^2},
\end{equation}
which incorporates a target density field $\rho_t$ and $p_0$ is the reference pressure \cite{fu2019isotropic}. 
In Zhu et al. \cite{zhu2021cad}, 
a constant pressure $\overline{p}_{ij} = p_0$ is applied as a homogeneous particle distribution 
can be obtained by applying the transport-velocity formulation \cite{litvinov2015towards, adami2013transport,zhang2017generalized} 
with a constant density and background pressure. 
Then,  the particle evolution is defined by 
\begin{equation}\label{update-position}
\mathbf{r^{n+1}}=\mathbf{r^n} + \text{d} \mathbf{r} =\mathbf{r^n} + \frac{1}{2} \frac{\text{d}\mathbf{v_i}}{\text{d}t}^n \Delta t^2, 
\end{equation}
where $\mathbf a^n = \mathbf{F}_p^n$. 
Note that
only the instant acceleration is considered for the evolution and 
particle velocity is set to zero at the beginning of each time step  
to achieve a fully stationary state following 
Ref. \cite{fu2019isotropic, adami2013transport, zhang2017generalized}.
For numerical stability, 
the time-step size $\Delta t$ is constrained by 
the body force criterion
\begin{equation}\label{time-step}
\Delta t \leq 0.25 \sqrt{\frac{h} {\left| \text d \mathbf{v} / \text d t\right|} }.
\end{equation}
%
\subsection{Surface bounding method}\label{sec:surface-bounding}
\begin{figure*}[htb!]
	\centering
	\includegraphics[trim = 2cm 8cm 3cm 5cm, clip,width=.6\textwidth]{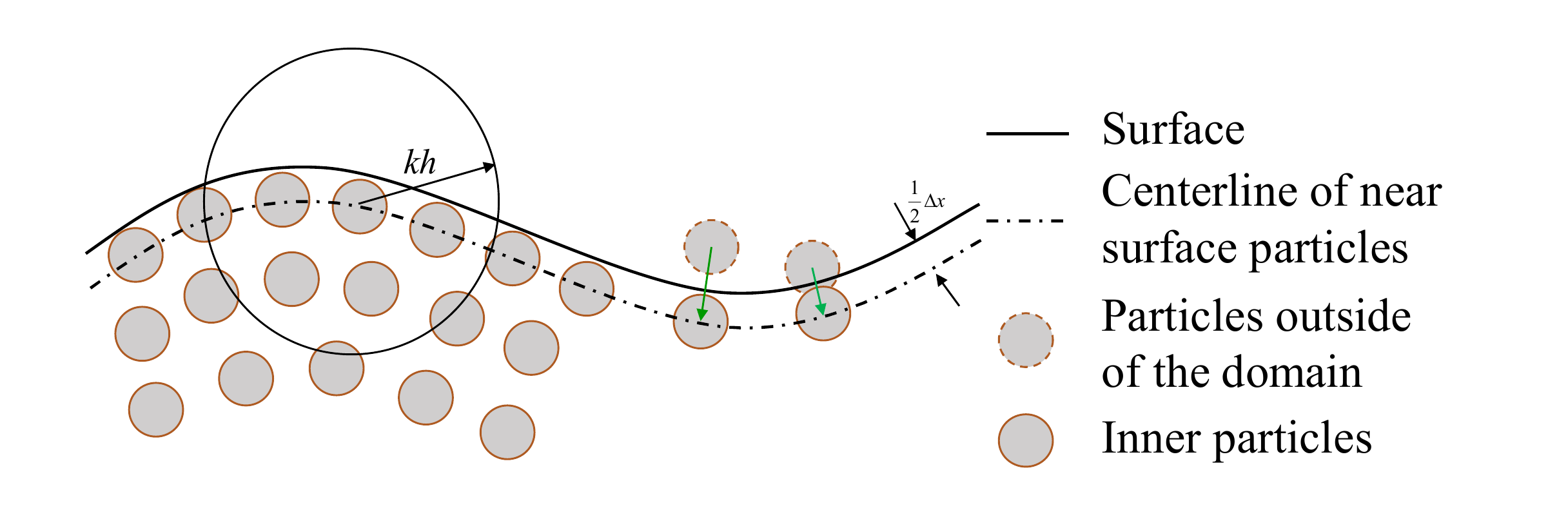}
	\caption{An illustration of surface particles bounding.}
	\label{figs:surface-bounding}
\end{figure*}

To achieve the body-fitted feature,
a suitable boundary condition treatment is required. 
In Ref. \cite{fu2019isotropic},
a dynamic ghost-particle method enforcing symmetry conditions at all domain boundaries is adopted.
However, 
it is challenge to construct ghost particles for complex geometries.
Zhu et al. \cite{zhu2021cad} proposed a simple surface particle bounding method.
Specifically,
the level-set value $\phi_i$ and normal direction $\mathbf N_i$ of each particle are first interpolated from the background mesh using trilinear interpolation.
Then, particles position is updated according to 
\begin{equation}\label{constrain}
\mathbf r_i= 
\begin{cases}
\mathbf r_i -  \left( \phi_i + \frac{1}{2} \Delta x \right)  \mathbf N_i  \quad  &\phi_i \ge - \frac{1}{2} \Delta x,\\
\mathbf r_i  \quad  &\text{otherwise}, \\
\end{cases}
\end{equation}
where $\Delta x$ denotes the initial particle spacing.
Figure \ref{figs:surface-bounding} presents the illustration of surface particles bounding.
When the particle locates outside of the geometry,
it will be enforced back on the surface following the normal direction and 
a body-fitted particle distribution can be achieved accordingly. 
Note that, 
the surface particles are relocated at $\phi = -\frac{1}{2}\Delta x $ 
instead of $\phi = 0 $ implying that the material interface assumed to be located at $\phi = 0 $ . 

\begin{figure*}
	\centering
	\begin{subfigure}{0.8\textwidth}
		\includegraphics[trim = 1mm  15cm 1mm  1mm, clip, width=\textwidth]{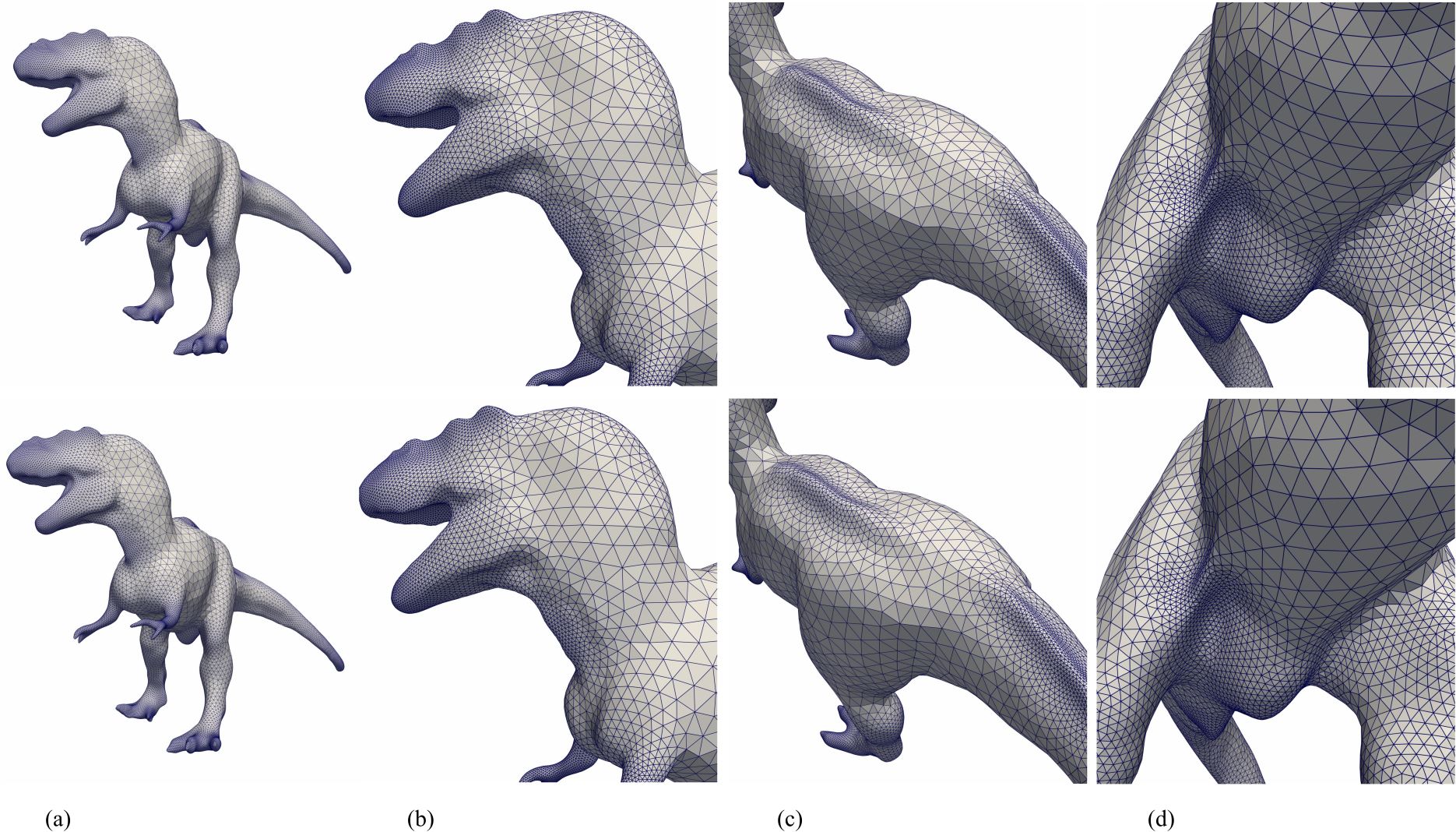}
		\includegraphics[trim = 1mm  15cm 1mm  5mm, clip, width=\textwidth]{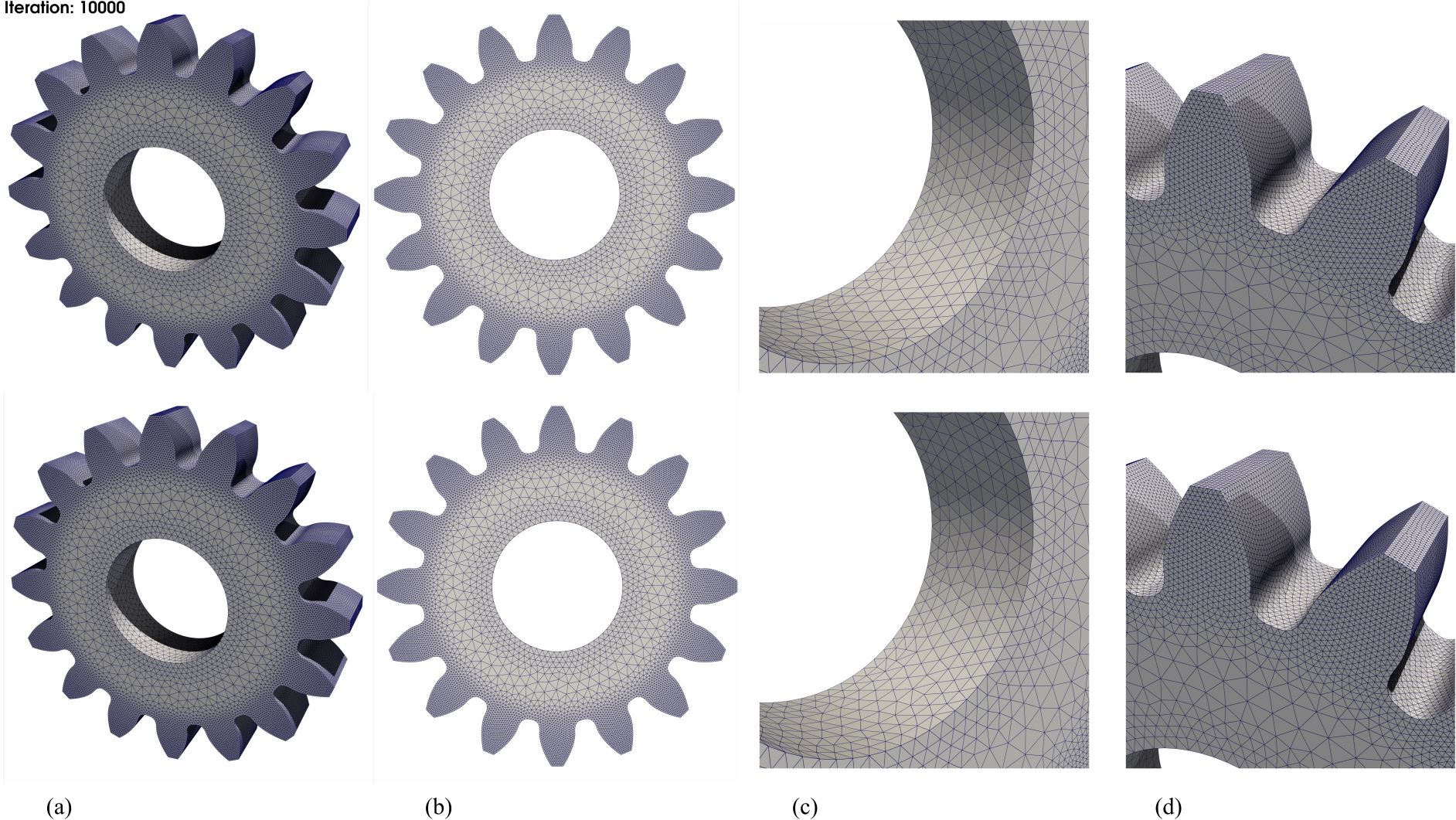}
		\caption{Mesh generation.}
		\label{figs:meshgeneration}
	\end{subfigure}
	\newline
	\begin{subfigure}{0.8\textwidth}
		\includegraphics[trim = 1mm 2mm 1mm 1mm, clip,width=.45\textwidth]{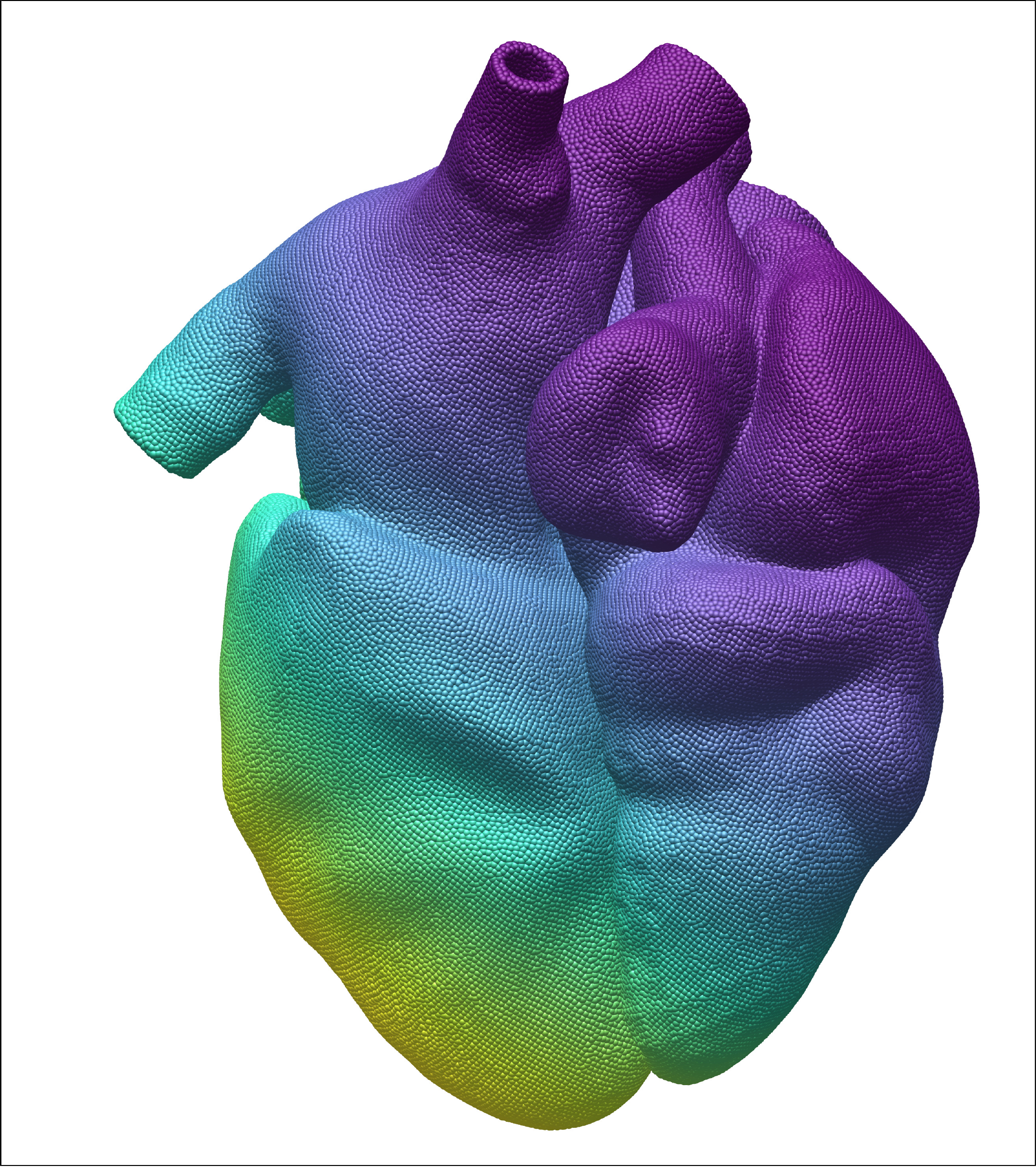}
		\includegraphics[trim = 1mm 2mm 1mm 1mm, clip,width=.45\textwidth]{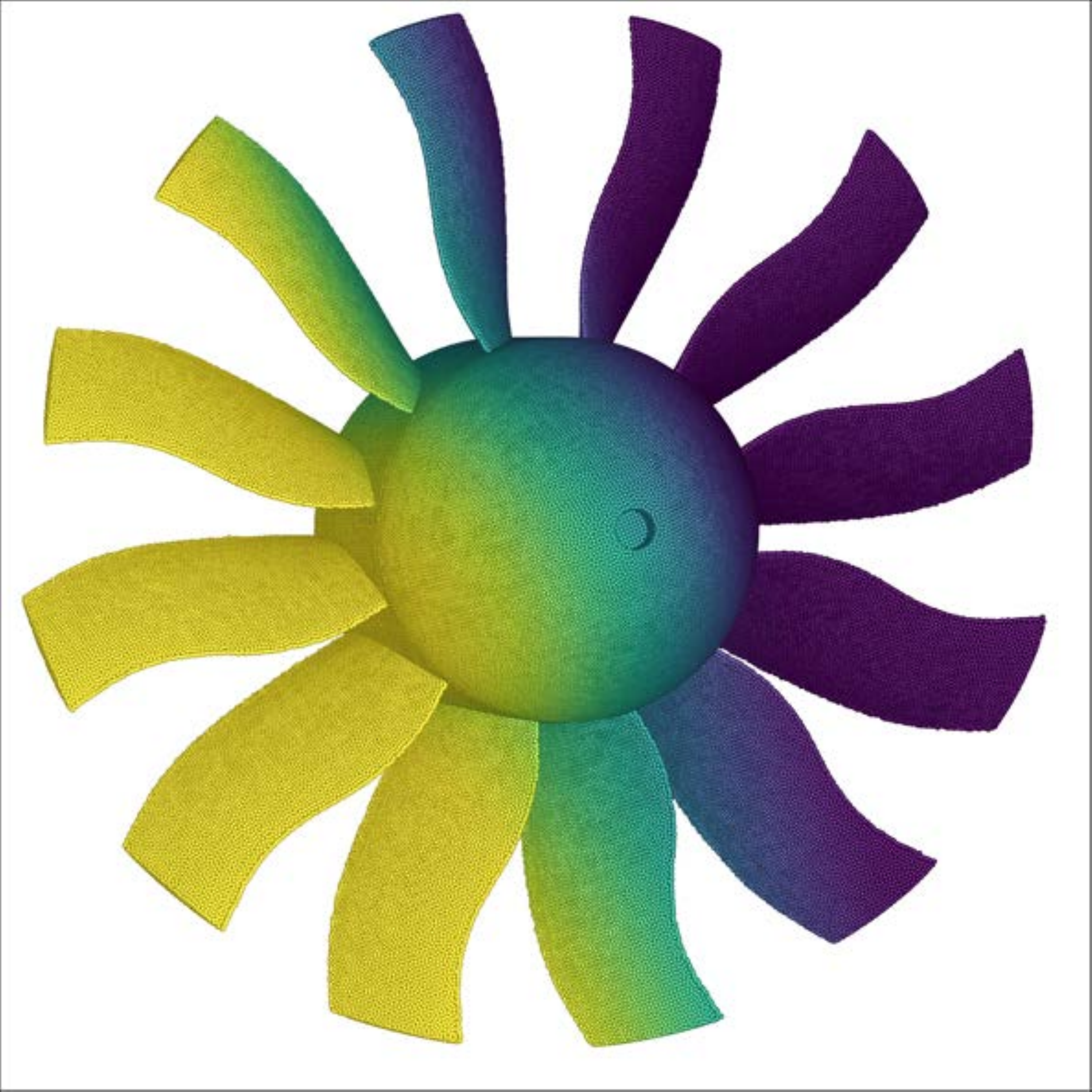}
		\caption{Particle generation.}
		\label{figs:particlegeneration}
	\end{subfigure}
	\caption{Mesh and particle generations for three-dimensional body with complex geometry. 
		(a) Delaunay triangulation of the unstructure for tyra and tear body \cite{ji2021feature}. 
		(b) Particle distribution of anotomy heart and propeller \cite{zhu2021cad}.}
	\label{figs:meshandparticle}
\end{figure*}

Figure \ref{figs:meshandparticle} portrays the generated unstructure mesh for tyra and gear body, 
and particle model for anotomy heart and propeller. 
It is obvious that the particle-relaxation generates high-quality globally optimized adaptive isotropic meshes 
and well-regularized particle distribution for three-dimensional body with high geometric complexity.
%
%
%
\section{Conclusion}\label{sec:conclusion}
In this paper, 
we present a concise review of SPH method on methodology development and recent achievement 
with highlights of aspects including numerical algorithms for fluid dynamics, solid mechanics 
and FSI, and novel applications in mesh and particle generations. 
Fundamentals and theory of the SPH method are first summarized. 
Recent developments of Riemann-based SPH method are presented with key aspects of 
Riemann-solver with dissipation limiter and high-order data reconstructions with 
MUSCL, WENO and MOOD schemes. 
Techniques for particle neighbor searching and efficient update of particle configuration 
are recalled. 
Concerning the total Lagrangian formulations, 
stablized schemes, steady state solution and hourglass control algorithms 
are reported. 
For FSI coupling, 
treatments of FSI interface and discretization schemes in multi-resolution scenario 
are surveyed. 
Last but not least, 
recent novel SPH applications in mesh and particle generations are reviewed. 
Abundant validations and benchmark test for demonstrating the computational accuracy, convergence,
efficiency and stability are also supplied in this survey. 
%
%
\section*{Declaration of competing interest }
The authors declare that they have no known competing financial interests 
or personal relationships that could have appeared to influence the work reported in this paper.
%
%
\section*{Acknowledgements}
C. Zhang and X.Y. Hu would like to express their gratitude to Deutsche Forschungsgemeinschaft (DFG) 
for their sponsorship of this research under grant number DFG HU1527/12-4. 
%
%
\bibliography{mybibfile}
%
%
\end{document}